\documentclass[]{aastex631}
\usepackage{graphicx}
\usepackage{subfigure}
\usepackage{multirow}
\usepackage{booktabs}
\usepackage{xcolor}
\usepackage{hyperref}
\shorttitle{The multi-layer Nature of Molecular Gas toward the Cygnus Region}
\shortauthors{Mwisp collaboration et al.}
\graphicspath{{./}{figures/}}
\begin{document}

\title{The Multilayer Nature of Molecular Gas toward the Cygnus Region}

\author[0009-0002-2379-4395]{Shiyu Zhang}
\affiliation{Purple Mountain Observatory and Key Laboratory of \\
Radio Astronomy, Chinese Academy of Sciences, Nanjing 210034, \\
People’s Republic of China}
\affiliation{School of Astronomy and Space Science, University of \\
Science and Technology of China, 96 Jinzhai Road, Hefei 230026, \\
People’s Republic of China}
%\footnote{syzhang@pmo.ac.cn}

\author[0000-0002-0197-470X]{Yang Su$^\dagger$}
\affiliation{Purple Mountain Observatory and Key Laboratory of \\
Radio Astronomy, Chinese Academy of Sciences, Nanjing 210034, \\
People’s Republic of China}
\affiliation{School of Astronomy and Space Science, University of \\
Science and Technology of China, 96 Jinzhai Road, Hefei 230026, \\
People’s Republic of China}
%\collaboration{6}{(AAS Journals Data Editors)}
%\footnote{yangsu@pmo.ac.cn}
\author[0000-0003-3151-8964]{Xuepeng Chen}
\affiliation{Purple Mountain Observatory and Key Laboratory of \\
Radio Astronomy, Chinese Academy of Sciences, Nanjing 210034, \\
People’s Republic of China}
\affiliation{School of Astronomy and Space Science, University of \\
Science and Technology of China, 96 Jinzhai Road, Hefei 230026, \\
People’s Republic of China}

\author[0000-0001-8060-1321]{Min Fang}
\affiliation{Purple Mountain Observatory and Key Laboratory of \\
Radio Astronomy, Chinese Academy of Sciences, Nanjing 210034, \\
People’s Republic of China}
\affiliation{School of Astronomy and Space Science, University of \\
Science and Technology of China, 96 Jinzhai Road, Hefei 230026, \\
People’s Republic of China}

\author[0000-0003-4586-7751]{Qingzeng Yan}
\affiliation{Purple Mountain Observatory and Key Laboratory of \\
Radio Astronomy, Chinese Academy of Sciences, Nanjing 210034, \\
People’s Republic of China}

\author[0000-0003-2549-7247]{Shaobo Zhang}
\affiliation{Purple Mountain Observatory and Key Laboratory of \\
Radio Astronomy, Chinese Academy of Sciences, Nanjing 210034, \\
People’s Republic of China}

\author[0000-0002-3904-1622]{Yan Sun}
\affiliation{Purple Mountain Observatory and Key Laboratory of \\
Radio Astronomy, Chinese Academy of Sciences, Nanjing 210034, \\
People’s Republic of China}
\affiliation{School of Astronomy and Space Science, University of \\
Science and Technology of China, 96 Jinzhai Road, Hefei 230026, \\
People’s Republic of China}

\author[0000-0003-2536-3142]{Xiaolong Wang} 
\altaffiliation{Physics Postdoctoral Research Station at Hebei Normal University}
\affiliation{Department of Physics, Hebei Normal University, \\
Shijiazhuang 050024, 
People’s Republic of China} 

\author[0000-0003-1714-0600]{Haoran Feng}
\affiliation{Purple Mountain Observatory and Key Laboratory of \\
Radio Astronomy, Chinese Academy of Sciences, Nanjing 210034, \\
People’s Republic of China}
\affiliation{School of Astronomy and Space Science, University of \\
Science and Technology of China, 96 Jinzhai Road, Hefei 230026, \\
People’s Republic of China}

\author[0000-0002-8051-5228]{Yuehui Ma}
\affiliation{Purple Mountain Observatory and Key Laboratory of \\
Radio Astronomy, Chinese Academy of Sciences, Nanjing 210034, \\
People’s Republic of China}

\author[0000-0002-6388-649X]{Miaomiao Zhang}
\affiliation{Purple Mountain Observatory and Key Laboratory of \\
Radio Astronomy, Chinese Academy of Sciences, Nanjing 210034, \\
People’s Republic of China}

\author[0000-0002-7413-7574]{Zi Zhuang}
\affiliation{Purple Mountain Observatory and Key Laboratory of \\
Radio Astronomy, Chinese Academy of Sciences, Nanjing 210034, \\
People’s Republic of China}
\affiliation{School of Astronomy and Space Science, University of \\
Science and Technology of China, 96 Jinzhai Road, Hefei 230026, \\
People’s Republic of China}

\author[0000-0003-2418-3350]{Xin Zhou}
\affiliation{Purple Mountain Observatory and Key Laboratory of \\
Radio Astronomy, Chinese Academy of Sciences, Nanjing 210034, \\
People’s Republic of China}

\author[0000-0003-0849-0692]{Zhiwei Chen}
\affiliation{Purple Mountain Observatory and Key Laboratory of \\
Radio Astronomy, Chinese Academy of Sciences, Nanjing 210034, \\
People’s Republic of China}

\author[0000-0001-7768-7320]{Ji Yang}
\affiliation{Purple Mountain Observatory and Key Laboratory of \\
Radio Astronomy, Chinese Academy of Sciences, Nanjing 210034, \\
People’s Republic of China}

\correspondingauthor{Yang Su} 
\email{yangsu@pmo.ac.cn}



\begin{abstract}
%On the age of Gaia, with large sky coverage unbiased survey in three CO isotopes, we work on the long term difficulties of near-far ambiguities entangled by crowded velocities and distance biases. Cygnus is a famous star formation region with a bulk of molecular gas and large extinction. Further decomposition is necessary to reveal large scale structures in this direction. We do Gaussian decomposition using $\rm ^{13}CO$ data to the large complexes, and take advantage of parallaxes and G band extinction from Gaia DR3, measure the distance of identified structures based on background extinction elimination (BEEP). An uniform $\rm ^{13}CO$ traced gas catalogue is derived, based on which multiple layers of clouds has been found toward Cygnus. 
%A new view of the Cygnus X region as well as foreground Cygnus Rift is presented. 

We study the physical properties and 3D distribution of molecular clouds (MCs) toward the Cygnus region using the MWISP CO survey and Gaia DR3 data. Based on Gaussian decomposition and clustering for $\rm ^{13}CO$ lines, over 70\% of the fluxes are recovered. With the identification result of $\rm ^{13}CO$ structures, two models are designed to measure the distances of the molecular gas in velocity crowding regions. 
%aimed at distance measurement to molecular gas in velocity crowding regions. 
The distances of more than 200 large $\rm ^{13}CO$ structures are obtained toward the 150 $\rm deg^{2}$ region. Additionally, tens of the identified MC structures coincide well with masers and/or intense mid-IR emission. We find multiple gas layers toward the region: (1) the extensive gas structures composing the Cygnus Rift from 700 pc to 1 kpc across the whole region; (2) the $\sim$ 1.3 kpc gas layer mainly in the Cygnus X South region; and (3) the 1.5 kpc dense filament at the Cygnus X North region and many cometary clouds shaped by Cygnus OB2. We also note that the spatial distribution of young stellar object candidates is generally consistent with the molecular gas structures. 
%as well as the cometary MCs directly shaped by Cygnus OB associations. 
The total molecular mass of the Cygnus region is estimated to be
%$\sim 1.13\times10^{6}~M_{\odot}$ by LTE, and 
$\sim 2.7\times10^{6}~M_{\odot}$ assuming an X-factor ratio $X_{\rm CO} = 2 \times 10^{20} \rm cm^{-2} (K\cdot km\cdot s^{-1})^{-1}$. The foreground Cygnus Rift contributes $\sim$25\% of the molecular mass in the whole region. Our work presents a new 3D view of the MCs' distribution toward the Cygnus X region, as well as the exact molecular gas mass distribution in the foreground Cygnus Rift. 
\end{abstract}


%% subSection for cross-referencing in a \ref command.
%% cross-references to Sections, equations, tables, and figures.

\section{Introduction} \label{sec:intro}

CO surveys are of great importance and helpful for studying MCs directly and coordinating Galactic emission at multiple wavelength bands \citep{2015ARA&A..53..583H}. In particular, stars are born in the densest parts of MCs, therefore, studies of MCs can accelerate our understanding of the link between star formation and the surrounding molecular gas environment, as well as the large scale structures of the Milky Way \citep{2001ApJ...547..792D, 2017A&A...601A.124S, 2017PASJ...69...78U}. 

As one of the most massive nearby star formation regions (SFRs) \citep{2008hsf1.book...36R}, the Cygnus region harbors giant MC complexes \citep[e.g., DR21,][]{2010A&A...520A..49S, 2022ApJ...927..106C} and several OB associations \citep[e.g., the well-known Cygnus OB2,][]{1991AJ....101.1408M, 2000A&A...360..539K, 2002A&A...389..874C, 2003ApJ...597..957H, 2010ApJ...713..871W, 2015MNRAS.449..741W}. 
%which includes over 100 O stars and thousands of B stars. 
With Cygnus OB2 in the center, Cygnus X region is divided into the northern and southern parts by \citet{2006A&A...458..855S}. 
Hundreds of OB stars toward this region indicate intense star formation activity therein, including extended ionized features from HII regions and supernova remnants \citep[SNRs;][]{1991A&A...241..551W, 2014ApJS..212....1A}, interstellar bubbles \citep{1981ApJ...250..645A, 1994A&A...291..295H}, and outflows \citep{2013A&A...558A.125D, 2020ApJS..248...15Z}. 

We focus on molecular gas in the whole Cygnus region, especially the gas emission within 1 kpc; including Cygnus X \citep{2006A&A...458..855S}, North America/Pelican \citep[NAP;][]{2014AJ....147...46Z}, and other interesting regions. The Cygnus X region is prominent by its strong and extended Galactic radio continuum emission \citep{1966ApJ...144..937D, 1991A&A...241..551W}. The multiple wavelength studies are also fruitful, e.g. CO surveys \citep{1987ApJ...322..706D, 2006A&A...458..855S, 2012A&A...541A..79G, 2018ApJS..235....9Y}, CII \citep{2023ApJ...951...39B, schneider_ionized_2023}, millimeter continuum \citep{2007A&A...476.1243M, 2010A&A...524A..18B}, and infrared \citep{1998ApJ...494L.199E, 2010ApJ...720..679B, 2012A&A...543L...3H, 2016A&A...591A..40S, 2019ApJS..241....1C}. Recently, many new works have also been presented toward the region \citep[e.g.,][]{2019ApJ...883..156T, 2020A&A...644A..62C, 2021A&A...651A..87O, 2021MNRAS.508.2370Q, 2022MNRAS.515..687Q, 2022A&A...665A..63B, 2022A&A...658A.166D, 2023A&A...678A.130G}. 
%all looking deep inside the Cygnus X area. 
%Using high resolution sub-millimetre data, \citet{2022ApJ...927..106C} and \citet{ 2023arXiv230510795L}  study hyperfine filamentary structures of DR21. 
The NAP region is located adjacent to the massive star-forming regions of Cygnus X in projection. L935 is the densest dark cloud in this region, which separates the North America and Pelican nebulae (see Figure \ref{fig:fig1}). Evidence of the star formation process across the complex has been revealed by studying MCs and young stellar objects \citep[YSOs;][]{2014AJ....147...46Z, 2020ApJ...904..146F}. 

Although the Cygnus region has been extensively studied in multiwavelengths, the distances and properties of clouds in the region are not well determined in a global view. Distance uncertainty is one of the major problems in studying the various star forming regions and gas structures/properties in this direction.  

The distances of MCs have been subject to considerable debate with several difficulties. Firstly, it is hard to derive the kinematic distances of clouds because of the near--far distance ambiguities in the first quadrant \citep{2021A&A...655A..64M}. 
In fact, the kinematic distances of MCs produce a large uncertainty when the velocity field of gas is crowding in a small velocity range \citep{2020ApJ...898...80Y}. 
On the other hand, due to the large degree of overlap between the inner and outer Galaxy along the line of sight (LOS), coherent MCs are difficult to distinguish in position--position--velocity (PPV) space. 
%Cause near-far ambiguities within the solar circle \citep{mertsch_bayesian_2021}, which aggravates biases when applying kinematic distance. 
In particular, toward the Cygnus X region, a collective cloud emission across the tangential point overlaps with each other from several hundred parsecs to 2 kpc and even farther in LSR velocities close to zero and velocity gradient smaller than the typical velocity dispersion of interstellar gas \citep{2008hsf1.book...36R}. 

Some methods have been used to deal with the kinematic distance ambiguities. Different layers may be demonstrated based on the HI self-absorption features toward the Cygnus X \citep{2012A&A...541A..79G}. The distances of some massive star formation regions (MSFRs) have been precisely determined by the trigonometric parallax of the associated masers \citep{2012Rygl, 2016SciA....2E0878X}. These methods are limited by the small amount of observed samples and can only focus on some small specific areas, thus lacking an overall understanding of the whole region. 

%(->)The reddening from interstellar medium(dust, atomic, molecular and ionized gas) to stars allows researchers to determine the distance of clouds based on spectroscopic studies. In the presence of dense molecular gas and dust, the luminosity of stars will decrease. The extinctions of the star can be obtained by fitting the color excess of different bands, while their distances are properly inferred from parallaxes. 

%(->)The indirect method measuring distance is suffered from the size of stars samples, the precision of parallaxes, and severely affected by the high extinction in some regions. 
Before Gaia’s age, large distance uncertainties of MCs by photometric methods prevented us from determining the exact 3D distribution of molecular gas.
Thanks to the release and convenient acquisition of Gaia DR3 data, huge amounts of stars with parallaxes and extinctions make it possible to accurately measure the distance of MCs. For example, all sky extinction maps are derived by different models based on dust properties \citep{
2019MNRAS.483.4277C, 2019ApJ...887...93G, 2019A&A...625A.135L, hottier_fedred_2020}. In addition, %sacrificing angular resolution for increased distance accuracy towards the MCs \citep{zucker_solar_2022}, 
extinction jump models are applied to distance measurements of MCs \citep{2019A&A...624A...6Y, 2019ApJ...879..125Z, 2021ApJS..256...46S, 2022MNRAS.511.2302G}, and SNRs \citep{2019MNRAS.488.3129Y, 2020ApJ...891..137Z}. 
Toward the Cygnus X region, \citet{2021MNRAS.502.6080O} revealed that the OB2 association consists of several substructures, ranging from 1.2 to 1.7 kpc based on Gaia DR2 data. 
The above evidences suggest that the gas and the SFRs toward this direction are chance superpositions of several complexes along the LOS.

%(see detailed results in Section 4.3.1-4.3.3).

Finally, ultrahigh energy (UHE) cosmic rays' emissions are prevalent in our Galaxy \citep{cao_ultrahigh-energy_2021, 2022ApJ...931L..30B}. 
%also indicates that intense physical processes are ongoing in this region, which is a natural experimental site for high energy astrophysics \citep{cao_ultrahigh-energy_2021, banik_probing_2022, 2023arXiv230517030C}. 
Several UHE sources are detected toward the Cygnus X region \citep{2024ApJS..271...25C}. The origin of these UHE sources is still unknown. Are they from hadronic processes from the interaction of high velocity winds of massive stars (e.g. Cygnus OB2) and/or SNR shock (e.g. $\gamma$-Cygni) with the surrounding dense molecular gas? 
%the observed high energy photons of gamma rays were accelerated. 
Evaluating the molecular gas distribution toward the Cygnus region might be helpful to reveal the possible origin of UHE emission. 

This paper is structured as follows. 
We will introduce the Milky Way Imaging Scroll Painting (MWISP) CO survey and the Gaia DR3 in Section \ref{sec:data}. The data processing methods are described in detail, including Gaussian decomposition in Section \ref{subsec:gaussian decomposition} and the clustering algorithm in Section \ref{subsec:clustering}. In Section \ref{subsec:distance result}, we determined the distances and physical properties of identified $\rm ^{13}CO$ structures. Based on our new distance measurement of MCs, we introduce the physical properties and the 3D distribution of gas structures for several subregions in Section \ref{subsec:subregions}, and later, in Section \ref{subsec:discuss}, we mainly focus on molecular gas in the Cygnus Rift, and discuss the big picture of MCs in different layers and estimate the total molecular mass. Finally, we give our summary in Section \ref{subsec:sum}. More details of techniques can be seen in Appendixes \ref{app:appa}--\ref{app:appf}.

\section{Data} \label{sec:data}

\subsection{CO data}
The MWISP \citep{2019ApJS..240....9S} project is an ongoing CO survey by using the PMO 13.7 m millimeter-wavelength telescope at Delingha, China.  
The survey observes ${}^{12}\mathrm{CO}$, ${}^{13}\mathrm{CO}$, and $\rm C^{18}O$ simultaneously toward the northern Galactic plane. The first epoch (MWISP I) has been completed from 2011 to 2021 for the whole region of $\rm 10^{\circ} \leqslant $ $l$ $ \leqslant 230^{\circ}$,  $|b| \leqslant5^{\circ}$. The MWISP II is launching toward $\rm 5^{\circ} \leqslant$ $|b| \leqslant10^{\circ}$. 

The PMO 13.7 m telescope uses the $3\times 3$ multi-beam side band-separating Superconducting Spectroscopic Array Receiver system \citep[see details in][]{2012ITTST...2..593S}.
Briefly, the total bandwidth of the receiver is 1 GHz with 16,384 channels, providing a frequency interval of 61 kHz and covering a velocity range of 2700 $\rm km~s^{-1}$. The channel separations (rms level) of the data are $\rm 0.158~km~s^{-1}$ ($\sim$ 0.48 K) for $\rm ^{12}CO$ and $\rm \sim0.166~km~s^{-1}$ ($\sim$ 0.25 K) for $\rm ^{13}CO$ and $\rm C^{18}O$, respectively. 
With moderate spatial resolution ($\sim 51^{''}$), the three-dimensional (3D) FITS data cubes of each cell ($30^{'}\times30^{'}$) were made with a grid spacing of $30^{''}$.

A preliminary analysis on the noise characteristics \citep{2021RAA....21..304C} has been done to increase the quality of data, including removing bad channels, decreasing edge effects, and correcting baseline distortion.

In this paper, the data cube to Cygnus is clipped in the following range: $l$, $72^{\circ} \sim 87^{\circ}$; $b$, $  -5^{\circ}.1 \sim 5^{\circ}.1$; $v$, $-100 \sim 50$ $\rm km~s^{-1}$. We resampled the velocity channels of all three lines to 0.2 $\rm km~s^{-1}$ (the corresponding rms level is 0.42 K for $\rm ^{12}CO$, 0.22 K for $\rm ^{13}CO$ and $\rm C^{18}O$) to reduce the bias in data processing. The resultant data are in  $-45 \sim 40$ $\rm km~s^{-1}$, covering the major emission of Cygnus (see Figure \ref{fig:fig1} and \ref{fig:fig2}). A dataset of $\rm ^{12}CO$ and $\rm ^{13}CO$ used in this work are available at \href{https://doi.org/10.57760/sciencedb.16716}{doi: 10.57760/sciencedb.16716}. 

\subsection{Gaia DR3}
We use the photometric data and parallaxes from Gaia DR3, which was released in 2022 June 13 \citep{2023A&A...674A..27A, 2023A&A...674A..26C,2023A&A...674A...2D}.
In total, 1.8 billion objects have source classification and probabilities in DR3. 
%As an upgrade of astrophysical parameters, the DR3 data have especially for extinction, from Gaia EDR3 \citep{2022A&A...658A..91A}. 
Over 470 million sources have astrophysical characterizations from General Stellar Parameterizer from Photometry (GSP-Phot) results for apparent magnitude $G$ $\leqslant$ 19. In our CO coverage, 1,730,905 stars are included with a parallax over error larger than 5. The stars with very small $A_{G}$ errors ($\leqslant$ 0.01 mag; see details in Section \ref{subsec:undertaintyb}) are discarded. Finally, 1,362,839 stars are used for following MC distance measurement. 

Assuming a constant $R_{0}$ = 3.1, the interstellar extinction was applied to the model grid according to \citet {1999PASP..111...63F}. We take $A_{G}$ from the GSP-Phot results. The reddened spectral energy distributions are integrated over Gaia $G$ passband, and the derived magnitude can be compared to the corresponding value without extinction \citep{2023A&A...674A..27A}. For estimating the geometric distances from parallaxes, we use a simple Monte Carlo sampling to inverse the parallaxes. Statistical comparisons of distances with different methods have been done by following the work of \citet{2018A&A...616A...9L}. We find all methods give consistent results (see Appendix \ref{app:appc} for detailed analysis).

\section{Cloud identification} \label{sec:cloud identification}
MCs often display extended and irregular morphology. 
%They are clumpy and fragmented with hierarchical structures. 
Some clouds show intricate hierarchical morphology, characterized by filamentary networks and clumpy structures. So how to describe or define the structure of an MC needs to be explored. 

Many agglomerative clustering algorithms have been proposed and applied to construct MC samples from various CO surveys. 
The Dendrogram \citep{2016ApJ...822...52R} is applied to CO data from CfA-Chile 1.2m survey \citep{2001ApJ...547..792D}. SCIMES is used in various CO lines surveys, for example, $\rm ^{12}CO$ (1--0) from the MWISP \citep{2021ApJS..254....3M}, $\rm ^{12}CO$ (3--2) from the James Clerk Maxwell Telescope CO (3--2) High-Resolution Survey \citep{2019MNRAS.483.4291C}, and $\rm ^{13}CO$ (2--1) from the SEDIGISM \citep{2021MNRAS.500.3027D}. Recently, \citet{2021A&A...645A.129Y} provided a catalog of MCs based on DBSCAN and $\rm ^{12}CO$ data in the MWISP. 

Different from the above methods, \citet{2017ApJ...834...57M} used a hierarchical cluster identification method to extract MC samples based on Gaussian decomposition. \citet{2013A&A...554A..55H} characterized $\rm C^{18}O$ velocity coherent components in PPV space by using the friends in velocity algorithm. And  \citet{2019MNRAS.485.2457H} used agglomerative clustering to organize nested structures (ACORNS). Due to the complicated velocity structures in Cygnus, we adopt the ACORNS algorithm based on Gaussian decomposition, and apply it to extract MC structures based on the $\rm ^{13}CO$ emissions from the MWISP survey (Section \ref{subsec:clustering}).   
%the new strategies is more suitable to cloud identification, we will discuss it in Section 3.2. 

\subsection{Gaussian decomposition} \label{subsec:gaussian decomposition}

Gaussian decomposition has been widely applied to the fitting of various spectral lines, e.g., $\rm HI$ \citep{2019A&A...626A.101M, 2020ApJ...902..120P};  
$\rm ^{12} CO$ \citep{2017ApJ...834...57M}; $\rm ^{13} CO$ \citep{2020A&A...633A..14R}; $\rm C^{18}O$ \citep{2018MNRAS.479.1722C}; 
$\rm HNCO$ \citep{2019MNRAS.485.2457H}; $\rm OH$ \citep{2021ApJ...923..261P}; $\rm NH_{3}$ \citep{2020ApJ...892L..32S}, etc.
There are several manual tools to fit the spectra, such as Pyspeckit \citep{2011ascl.soft09001G, 2022AJ....163..291G}, ScousePy \citep{2019MNRAS.485.2457H}, etc.
And automatic fitting methods are also developed, such as GaussPy+ \citep{2019A&A...628A..78R},  Amoeba \citep{2021ApJ...923..261P}, and other methods \citep[e.g.,][]{2020ApJ...892L..32S}.  

Manual fitting methods are often applied to small data sets \citep{2019MNRAS.485.2457H}, especially with prior understanding of the data. 
However, once the amount of data becomes large, it is really a time-consuming work for manual fitting. Additionally, the fitting results might be biased because of human factors. Therefore, automatic fitting methods are better suited for large-scale data analysis, such as the MWISP survey. 

In the case of sufficient computing resources, the fitting results can be quickly obtained. Then, one or more sets of fitting parameters for each pixel are obtained, such as centroid velocity $v_{0,i}$, full width at half-maximum (FWHM=$\sqrt{8\log(2)}\sigma_{i}$), and peak intensity $A_{i}$. 
In this way, the spectral line data of each pixel can be described by several parameters, leading to a clear presentation of line intensities and spatial relations among different velocity components. For example, \citet{2020NatAs...4.1064H} has obtained the velocity structure of molecular gas at different scales by fitting the centroid velocities of multiple lines, and found the fluctuation and periodicity everywhere. Without spectral fitting, such interesting results cannot be obtained from the original PPV data due to the effects of line broadening and blending velocity components. 

In this work, we use GaussPy+ module proposed and improved by \citet{2019A&A...628A..78R} to fit and decompose the $\rm ^{13} CO$ lines. 
The $\rm ^{13} CO$ data are chosen because it is usually optically thin \citep[e.g.,][]{2023AJ....165..106W, 2023ApJ...958....7Y} and more favorable to trace the inner structure of the MCs. Moreover, \citet{2022ApJS..261...37Y} shows that the $\rm ^{13}CO$ emitting area of a large-scale cloud can cover $\sim$30\% of the corresponding $\rm ^{12}CO$ area. It indicates that $\rm ^{13}CO$ is actually a good tracer to reveal large MCs \citep[also see][]{2020ApJ...893...91S, 2023AJ....165..106W}. In the same time, $\rm ^{13} CO$ spectral lines are often relatively velocity separated compared to the $\rm ^{12} CO$ line profile for different MCs. All these features can largely mitigate overfitting problems. 

The study on giant MCs in Orion B \citep{2018A&A...610A..12B} proves that $J$ = 1--0 lines of three isotopologues of CO are good at revealing distinct density regimes. In a data-driven approach, \citet{2021A&A...645A..27G} found that the $\rm ^{13} CO$ line from Orion B cloud is effective for the estimation of the column density in various environments, especially for translucent gas ($2 \leqslant Av \leqslant 5$). For clouds traced by $\rm ^{13} CO$ emission, its extinction is neither too large nor too small. Actually, the MWISP survey shows that $\rm ^{13} CO$ emission can trace the molecular gas in a range of $ N_{\rm H2} \sim3 \times 10^{21}$ $\rm cm^{-2}$ to several $ 10^{22}$ $\rm cm^{-2}$ \citep{2021ApJS..254....3M, 2023AJ....165..106W}. As a result, it is helpful for the distance measurement based on MC samples identified from $\rm ^{13} CO$ emission. Finally, different molecular structures can be well distinguished for velocity-crowded regions using $\rm ^{13} CO$ data. 

Based on our experiments, we adopt the following parameters:

\begin{enumerate}
\item $Signal$-$to$-$noise$ $ratio$. We set the signal to $\rm 5~rms$ to skip those weak emission. The signals with a level  $\gtrsim 1.2$ K are remained. 
\item $Minimum$ $FWHM$. We set it to be at least 2 channels ($\rm 0.4 ~km~s^{-1}$) for the MWISP data. 
\item $Maximum$ $FWHM$. We did not set this parameter before fitting. 
%to avoid systematically smaller line widths. However, some samples with too large FWHM are discarded since they are not physical;
\item $Smoothing$ $parameter$. We take the optimal smoothing parameters \citep[i.e., $\rm \alpha_{1}=2.18$, $\rm \alpha_{2}=4.94$,][]{2020A&A...633A..14R} for the MWISP. 
\end{enumerate}

For each pixel at sky position, the brightness temperature $T_{\mathrm{B}}^{\prime}(l,b,v)$ is described as the sum of all Gaussian components, and the total intensity $W_{\rm CO}(l,b)$ as a function of velocity $v$ is in the following form:

\begin{equation}
T^{\prime}_{B}(l,b,v)=\sum^{N}_{i=1} A_{i} e^  -\frac {\left(v-v_{0,i}\right)^{2}}  {2 \sigma_{i}^{2}}
%T_{\mathrm{B}}^{\prime}(v)=\sum_{i=1}^{N} A_{i} \exp \left(\left(v-v_{i}\right)^{2} / 2 \sigma_{i}^{2}\right)  \\
\end{equation}
and 
\begin{equation}
%I(v)=\sum_{i}^{N}A_{i}\frac{1}{\sigma_{i}\sqrt{2\pi}} e^{\frac{\left(v-v_{0,i}\right)^{2}}{2\sigma_{i}^{2}}}
%I(v)=\sum_{i}^{N} A_{i} \frac{1}{\sigma_{i} \sqrt{2 \pi}} e^{\frac{\left(v-v_{0, i}\right)^{2}}{2 \sigma_{i}^{2}}}
W_{\rm CO}(l,b)=\sum_{i}^{N}\int A_{i} e^{-\frac {\left(v-v_{0,i}\right)^{2}}{2 \sigma_{i}^{2}}}\mathrm{d}v=\sqrt{2\pi}\sum_{i}^{N}A_{i}\sigma_{i}
\end{equation}
where $A_{i}$, $v_{0,i}$ and $\sigma_{i}$ are the amplitude, centroid velocity, and width of each Gaussian component, respectively.

Our data cube includes 1801 * 1225 (2,206,225) pixels, among which 17.5\% of the pixels have been fitted by GaussPy+. According to the fitted results, we can discern some MC structures by combining the pixels with a coherent single velocity component (see Figure \ref{fig:fig2}g). 
%(386310),(307973),54510,19371
79.7\% of pixels have only one velocity component along the LOS in the fitted pixels. 14.1\% pixels have two, and 5\% pixels have three components. Only 1.2\% pixels have more than three components among fitted pixels. The pixels with two components are mainly located in the boundaries of different MC structures (see Figure \ref{fig:fig2}g), showing superposition between them. The pixels with multiple components ($\geqslant3$) are located near the Galactic disk, revealing the crowded velocities therein. 

For the whole map, we apply a moment masking criteria ((1) pixels within the $\rm ^{12}CO$ emission structure and (2) pixels with three consecutive channels larger than 3 times the noise rms) to estimate the total valid flux of the raw data. The identification of $\rm ^{12}CO$ emission was following DBSCAN algorithm \citep[see details in][]{2021ApJ...922....8Y}. The total recovered flux from Gaussian reconstruction makes up 74.3\% of the flux of the raw data. 
And we reconstruct the integrated intensity map from the Gaussian fitting shown in Figure \ref{fig:fig2}a. The distribution of other Gaussian parameters (e.g., centroid velocities, velocity dispersion, fitted components numbers, and residuals between the reconstructed map and raw image) is presented in Figure \ref{fig:fig2} c--h.

\subsection{Clustering} \label{subsec:clustering}

As mentioned above, many MCs in the Cygnus region display hierarchical structures and nested velocity components along the LOS. 
And \citet{2019MNRAS.485.2457H} summarized that a single MC structure has the features to be coherent in both space, velocity, and velocity dispersion. In other words, MC structures separated by ACORNS are different from each other in physical properties and statistics. 
%which probably reflects the rationality of cloud identification result. 
For the Gaussian decomposed $\rm ^{13} CO$ data cube, ACORNS can effectively avoid line blending effects along the LOS, and successfully separate molecular gas emission toward the Cygnus region into substructures. However, the definition of MCs has been debated for decades and as indicated in some simulations \citep[e.g.,][]{2017MNRAS.472..647Z, 2018MNRAS.479.1722C}. We will give some discussions in Section \ref{subsec:ppvp}.

We made parametric tuning on the algorithm implemented in the MWISP data. In order to avoid the influence of noise and highlight the major structure, we only focus on the structures with relatively strong emission and large angular areas. As the signal is set to 5 times the noise RMS in Section \ref{subsec:gaussian decomposition}, 
the decomposition results are reliable for clustering. In order to further develop the hierarchy and to reduce overdecomposition, we specify the ``relax" step. After manually comparing the identified cloud structure with the coherent characteristics of the raw data, a group of parameters are adopted in ACORNS here:

\begin{enumerate}
\item $min\_radius$. We set it to be a bit smaller than 2 pixel. It ensures that the smallest structure in the clustering result has $\geqslant 9$ pixels. 
\item $velo\_link$. We set it to 0.2 $\rm km~s^{-1}$ based on resampled data. 
\item $dv\_link$. The line width link parameter is set to 0.4 $\rm km~s^{-1}$.
\item The coefficients of the relax step are set to [3, 2, 0.5] times the cluster criteria (as default if ``relax'' was activated). 
\item The stop criteria is set to 3. 
\end{enumerate}

Then, the algorithm further eliminates some small structures that are not merged into the adjacent large structures. Finally, the MC samples with coherent $\rm ^{13}CO$ emission are constructed based on the above steps. 
%As a result, the total flux of our final clouds table has a loss compared %to the Gaussian decomposition result, and find it about 2.5\% smaller. 

After adding another loop to the clustering process (details in Appendix \ref{app:appe}), 72.6\% of the flux is recovered compared to the raw data. It indicates that a large proportion of flux restored by both Gaussian decomposition (Figure \ref{fig:fig2}a) and clustering (Figure \ref{fig:fig2}b).  
We find that only a small fraction of emission ($\approx$ 1.7\%) fails to merge into branches. 

Different from the other methods (e.g. SCIMES, Dendrogram, DBSCAN, etc.), we mainly use the results of Gaussian fitting and the ACORNS clustering algorithm to produce the cloud table, so the definition of cloud boundary is different in the spatial projection and in the direction of velocity. 
In the plane perpendicular to the LOS, the boundary of the cloud can be determined by aggregating all pixel locations. 
In the forest of clusters, each tree corresponds to a cloud structure. For a tree with a hierarchical structure (with branches and leaves for large-scale structures), the cloud is a complex that is composed of discrete, nonoverlapping substructures. For a tree with no hierarchy, the cloud is a uniform and coherent structure. 
In either case, the positions of the outermost pixels form the boundary of the cloud in the projection plane. In the LOS direction, the 
Gaussian fitted line width of each component corresponds to the ``velocity thickness'' of the cloud. For simplicity, we take the position of the line centroid velocity $\pm3\sigma_{v}$ (see equation (9)) as the range of cloud in the velocity direction. According to the above clustering criterion, the velocity boundary of adjacent pixels is also similar, so that the cloud in PPV space has a relatively smooth contour.

\subsection{Check identified structures} \label{subsec:check}

For comparisons, we plot an average spectrum that illustrates the total emission intensity reproduced by GaussPy+ (blue line in Figure \ref{fig:fig3} ) as well as clouds clustering (red line in Figure \ref{fig:fig3} ).
The total integrated intensity reconstructed by fitting components and clustering accounts for 81.5\% and 79.5\% of flux (black line in Figure \ref{fig:fig3} ), respectively, for all fitted pixels based on Gaussian decomposition (see Section \ref{subsec:gaussian decomposition}). 
%of the raw data (see black and blue lines in Figure 2).} 
Note that the flux of all fitted pixels here is $\sim$ 10\% smaller than the flux of the whole region. 

In the Cygnus X North region, the widely studied filament DR21 with HII regions is extracted, as well as other globular structures with brightness temperatures in the vicinity (see Section \ref{subsec:cygn}). In the Cygnus X South region, a large cloud is identified with a straight clubbed body, and waterfall-like arms adjoin in its center (later on, we call it L889; see Section \ref{subsec:cygs}). 
The filamentary cloud L914 is divided into two segments because the centroid velocity turns sharply between the two MC structures (see Section \ref{subsec:l914}). As a typical subregion with relatively simple velocity components, the identified cloud of L914 after clustering restores about 77\% flux of the raw data in a moment masking method (larger than the average 72.6\%).

\subsection{Cloud parameters}

Based on cloud identification from $\rm ^{13}CO$ emission, there is only one single Gaussian component along the LOS in each coherent structure. 
For a structure with pixels number $n_{\rm pixels}$, the integrated intensity of the $j$th component $W_{\rm CO}^{\rm cloud}(l_{j},b_{j})$ = $W_{\rm CO}^{j}$. 
Here, we define intensity-weighted coordinates of identified structures, with the following: 
\begin{equation}
l_{0}=\frac{\sum_{j}W_{\rm CO}^{j}l_{j}}{\sum_{j}W_{\rm CO}^{j}}
\end{equation}  
and 
\begin{equation}
b_{0}=\frac{\sum_{j}W_{\rm CO}^{j}b_{j}}{\sum_{j}W_{\rm CO}^{j}}
\end{equation}
The total intensity of cloud is $W_{\rm CO}^{\rm cloud}=\sum_{j}W_{\rm CO}^{j}$.
The standard deviation along $l$ and $b$ is 
\begin{equation}
\sigma_{l}=\sqrt{\frac{\sum_{j}W_{\rm CO}^{j}\left(l_{j}^{2}-l_{0}^{2}\right)}{\sum_{j}W_{\rm CO}^{j}}}
\end{equation}
and 
\begin{equation}
\sigma_{b}=\sqrt{\frac{\sum_{j}W_{\rm CO}^{j}\left(b_{j}^{2}-b_{0}^{2}\right)}{\sum_{j}W_{\rm CO}^{j}}}
\end{equation}
The total brightness temperature of a cloud at $v$ is provided by summing of all pixels in the identified boundary:
\begin{equation}
T^{\rm cloud}_{B}(v)=\sum_{j} A_{j} e^  -\frac {\left(v-v_{0,j}\right)^{2}}  {2 \sigma_{j}^{2}}
\end{equation}
We then can compute the cloud's intensity-weighted mean velocity and velocity dispersion as 
\begin{equation}
v_{0}^{\rm cloud}=\frac{\int v T^{\rm cloud}_{B}(v) dv}{W_{\rm CO}^{\rm cloud}}
\end{equation}
and 
\begin{equation}
\sigma_{v}=\sqrt{\frac{\int v^{2}T^{\rm cloud}_{B}(v)dv}{W_{\rm CO}^{\rm cloud}}-{v_{0}^{\rm cloud}}^{2}}
\end{equation}
The angular size of a given cloud is straightforward, which is described in units of square arminutes: 
\begin{equation}
S_{\rm ang}=n_{\rm pixels} \delta l~ \delta b 
\end{equation}
and angular radius 
\begin{equation}
R_{\rm ang}=\sqrt{\frac{S_{\rm ang}}{\pi}}
\end{equation}
where $\delta l$, $\delta b$ is the grid size of data cube, respectively.
The peak intensity of the cloud $T_{\rm peak}$ is the maximum of derived amplitudes:
\begin{equation}
T_{\rm peak}\left(^{13} \mathrm{CO}\right)=max\{A_{1},A_{2},\cdots,A_{j},\cdots\}
\end{equation}
The column density of MCs here can be estimated by the abundance of $\rm ^{13} CO$ emission as well as the $\rm H_{2}$--$\rm ^{12}CO$ conversion factor. The first method uses a radiative transfer model under the local thermodynamic equilibrium (LTE) condition, assuming an equal excitation temperature for $\rm ^{12} CO$ and $\rm ^{13} CO$ lines, while the second method gives a direct relation between the line intensity and the $\rm H_{2}$ column density. 
For the identified $\rm ^{13} CO$ MC structures, assuming the emission is optically thin, the temperature of the cosmic microwave background radiation $T_{\rm bg} \approx$ 2.7 K. The summed column density of a given cloud can be described as follows \citep{1997ApJ...476..781B, 2009tra..book.....W}: 
\begin{equation}
N_{\rm tot}=C\times 2.42\times 10^{14} \frac{\tau(\rm ^{13} CO)}{1-e^{-\tau(\rm ^{13} CO)}}\times \frac{1+0.88\/T_{\rm ex}}{1-e^{-5.29/T_{\rm ex}}} W_{\rm ^{13} CO}^{\rm cloud}
\end{equation}
Here, $C$ is a ratio between the abundance of $\rm H_{2}$ and $\rm ^{13} CO$. We adopt $\rm 8\times  10^{5}$, from the empirical abundance ratio [$\rm H_{2}$/$\rm ^{12}CO$] = $1.1 \times 10^{4}$ in \citet{1982ApJ...262..590F} and $\rm ^{12}C$/$\rm ^{13}C$ relation from \citet{2005ApJ...634.1126M}. $\tau(\rm ^{13} CO)$ is the optical depth at the peak intensity $T_{\rm peak}$,
which is calculated by the excitation temperature $T_{\rm ex}$. The values of $T_{\rm ex}$ and $\tau_{13}$ can be written as follows \citep[see also][]{1998AJ....116..336N, 2010ApJ...721..686P, 2018ApJS..238...10L}: 
\begin{equation}
T_{\mathrm{ex}}=\frac{5.53}{\ln \left\{1+5.53 /\left[T_{\text{peak}}\left(^{12} \mathrm{CO}\right)+0.819\right]\right\}}
\end{equation}
and 
\begin{equation}
\tau_{13}=-\ln \left\{1-\frac{T_{\rm peak}}{5.29\left[1/\left(e^{5.29/T_{\rm ex}}-1\right)-0.164\right]}\right\}
\end{equation}
Finally, the average column density of each $\rm ^{13} CO$ cloud is 
\begin{equation}
N_{\rm mean}=\frac{N_{\rm tot}}{n_{\rm pixels}}
\end{equation}

\section{Distances and Properties of MCs} \label{subsec:distance result}

In the background eliminated extinction--parallax method \citep[i.e., BEEP;][]{yan_molecular_2019}, the unrelated extinction was removed to calculate distances of MCs by calibrating the stellar extinction toward MCs with the extinction of stars around them. The jump point in the $A_{G}$--$Distance$ map is detected using Markov Chain Monte Carlo (MCMC) method by a Bayesian modeling approach.  
There are five key parameters in the distance measurement: the location of the jump point ($Distance$), the extinction value of the foreground and background stars ($\mu_{1}$ and $\mu_{2}$), and the dispersion of the foreground and background extinctions ($\sigma_{1}$ and $\sigma_{2}$). A Gaussian distribution is used as the likelihood function of $A_{G}$ distribution. 
We set the jump points in 100--3000 pc with a uniform distribution, considering the limitation of Gaia's precision. The initial distance is set to the average value of the selected stars. The initial values and errors' transformation of other parameters follow \citet{2019A&A...624A...6Y}. 

We choose $emcee$ \citep{2013PASP..125..306F}, the affine-invariant ensemble sampler for our model, instead of $Pymc3$ and Gibbs samplings \citep{2019A&A...624A...6Y}. 
We also test the difference between $emcee$ and $Pymc3$ \citep{2016ascl.soft10016S} for MCMC sampling (see details in Appendix \ref{app:appb}). $emcee$ has better performance. 
%Pymc3 \citep{2016ascl.soft10016S} offers powerful sampling algorithms.  
%which uses Theano \citep{2016arXiv160502688T} to compute gradients via automatic differentiation. 
%The performance is excellent when implement to our work in most cases, 
%but not as stable as emcee (see details in Appendix B). 

\subsection{Selection of star samples}

%Following the work of \citet{yan_molecular_2019}, we take the advantage of dust extinction jump near the edge of MC structures.  We find some of the cloud samples found by DBSCAN \citep{yan_distances_2020} in the first quadrant are too large in the PPV space, especially for Cygnus region. When the boundary is not clearly and efficiently demarcated, structures include emission from others due to the overlap. As a result, the influence of other components may lead to inaccuracy of distance measurement given deviation of jump point. We will introduce our methods based on identifying $\rm ^{13} CO$ cloud structures and the idea of BEEP \citep{yan_molecular_2019}. Some improvements are made to solve the distance measurement puzzles in velocity crowding regions, with which combine the latest Gaia data release (Gaia DR3).

\citet{yan_molecular_2019} showed that the BEEP method is suitable for distance estimations of MCs at low Galactic latitudes. However, cloud identification in their procedures could introduce significant uncertainties for clouds in velocity crowding regions, especially in the first quadrant. As discussed in Section \ref{sec:cloud identification}, the cloud boundary in the projection toward Cygnus region cannot be well delineated by direct clustering (e.g., Dendrogram, DBSCAN, etc.) in PPV space. The identified structures might include emission from other clouds due to the overlap between neighboring structures in PPV space. The extent of both on-cloud and off-cloud stars is thus not well demarcated by the signal levels \citep{yan_molecular_2019} or identified boundaries \citep{2021ApJ...922....8Y}. As a result, the influence of other structures along the LOS could lead to an inaccurate distance measurement with a deviation of the jump point. We follow the procedures of BEEP, but with a different cloud identification method. Some improvements are made to solve the puzzles in the velocity crowding regions (see details in Section \ref{subsec:selectf}).

\subsubsection{Selection of on-cloud stars} \label{subsec:onstar}
After clustering based on Gaussian decomposition (see Section \ref{subsec:clustering}), we obtain coherent MC structures with exact boundaries from $\rm ^{13} CO$ emission. As seen in the work of \citet{2019A&A...624A...6Y}, the number of star samples within the cloud helps to improve the precision of the distance measurements. We thus include all stars inside the cloud boundary (see red and black area in Figure \ref{fig:fig4}).  
%with all stars selected the boundary derived from molecular cloud is made fully used, i.e., the morphological information of the cloud, to distinguish it from other neighboring structures. 

A further validation of the reliability of the selection is 
to consider those stars in the overlapping part (black region in Figure \ref{fig:fig4}). We propose that stars in the overlapping area have more of a contribution to the distance measurement than introduced uncertainties. This was also confirmed in a series of tests (see Appendix \ref{app:appb}), where we used all the on-cloud stars and another set of stars with the removal of the overlapping region. We found that most distances show good coincidence between the two scenarios. While the results from selecting all the source stars show smaller errors and more stable sampling.

\subsubsection{Selection of field stars} \label{subsec:selectf}

It is also crucial to construct samples of field stars for accurate distance measurement. That is, the selection of suitable field stars can be used for background elimination to highlight the jump caused by the target MC \citep{yan_molecular_2019}. 
It is difficult to find a clean reference region near the Galactic disk. Generally, to determine the jump point of $A_{G}$--$Distance$ map after background elimination, the extinction of field stars should be rather smaller than that of on-cloud stars, and the displacement from on-cloud stars should be close enough. That is so that reference stars can be fitted as an extinction background (hereafter ``background extinction'' means the unrelated extinction toward MCs) for on-source stars and the $A_{G}$ jump of the cloud structure can reach a threshold to be detected. 
Obviously, the above two conditions need to be balanced. So we designed Models A and B (see Figure \ref{fig:fig4}) to measure the distance of MCs. 

$Model$ $A$. To highlight the jump features in the $A_{G}$--$Distance$ map, we use the region of $\rm ^{12} CO$ emission free ($\rm ^{12} CO$ integrated intensity $\rm \leqslant 1~K~km ~s^{-1}$) as the reference region (see orange region in Figure \ref{fig:fig4}a). 
%while we make two outward expansions according to the border of the cloud, a loop %close to the molecular cloud is expanded by 5 arcmin from cloud border, %which takes into account that the distribution of molecular gas has an extension outward;
The outer boundary of the reference region (black dashed contour in Figure \ref{fig:fig4}) is limited to within $30'$ from the MC. Obviously, here, we considered the complicated features of MCs in the velocity crowding region based on $\rm ^{12} CO$ and $\rm ^{13} CO$. 
This is due to the fact that MCs with multiple velocity components are filled with $\rm ^{12} CO$ emission with which they can contribute nonnegligible extinction. On the other hand, enough field stars are needed for the accurate distance measurement. 
%Yet the position of the field stars cannot be too far from the molecular cloud in the projection, otherwise it is difficult to determine whether they have a similar extinction background with the stars on the cloud. 

This scheme allows more $\rm ^{13} CO$ MCs to do background elimination in complex regions. After correcting the field stars' effects, the distance of the cloud can be clearly identified (see examples in Figure \ref{fig:fig5}a). 
%Samples without jumps exhibit decoupling of the molecular gas from the dust distribution and also reflect the complexity of the extinction background, these samples are subsequently discarded. The drawback of this scheme also comes from the selection of field stars, shielding out the $\rm ^{12} CO$ emission makes some molecular cloud include less field stars, the fitting of the background extinction may not be reliable enough, and the field stars far from the molecular cloud boundary cannot fit the most accurate extinction background, nor do they make the most use of the molecular cloud boundary, which may cause uncertainties in the near and far distance ambiguities in the same line of sight. 
185 clouds (see Table \ref{tab:tab1}) out of 300 large-size clouds (angular sizes $\rm \geqslant 60 ~armin^{2}$) are measured by Model A. 

$Model$ $B$. Another scheme is more straightforward. We choose a ring-like area outside the $\rm ^{13} CO$ emission as the reference region. An empirical offset helps reveal the jump point. The inner and outer boundaries of the ring-like structure are $2.'5$ and $7.'5$ away from  $\rm ^{13} CO$ emission, respectively. We do not exclude field stars located within the $\rm ^{12} CO$ emission in the ring-like area (see Figures \ref{fig:fig5}b and \ref{fig:fig6}b). The number of field stars is then sufficient to capture the extinction characteristics of the background. The distance measurements from Model B are probably more reliable for some special cases (see Appendix \ref{app:appd}). 
 
%Regardless of the contamination from surrounding velocity components, if extinction from the target molecular cloud was prominent enough, after subtracting the influence of the background extinction, jumps can still be detected, thus obtain a distance with higher level of confidence. Compare with model A, this model can solve the near-far distance blurred problem on specific directions. 

Nevertheless, Model B demands that the target MC should have a higher column density than its surroundings. In other words, Model B only works when CO-dark extended gas has a small influence on the distance measurement. 
%, which result in more prominent extinction than vicinity. 
%As the background extinction revealed by model B is larger, 
Otherwise, after subtracting the background extinction, the jump value is too small to be detected. Generally, the results from Model B may be less accurate than those from Model A (see simulations in Appendix \ref{app:appb}).  
%For most of the molecular clouds with complex extinction environment can not meet the above conditions at the same time. This model rather apply to relatively isolated and large structures. 

In total, the distances of 120 MC structures from Model B are roughly consistent with those from Model A (see Table \ref{tab:tab1} and Figure \ref{fig:fig7}). Additionally, another 22 clouds are successfully measured by Model B, which are not measured by Model A.

%For individual cases, no enough field stars that shield $\rm ^{12} CO$ emission can be taken, the sample might overlap completely with the neighboring molecular clouds, i.e., model A fails, while model B can be measured successfully, which also verifies our assumptions discussed above. For most of the measured distance samples, model A give larger jump values and smaller sampling uncertainties, 

%For most of cases, distances derived from model A are accepted. We have to admit that accurate distance measurements are currently difficult and even subject to errors, because of uncertainties in the extinction values of stars, problems in the background extinction fitting method, cloud identification, and decoupling between molecular gas and dust. We expect more accurate data that providing more constraints on the distance measurements.

\subsubsection{On-cloud stars only} \label{subsec:onlyon}

As a comparison, we also measured a group of results without background elimination (see blue squares in Figure \ref{fig:fig7}). To reduce the contamination from different components along the same LOS, we remove the on-source stars in the overlapped region (from ACORNS). Since the background extinction is not subtracted, the distance measurements will be affected by other structures (traced by $\rm ^{12}CO$ emission) as well as the accumulated extinction along the LOS. So the obtained distances tend to jump to the position corresponding to the maximal $ \Delta A_{\rm G}$ (i.e., $\mu_{2}-\mu_{1}$), or to the average of multiple components. 
%We suggest for clouds that are relatively isolated, at the edge of cloud complexes, or in a large angular size, this scheme provide significant jumps with good correspondence. 
As a result, this scheme gives more distances to the MC samples, but the measurements are not accurate enough for the identified MC structures in the Galactic plane \citep{2019ApJ...879..125Z, 2021ApJS..256...46S, 2022MNRAS.511.2302G}.

%we include all on-cloud stars inside the cloud boundary in our following analysis. (<-)
%To derive an uniform clouds catalogue with measured distances in large samples, we treat the distance cutoff parameter ($D_{cut}$) following \citet{2021ApJ...922....8Y}. After that the pipe line can be started automatically based on-clouds identification algorithm in order of angular sizes. 

\subsection{Uncertainties}

\subsubsection{Uncertainty from our model}
%Molecular gas are associated with atomic gas and dust. Generally HI emission is more extensive than molecular gas with wider line width. 
%There are different distribution for multi-phase interstellar medium (i.e. dust, ionized gas) in complicated environment. Especially in the MSFRs, 
%molecular gas has been blew out by the drastic star formation activities. That is, some $\rm ^{12}CO$ free region might be consisted of extensive ionized gas.  Furthermore, there are intense HI envolope outside the ionized bubble. Both of them in which are CO free also contribute to the extinction. 

%Base on above discussions, it is much more difficult to choose off-cloud stars in complicated regions.

Two effects have a decisive influence on distance measurements of clouds. 
%(selection of on-cloud stars). 
One comes from the nested clustering algorithm. Considering the complex cloud structures, the criteria settings (see details in Section \ref{subsec:clustering}) can only extract the physical structure to some extent. On the other hand, a physical structure might be divided into different parts due to velocity gradients, variations of line widths, and nonuniform intensities. This effect can be retrieved by accurate distance measurement for overdecomposed complexes in the procedure. 
% might merge together

Another effect comes from the complicated extinction of the interstellar medium (ISM) environment. As discussed in Section \ref{subsec:gaussian decomposition}, $\rm ^{12} CO$ emission can trace more extended MC structure of the translucent molecular gas than $\rm ^{13} CO$ emission. Even though we have already considered the extinction effect of $\rm ^{12} CO$ gas in the Model A, the atomic gas and CO-dark gas are likely more space filled in the surrounding region. For the $\rm ^{13} CO$ free region, the envelope of the cloud might still contribute to the considerable extinction. We have to carefully deal with the influence from neighboring gas in complicated environment.   
%%As a result structures still show more than one jumps.
The separation of the different extinction environments cannot be thoroughly removed due to the multiphase gas structure in the ISM. Therefore, it is much more difficult to choose on-cloud and off-cloud regions traced by $\rm ^{13} CO$ emission in the complicated extinction environment. Nevertheless, our methods (Models A and B) should be better than detecting extinction jumps directly by considering the MCs' morphology of $\rm ^{13}CO$ (see Section \ref{subsec:clustering}) and background elimination (see Section \ref{subsec:selectf}).  
%so that the superposition can't be thoroughly removed.  

%with different velocity components
How do we check whether identified MC structures in the clustering are from a physical cloud? For some large-scale structures, we can measure the different subregions to explore the possible connection between them.  
In Appendix \ref{app:appa}, some subregions (boxes or contours in Figures \ref{fig:figa1} and \ref{fig:figa2}) are chosen to do distance measurements. We find that most of them are at similar distances (see lower left panel in Figures \ref{fig:figa1} and \ref{fig:figa2}).
%whether all have similar jump locations of extinction. 
The ambiguities from neighboring clouds cannot be totally removed, especially for small structures overlapping with adjacent extensive structures. 
%As examples shown in Appendix A, some selected squares and secondary trunks are contaminated by clouds from other layers. 
%The overlaps often exist in cloud edges, approves the extinction contamination from neighbours, which might not be seen in our CO line data. 

Besides the 5\%$\sim$10\% system uncertainties within 3 kpc from Gaia data, the sampling process itself also introduces measurement dispersion. As mentioned above, many effects can cause uncertainties of distance measurements, e.g., fewer on-cloud stars, smaller $ \Delta A_{\rm G}$ of background eliminated extinction, and multiple jumps along a certain LOS. We first remove the clouds with a large dispersion of distance caused by the lacking of on-cloud stars ($\leqslant 20$). Then, some clouds have a small difference between both sides of the jump point (i.e., $ \Delta A_{\rm G}$ smaller than 0.2 mag).  
%the samples with small $ \Delta A_{\rm G}$ is not prominent relative to background extinction. 
As for the detection of multiple jumps, this problem will be further discussed in a future paper. In this paper, combining the results of Model A and B can largely reduce this problem. 

For some large-scale MC structures, their distances are not detected by either Model A or B. It could be due to the severe extinction environment in front of them. The reddening from dust might be more extensive or saturated by Gaia's observation limit for the foreground dark clouds. 

In our model, we assume that the cloud is a simple screen perpendicular to our sight lines. 
We thus ignore the clouds' thickness and possible inclination to us. Among MCs with a measured distance, some display an abnormal feature near the jump point (e.g., the ascending slope in $A_{G}$--$Distance$ map). In particular, for some large-scale MC structures with a large number of stars (e.g. L889 in the upper right panel in Figure \ref{fig:figa2}), it might suggest a continuous medium or cloud inclination along the LOS. These effects can cause the deviations of the jumps we determined.

\subsubsection{Uncertainty from baseline elimination} \label{subsec:undertaintyb}
The statistical distributions of stars in the foreground and/or background of the cloud follow a Gaussian distribution. In some of LOS, we found that the distribution of $A_{G}$ is Gaussian with an increasing baseline near the jump point. The reddening of stars increases with a gentle slope when the distance increases. Note that the off-cloud region is not extinction free. 
%And interstellar medium in the vicinity of MC attribute to the reddening, such as dust, HI or ionized gas. 
We just identify the jump point of this cloud when the jump is much steeper to the off-cloud region. 

Based on the monotonic fitting method \citep{kruskal_nonmetric_1964,de_leeuw_correctness_1977} and using a module called scikit-learn in Python, we performed the isotonic regression to fit the baseline of background extinction weighted by the inverse-variance of $A_{G}$ \citep{yan_molecular_2019}. The stars with high weights (standard deviations $\leqslant 0.01$ mag) are removed in this work. As the uncertainty of extinction in the $G$ band of Gaia is not a direct measurement, the posterior distribution of $A_{G}$ with very small values could be unphysical. On the other hand, the high weights of individual sources might introduce deviations of fitting, especially for the monotonic fitting to the off-cloud stars \citep{yan_molecular_2019}. Nevertheless, the scarcity of stars and the large dispersion of $A_{G}$ in the far end cause a large bias to our results. We found that $\Delta A_{\rm G} \geqslant 0.2$ mag can mitigate the problems.  

$A_{G}$ is derived from $ A_{0}$ in the Fitzpatrick extinction law \citep{1999PASP..111...63F} and absolute $M_{G}$ magnitude from isochrones \citep{2023A&A...674A..27A}. 
The median value of uncertainties of $A_{G}$ and $\rm A_{0}$ is around $0.06\sim0.07$ mag, considering the inflation effect. Although Gaia collaboration did not apply corrections for GSP-Phot uncertainties, we use the inflated value for safety, i.e. the lower limit $\Delta A_{\rm G}=0.2$ mag ($3\sigma$ confidence level) for identifying extinction jumps.
For Model B, the larger $A_{G}$ values in the reference region make a smaller $ \Delta A_{\rm G}$ after background elimination, we thus set $\Delta A_{\rm G}=0.15$ mag ($2\sigma$). 
Considering the fluctuations of $A_{G}$ from random errors, biases from monotonically increasing fit, and the variations of stars' $A_{G}$ dispersion with increased distance, the larger $\Delta A_{\rm G}$ from MC structures helps to reduce false detection. 
%Prominent structures with higher confidence levels are kept.  
 
Figure \ref{fig:fig7} shows the distance results of $\rm ^{13}CO$ structures (see Sections \ref{subsec:selectf} and \ref{subsec:onlyon}). For most MC samples, the distance measurements match well in Models A and B.  %The differences bigger than (set as 100 pc or the summation of their errors from MCMC if larger) were checked once more. 
%We find samples in either model with smaller differences($\leq 150 ~pc$) revealed the same jump. The deviations come from that sampling of model B are not sharp, it could be more sensitive to $A_{\rm G}$ dispersion changes as their $ \Delta A_{\rm G}$ are smaller. 
We check all the samples and confirm the jump points presented by Models A and B are the same one. 

We find 75\% of samples from Model A have a slightly smaller distance compared to those from Model B. We adopt the result from Model A in the following analysis because of the more clear jump in Model A (see example in Figure \ref{fig:fig5}).
The differences come from a sampling process due to different $\Delta A_{\rm G}$ for on-cloud stars after a different background elimination (see simulations in Appendix \ref{app:appb}).

%as model B give a more reliable jump location(to which layer), while model A give more precise distances. 

There are two exceptions, G076.74$-$0.43 (ID 30 in Table \ref{tab:tab1} and Figure \ref{fig:fig6}) and G084.70+0.44 (ID 181 in Table \ref{tab:tab1}). For cloud G076.74$-$0.43, the distance difference from Models A and B is larger than 200 pc.  
In fact, we find two extinction layers toward the direction (see details in Section \ref{subsec:cygs}). When reference stars are chosen farther away from the target cloud (e.g., Model A), the measured distance tends to be affected by other structures along LOS. In Figure \ref{fig:fig6}a, after the baseline elimination, extinctions of on-cloud stars fluctuate along the $A_{G}$--$Distance$ map, indicating the inconsistency of $A_{G}$--$Distance$ relation between on-cloud and off-cloud stars. So the baseline subtracted data from Model A might suffer from fluctuations in the sampling result. And the distance of the cloud should be $\rm 1041^{+67}_{-68}$ pc from Model B. Similar to cloud G076.74$-$0.43, the distance difference of G084.70+0.44 from Models A and B is 137 pc, but Model B has a better background elimination, and presents a smaller sampling uncertainty. 

For more special cases with complex background extinction, Model A is invalid because of insufficient field stars in the nearby region. Model B provides distance measurements for another 22 MC structures (see Appendix \ref{app:appd}). Among these cases, some structures likely suffer from severe contamination from neighboring clouds. 

\subsection{Molecular cloud parameters} \label{subsec:mcp}

%Generally the distances of those samples with large scale, as well as less overlapping can be well determined based on our model. Over two hundred clouds with precise distance using MCMC sampling by emcee based on Gaia parallaxes and photometry. Among them, 120 clouds are measured both by model A and B (class I), distances of them are used in Figure 7 $\sim$ Figure 16. Another 22 clouds are only measured by model B, their distances are also reliable, combined with class I samples, we construct the big picture of multiple layers in Cygnus (see Figure 17). Besides, there are over 60 clouds are alone measured by model A, we remind that these results should be used with caution. (<-)

We summarize the parameters of 207 clouds with distances in Table \ref{tab:tab1}. 120 clouds are measured by both Models A and B (class I). Their distances are used in Figures \ref{fig:fig7} to \ref{fig:fig16}. 22 clouds are only measured by Model B (class II), and over 60 clouds are alone measured by Model A (class III). Generally,  the dust-based distances to specific MCs might introduce $\approx$ 5\% uncertainties out to at least 2 kpc from Gaia data \citep[see the recent review from][]{2023ASPC..534...43Z}. In all the following discussions, we added a 5\% system uncertainty to the MC distances. 

In the left panel of Figure \ref{fig:fig7}, we plot the distance distribution of the 120 class I clouds. There is a gradient from east to west for the molecular gas traced by $\rm ^{13}CO$ emission within 1 kpc. For example, the clouds near NAP are located at $\rm 759\pm57 ~pc$, while the western clouds near L889 are systematically larger than 800 pc. In Cygnus X South region, we also find a middle gas layer in $\sim1.3$ kpc. 

Because of the ``extinction wall'' effect \citep{1993BaltA...2..171S}, we fail to obtain the distance of some clouds in the Cygnus X SFR, 
%as well as the background emission of North America(NGC 7000) and L889, 
including DR21, DR20, and the clouds behind L889 (see details in Section \ref{subsec:cygn}--\ref{subsec:cygs}). After removing the foreground clouds, the distance of the rest of clouds can be determined based on the spatial coincidence between the $\rm ^{13}CO$ structures and the corresponding masers/photodissociation (PDR) interfaces (see details in Cygnus North Filament from $\rm ^{13}CO$ in Section \ref{subsec:cygn} and PDR interfaces traced by intense infrared emission in Section \ref{subsec:cygs}). 
%it's possible to reveal the Cygnus X region and other more distant structures. 

%\citet{2019ApJ...879..125Z} and \citet{chen_three-dimensional_2019}'s method also measured distances around these two regions with Gaia data, but methods based on dust properties might neglect the complexity of this region, which could lead to systematic biases in the measurements (such as the problems mentioned in \citet{2019ApJ...879..125Z}'s work). In other words, owing to the lack of velocity information, it is hard to determine the specific measurement object, but simply looking for jumps in a direction may yield averaged results for different components, as a result discourage further analysis of the structure. We will discuss this later by comparing the results of different distance measurements. 
%The direct use of BEEP \citep{yan_molecular_2019} only get the distances of small isolated clouds at edges, since the majority complex is connected together by DBSCAN in $\rm ^{12} CO$ data cube and failed to be measured on the conditions of background elimination. (on-cloud only)

Once the distances of clouds are determined, we can derive the physical parameters of these MCs, such as effective radius, mass, and surface density. 
Physical radius can be calculated from the angular size $R_{\rm ang}$ and cloud distance $d$: 
%\begin{equation}
%R_{\rm eff}=\frac{1}{2}d \sqrt{{R_{\rm ang}}^{2}-{\theta_{\rm beam}}^{2}}
%\end{equation}
\begin{equation}
R= d\tan \left(R_{\rm ang} \right)
\end{equation}
%$\theta_{\rm beam}$ is the HPBW of telescope, 
Then, the mass and surface density of $\rm ^{13} CO $ clouds can be estimated by 
\begin{equation}
M=N_{\rm tot} d^{2} \Omega \mu m_{\rm H}
\end{equation}
where $\Omega$ is the solid angle of each pixel; $\mu$ = 2.8 is the atomic weight of molecular hydrogen. And the surface densities of cloud structures is 
\begin{equation}
\Sigma_{\rm cloud}=\frac{M}{\pi {R}^{2}}
\end{equation}
We compute the virial parameter to characterize the dynamical state of clouds or cores; the virial mass is in the form 
\begin{equation}
M_{vir}=5\frac{R \sigma_{v}^{2}}{G}
\end{equation}
where $G$ is the gravitational constant; $\sigma_{v}$ is the velocity dispersion of each cloud, which is provided in equation (9), assuming the $\rm ^{13} CO$ cloud has a uniform density profile, and the virial parameter can be described as 
\begin{equation}
\alpha_{vir}=\frac{5\sigma_{v}^{2} R}{GM}
\end{equation}
The cloud distance perpendicular to the Galactic disk is directly described by 
\begin{equation}
z=d \sin (b_{0})
\end{equation}

%We suggest lots of MCs with small angular sizes are ignored in this work, for we only highlight those major structures. Distance measurement to them can still be implemented if they were relatively isolated though few on-cloud stars are included. 

%More stars are included by morphological dilation, as $\rm ^{13} CO$ only trace high density gas, diffuse gas in its vicinity also contribute to the extinction for small structures.
%We control the width of dilation by a function SQUARE, which generates a flat, square-shaped structuring element, returns a structuring element consisting every pixel belongs to the neighborhood. So with a bigger width, the structuring element is larger and more stars are involved. If the extinction of included stars are affected by this structure, better distance measurement result can be evaluated. For small structures, the effect of expansion is obvious, and the jump point becomes clear by complement of stars near jump point. So the location of jump can be confined in a relatively small range. 
%Dilation is a quick reduction to the $\rm ^{13} CO$ clouds boundary. Using $\rm ^{12} CO$ emission properly will promote the measurement to these structures, however, we don't highlight it in this work.

\section{subregions toward Cygnus} \label{subsec:subregions}
\subsection{Cyg-North region} \label{subsec:cygn}

Based on the measured distances of MCs, we find at least two layers of gas emission in the foreground toward the Cygnus X North region (see Figures \ref{fig:fig8}--\ref{fig:fig10}). One is at $\sim$800 pc and another is at $\sim$1 kpc. Combining the other information (see below), many $\rm ^{13}CO$ structures are located in more distant regions. 

$800 ~pc ~gas ~layer$. The velocities of molecular gas in this layer are mainly concentrated in 1--7 $\rm km~s^{-1}$ (see contours in Figure \ref{fig:fig8}b and orange dots in Figure \ref{fig:fig9}), including the filament L914 (see Section \ref{subsec:l914} and Figure \ref{fig:fig15}).
In the region of $l$ = [$82^{\circ}$, $84^{\circ}$] and $b$ = [$1.5^{\circ}$, $2.5^{\circ}$], some MCs are at higher velocities intervals ($\rm 8 \leqslant v \leqslant 13 ~km~s^{-1}$). 
Including those NAP clouds in the velocity interval of [-6, 6] $\rm km~s^{-1}$ (Section \ref{subsec:nap}), we find that a large molecular gas loop is located at $\sim$ 800 pc (see below in Figures \ref{fig:fig8} and \ref{fig:fig17}). 

$1 ~kpc ~gas ~layer$. In Figure \ref{fig:fig8}a--c, we find a gas layer at $\sim$ 1 kpc.  MCs in this layer have a large velocity extent from $-4$ to 14 $\rm km~s^{-1}$ (black dots in Figure \ref{fig:fig9}), inferring a large velocity dispersion between clouds. We identify them as an MC complex from ($l$ = $81^{\circ}$, $b$ = $0.8^{\circ}$) to ($l$ = $79^{\circ}$, $b$ = $-1.8^{\circ}$). 
%Contours are clouds located in foreground layers, grey scale denotes the filament in star formation region, which is also shown in Figure 10 (b)).  
We note that the ``twins cluster'' structures are located in this layer (marked as clusters a and b in Figures \ref{fig:fig9} and \ref{fig:fig10}b, also see IDs 123 and 113 in Table \ref{tab:tab1}). The two clusters' distances are $962 ^{+64}_{-67}$ pc and $1024^{+77}_{-73}$ pc, respectively.
%considering the uncertainties they should be related together.
We note that the two structures have almost the same physical properties but very different velocities. Both of them also have higher bright temperature than other clouds in the 800 pc and 1 kpc gas layer. What causes the twin clouds' velocity to differ by 15 $\rm km~s^{-1}$ is still unknown. Further studies would be interesting to find out the kinematic origin of them.

$Background$ $layer$ $for$ $more$ $distant$ $gas$ $at$ $\sim1.5 ~kpc$. The distances of many MC structures with intense $\rm ^{13}CO$ emission have failed to be determined because of the large extinction from the above two layers.  We attribute these filamentary structures with fluctuated velocities from $-9$ to 1 $\rm km~s^{-1}$ (see Figure \ref{fig:fig8}a) as the Cygnus North Filament. The Cygnus North Filament is very prominent in the north of Cygnus OB2. DR21(OH), DR21, DR23, DR22 are all located along the dense CO filament. The filament can be divided into different substructures in a clustering result due to velocity gradient (see Figure \ref{fig:fig8}d). DR21 and DR21(OH) at $\sim-3$ $\rm km~s^{-1}$ are on the most bright DR21 filament (see the intensive $\rm C^{18}O$ emission in gray scale of Figure \ref{fig:fig8}a). The distance of the Cygnus North Filament is estimated to be 1.5 kpc by a methanol maser measurement from DR21 and DR20 \citep{2012Rygl}. We also find that some dense molecular gas with slightly high velocities are different from MCs at 800 and 1 kpc layers. For example, the MC associated with W 75N (see m11 in Table \ref{tab:tab2}) is overlapped with DR21 filament, but at a nearer distance of 1.3 kpc \citep{2012Rygl}. 

Furthermore, we also find several corresponding structures between $\rm ^{13}CO$ emission and bright infrared features (see Figure \ref{fig:fig8}e). Among them, the two globules \citep[see][]{2006A&A...458..855S, 2016A&A...591A..40S} and DR17 with large velocities (from 7 to 20 $\rm km~s^{-1}$) are probably associated with nearby star-forming regions (e.g., Cygnus OB2). 
We assume these MCs, together with clouds on the Cygnus North Filament, range from 1.3 to 1.7 kpc based on the association between MCs and masers or OB stars \citep{2012Rygl, 2021MNRAS.508.2370Q}. Tentatively, we put them together as a background layer with an averaged distance at $\sim1.5$ kpc.

In Figure \ref{fig:fig9} we plot the $\rm ^{13}CO$ bright temperature, velocity dispersion, and column density of MCs in different gas layers. 
The bright temperature and column density of the molecular gas at $\sim$1.5 kpc are obviously larger than those for the gas in 800 pc and 1 kpc gas layers.

The 3D illustration in Figure \ref{fig:fig10} denotes the distribution of three layers toward the Cygnus X North region. Again, the extended molecular gas in the 800 pc and 1 kpc layers is overlapped in front of the $\sim1.5$ kpc MCs, leading to the failure of distance measurements of these MCs at larger distances based on our models.

\subsection{Cyg-South region} \label{subsec:cygs}
The molecular gas in $\sim4^{\circ}\times4^{\circ}$ region of Cygnus X South is divided into three velocity intervals: the mid-interval ($\rm -3\sim3~ km~s^{-1}$), minus interval ($\rm -10\sim-3~ km~s^{-1}$), and positive interval ($\rm 3\sim12~ km~s^{-1}$) based on our clouds' identification. These results can be compared with the work from \citet{2006A&A...458..855S}. We also identified at least three gas layers toward this direction. 

$950 ~pc ~gas ~layer$. In Figure \ref{fig:fig11}a, we found that a bulk of molecular gas are at a distance of $\sim950$ pc (e.g. the prominent emission from giant molecular complex L889). 
The mean bright temperature of the clouds in this layer is low, which is different from the gas associated with the star-forming regions (see the background layer below). 
%a nearer distance rather than in background star-forming regions. 
From Figures \ref{fig:fig11} and \ref{fig:fig12}, most clouds in the Cygnus X South region are near 0 $\rm km~s^{-1}$, which causes great difficulty to distinguish them. The clouds in $\sim950$ pc layer are distributed in all velocity intervals. Figure \ref{fig:fig11}f shows agreement between our result and the 3D dust maps by \citet{2019ApJ...887...93G}. As the largest $\rm ^{13}CO$ structure in Table \ref{tab:tab1} (ID 57 for L889), we will further discuss it in Appendix \ref{app:appa}. 

$1.3 ~kpc ~gas ~layer$. We also identify a group of clouds with larger distances, which construct a layer of molecular gas at $\sim$1.3 kpc toward the Cygnus X South. The velocities of these clouds are distributed in [$-12$, 12] $\rm km~s^{-1}$.  
The spatial distribution of these 1.3 kpc clouds with smaller sizes are relatively discrete from those in the 950 pc gas layer. Because of the great extinction caused by the 950 pc layer, the clouds with measured distances in the 1.3 kpc layer are mainly found in the region with weak CO emission from L889 structure (see Figure \ref{fig:fig11}ac). 
%That is, possible structures in this layer might exist behind L889.  

Interestingly, we find a cloud (see cloud 94 in Table \ref{tab:tab1}) is located at $1271^{+116}_{-115}$ pc, which agrees well with the maser G079.73+0.99 at 1.36 kpc \citep{2012Rygl}. Another cloud associated with HII region DR7 (see cloud 86 in Table \ref{tab:tab1}) is at $1247^{+113}_{-100}$ pc based on our models. 

For the DR13 cloud (see cloud 48 in Table \ref{tab:tab1}) with bright CO and infrared emission, its distance by Model B is also at $\sim$1.3 kpc. Next to DR13, the cloud G077.9$-$1.1 with bright CO emission (see cloud 44 in Table \ref{tab:tab1}) matches the IR dark area very well (see Figure \ref{fig:fig11} a and e). Our distance measurement confirms it is at $1215^{+90}_{-87}$ pc. We suppose that cloud G077.9$-$1.1 is slightly in front of DR13 cloud, leading to the matched morphology between the molecular gas structure and IR dark area. These results demonstrate the accuracy and effectiveness of our methods. 
%and lies next to each other in 3D space. 

$Background ~layer ~for ~more ~distant ~gas ~associated ~with ~PDR ~interfaces$. For the clouds farther away than $\gtrsim$ 1.4 kpc toward the Cygnus X South region, only two MCs can be determined by Model B (IDs 19 and 67 in Table \ref{tab:tab1}). However, at least tens of dense CO clouds with cometary, oval-shaped, and irregular structures coincide well with bright $8\mu$m emission. We also find good agreement in physical properties of these clouds (e.g., $\rm ^{13}CO$ peak temperature and column density; see blue dots in Figure \ref{fig:fig12}). 

We note that the cometary clouds (see Table \ref{tab:tab3}) show a head-to-tail morphology pointing away from the Cygnus OB2, indicating the interaction between the molecular gas and intense UV radiation or stellar wind from nearby OB stars. These results are similar to those from \citet{2007A&A...474..873S}. 
One of those cometary clouds \citep[see m7 in Table \ref{tab:tab2} and p11 in Table \ref{tab:tab3}; also named globules in][]{2006A&A...458..855S, 2016A&A...591A..40S} is exactly located at 1.61 kpc based on maser measurement of G079.87+01.17 \citep[see IRAS 20286+4105 in][]{2013ApJ...769...15X}. The coincidence of molecular gas and IRAS 20286+4105 demonstrates the above picture. 

For these cometary clouds, clouds A and B identified by \citet{2006A&A...458..855S} correspond to our cases p1 and p3 (see Table \ref{tab:tab3}). 
%The morphology of the cometary structures of CO emission supports the interaction between the stellar winds/UV radiation and the dense molecular gas. 
From our clouds' identification, we find another cloud C (p5 in Table \ref{tab:tab3}) is next to cloud B (see Figure \ref{fig:fig11} a and e). The high temperatures of the three clouds in the Cygnus X South region, as well as the high column densities, are different from clouds in 950 pc and 1.3 kpc gas layers (see Figure \ref{fig:fig12}). We suggest that all of them are probably shaped by nearby Cygnus OB2 and subgroups of OB1 associations.

%probably being shaped by Cygnus OB2 and subgroup of OB1 associations. The high temperature of the three clouds is probably related to We suspect gases in there being heated as they are all with relatively high bright temperature (see Figure 12 (a)). These clouds are also quite dense with high column densities , within them $\rm C^{18}O$ emission traced cores are illuminated (see Figure 11 (a)). 

%We have cloud by cloud comparison with the mid-IR emission, and find over ten correspondences in Cygnus South region. 
%The highly agreement with morphologies and locations as well expected physical properties, validates the cloud clustering result, and shows the potential to analyze interactions between interstellar medium with their density and velocity fields.
Besides the cometary clouds, the spatial morphology of some oval-shaped clouds also coincide with the structure of HII regions \citep[e.g. DR 6, DR 9 identified by][]{1966ApJ...144..937D}, indicating the physical association between them. 
%However some of these clouds seems to be nearer than 1.7 kpc.
For these oval-shaped clouds, we do not have enough evidences to associate them with a large-scale radiation field; thus, their distances cannot be confirmed. Here, we temporarily put the oval-shaped and irregular MCs in the background layer (see hollow circles in Figure \ref{fig:fig12}).

%Two globular clouds are found in minus velocities and almost no determined emission from positive velocities,  
%The large inconsistency between mid-IR emission and CO emission show that 
We conclude that L889 and the nearby clouds in 950 pc layer contain the majority of molecular gas with low peak temperature and column density. 
%This layer, together with 800 pc layer mentioned in Cygnus North region (Section 4.3.1), constructs the well-known Cygnus Rift. We suggest that the Cygnus Rift probably extend from $\sim$770~pc (NAP see below) 
%We note that the molecular gas in L889 is a bit further than previous thought, according to our new results. 
Toward the Cygnus X South region, parts of MCs in the 1.3 kpc layer are associated with the nearer HII regions and star-forming regions (e.g., DR 13). We propose MCs associated with W75 N in the Cygnus X North also belong to the 1.3 kpc layer. The clouds embedded in the Cygnus X South SFR are mainly in a narrow velocity interval $\lesssim 3$ $\rm km~s^{-1}$, mixing up with 950 pc and 1.3 kpc gas layers. It is different from velocity distributions in the Cygnus X North region (see Figure \ref{fig:fig9}). These clouds with high column densities and temperatures are being affected by stellar feedbacks from massive OB associations. 

%We conclude that a majority of cold gas are within 1.3 kpc. not in the same place, the latter are confirmed to be within 1 kpc, a bit further than previous think of emission in Cygnus Rift. 

\subsection{North America/Pelican (NAP) region} \label{subsec:nap}

Our identified clouds in NAP have multiple velocity components from $-6$ to $\rm 6~ km~s^{-1}$ (see Figure \ref{fig:fig13}b), which is also shown by \citet{2020ApJ...899..128K} using the moment 1 map from FCRAO $\rm ^{13} CO$. We also present $\rm C^{18}O$ emission in this area (gray scale in Figure \ref{fig:fig13}a).  %draw the isotopes together, $\rm C^{18}O$ emission is found in $\rm ^{13} CO$ bright area, which correspond to the dense cores. 

In Figure \ref{fig:fig14}a, we plot the measured distances of different trees from ACORNS of the NAP (see cloud IDs $169\sim195$ in Table \ref{tab:tab1}). We find that the MC complex is located at about $759\pm57$ pc, which is roughly consistent with previous studies based on Gaia EDR3 YSO samples in the NAP region \citep[ $\sim$785 pc in][]{2020RNAAS...4..224K}. The group E (l $\sim84.^{\circ}8$, b $\sim-1.^{\circ}2$) identified by \citet{2020ApJ...899..128K} has a distance of $\sim$746 pc, which is slightly smaller than their average distance of $\sim$785 pc for the complex. We note that at least three cloud structures (IDs 173, 190, and 193 in Table \ref{tab:tab1}) are located at $\sim$740 pc toward the Group E. 
%(mass)NAP complex contains a bulk of molecular gases, they overlap with each other in an inter-velocity dispersion and introduce intense CO emissions.  
Figure \ref{fig:fig14}b presents a system deviation of distance measurements between our results and \citet{2020A&A...633A..51Z}. 
%\citet{2020RNAAS...4..224K} use data of Gaia EDR3, give mean parallaxes of 6 subgroups of YSO members divided in their earlier work, they range from $\sim740~pc$ to $\sim880~pc$ after converting to distances. Our distances agree well with \citet{2020RNAAS...4..224K}, which are systematically small than \citet{2020ApJ...899..128K} produced by Gaia DR2. Actually our results also present a system deviation from \citet{zucker_compendium_2020}, see Figure 14 (b), 
We propose that the differences come from the upgrade of both $A_{G}$ and parallaxes from DR2 to DR3 (systematic uncertainty), and different details between two models: (1) an on-cloud stars' selection based on different cloud decomposition to the stratified gas structure, and (2) a model with vs. without background elimination. Moreover, $A_{G}$ uncertainty is likely to be different along different LOS. For example, our distance measurements toward the Cygnus X region have no prominent systematic deviation compared with those of \citet{2020A&A...633A..51Z}. 
We suggest that the NAP complex in the ``800 pc gas layer'' (see Section \ref{subsec:cygn}) is part of the Cygnus Rift. 

%Among the 6 subgroups, young stellar objects (YSOs) in group E are spatially coherent with dense L935 in the complex, and group E YSOs are dominated by class I young stars \citep{2020ApJ...904..146F}, which indicates L935 is currently most active star formation site in NAP.

\subsection{L914 filament} \label{subsec:l914}
L914 (see cloud IDs 139 and 161 in Table \ref{tab:tab1}) is a filamentary dark cloud connecting the NAP region and the Cygnus X North region. The filament with a large velocity gradient along the spine has a rather small intrinsic velocity dispersion (see Figure \ref{fig:fig15}b). Bright $\rm C^{18}O$ emission also presents a filamentary structure from right to left (hereafter head to tail). We find that the main structure (head, ID 139 in Table \ref{tab:tab1}) has a distance of $787^{+45}_{-44}$ pc. Obviously, L914 and NAP clouds belong to the $\rm 800~pc$ gas layer of the Cygnus X North region. We also note that the substructure in the tail of L914 (ID 161 in Table \ref{tab:tab1}) is located at a smaller distance of $723^{+51}_{-47}$ pc. Considering the spatially continuous distribution of the molecular gas (especially for the coherent filamentary structure traced by $\rm C^{18}O$ in Figure \ref{fig:fig15}a), we propose that they belong to a single structure at $\sim$770 pc. There are also some interesting striations \citep{2008ApJ...680..428G} perpendicular to the spine of L914 \citep[see details in][]{Sun2024L914}. 

\subsection{S106} \label{subsec:s106}

S106 is a well-known HII region in the Cygnus X South. The dense cloud G076.33$-$0.74 (hereafter S106 cloud; see contours in Figure \ref{fig:fig16}) is considered to be shaped by nearby OB associations \citep[e.g., see NGC 6913 discussed by][]{2007A&A...474..873S}. Besides, the new results show that the OB stars of Group C identified by \citet{2021MNRAS.508.2370Q} could be the source shaping S106 cloud. 

Although the distance measurement to the S106 cloud failed for both Models A and B, we suspect  the cloud is located at least 1 kpc away. The 800 pc gas layer (see ID 28 in Table \ref{tab:tab1}) extends to the S106 region, which could be why previous measurement to S106 have a closer distance \citep[e.g., 600 pc in][]{1982ApJ...255...95S}. Actually, one nearby cloud (see cloud ID 30 in Table \ref{tab:tab1}; see Section \ref{subsec:cygs}) in the 950 pc gas layer is also overlapped with the S106 cloud. The heavy extinction effect of these clouds leads to the failure of distance measurement for the cloud. 
 
The maser G076.38$-$0.61 (see m3 in Table \ref{tab:tab2}) is located in S106 region at 1.3 kpc \citep{2013ApJ...769...15X}, suggesting molecular gas at 1.3 kpc layer in this direction. In addition to the S106 cloud, we also identify another cloud structure G076.39$-$0.66 (see red dots in Figure \ref{fig:fig16}) is superposed with the maser. This structure (see m3 in Table \ref{tab:tab2}) is presented as another velocity component, which spatially coincides with the maser G076.38$-$0.61. Similar to the S106 cloud, MC G076.39$-$0.66 has high peak temperature and intense $\rm ^{13}CO$ emission. Both structures could be the HMSFR where the maser located in, indicating the S106 cloud probably at the 1.3 kpc gas layer. 
%the good agreement with position and velocity approves it to be the HMSFR where the maser located in. 
However, the cometary morphology traced by $\rm ^{13}CO$ and infrared emission for the S106 cloud might be shaped by the nearby OB associations. Thus, other opinions suggest the distance to S106 cloud is at $\sim$1.7 kpc \citep{2021MNRAS.508.2370Q}, while the maser G076.38$-$0.61 might be associated with G076.39$-$0.66 at 1.3 kpc. 
%The HMSFR generating this maser should be in the 1.3 kpc layer, where no enough massive OB stars trigger the large IR-bright morphology related to S106. 

Interestingly, we find that a cloud G075.79+0.42 (see cloud ID 24 in Table \ref{tab:tab1}) is adjacent to the S106 cloud in the projection. The cloud is at $\sim800$ pc from Model A. In addition, another component at 1.7 kpc can be discerned in a new mulitjump model (in preparation), indicating the possible gas layer in there. 
It shows that, toward the region, there are complicated gas distributions at different distances. We will study the multijump features of molecular gas in a forthcoming paper. 
In Table \ref{tab:tab2}, we temporarily associate maser G076.38$-$0.61 with MC G076.39$-$0.66 (m3 in Table \ref{tab:tab2}). 
%The distance results to the neighbour clouds also show near-far ambiguities, where cloud G076.15-0.09 (see cloud id 27) has no $A_{\rm G}$ jump by model B, but a jump near 800 pc by model A, while cloud G075.79+0.42 (see cloud id 23) show multiple sampling components in model B, the last one near 1.7 kpc. We suspect most emission of those clouds are from Cygnus Rift, dust extinction contributes to further distance also exists in this direction. We could not exclude the possibility that the gases are associated with OB associations, as they coincide with those IR-bright region well shaped by stellar feed backs.

\subsection{Other interesting regions}
\begin{figure}[!htb]
\centering
%\hspace{-0.12\linewidth}
\includegraphics[width=1\linewidth]{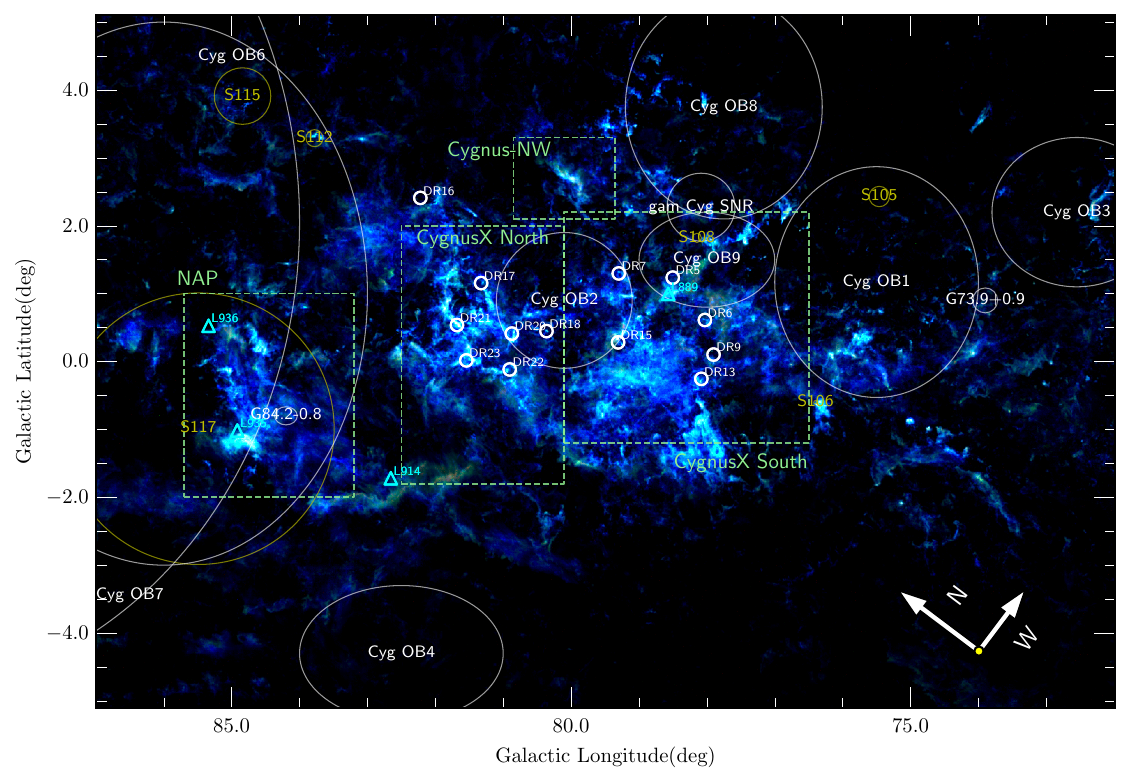}
\caption{A composite guide map colored in R ($\rm C^{18}O$), G ($\rm ^{13} CO$), B ($\rm ^{12} CO$) based on the MWISP data. 
%Among which blue indicates extensive $\rm ^{12} CO$ emission, green indicates translucent $\rm ^{13} CO$ emission, and red show dense $\rm C^{18}O$ emission. All single 
The map is made by moment masking method. That is, pixels with three continuous channels beyond 3 times that of the noise level have been kept. 
Many prominent structures are denoted on this map, including OB associations and SNR $\gamma$ Cygni (labeled with large circles or ellipses), dark clouds (triangles), and radio sources (white small circles) collected by \citet{1966ApJ...144..937D}, as well as several HII regions (yellow circles). The light green rectangles indicate some interesting subregions discussed in Section \ref{subsec:subregions}. 
\label{fig:fig1}}
\end{figure}

$Cygnus$-$NW$. The dark cloud complex L897 is located in the Cygnus Northwest (Cygnus-NW, see Figure \ref{fig:fig1}). The clouds associated with it have velocities ranging from $\sim$5 to 10 $\rm km~s^{-1}$. Four of them (IDs 97, 99, 102, and 104 in Table \ref{tab:tab1}) have been measured at a uniform distance (UD) of $\sim$ 1 kpc, reflecting the 1 kpc gas layer therein. We also note two structures (IDs 99 and 104 in Table \ref{tab:tab1}) have very small virial parameters ($\lesssim$ 1). Combined with the good agreement with the overdensity of YSOCs toward the Cygnus-NW clouds within 1 kpc (see Appendix \ref{app:appf}), we infer that these MCs are gravity bounded and that the dark cloud is physically associated with the ongoing star-forming region. 

$Cygnus ~OB3 ~and ~OB8$. Cygnus OB3 and OB8 are located at $\rm \sim1895~pc$ and $\sim$1726 pc, respectively \citep{2021MNRAS.508.2370Q}. 

Toward the Cygnus OB3, we identified a massive $\rm ^{13} CO$ cloud (ID 1 in Table \ref{tab:tab1}) at $\rm \sim1.8~kpc$, which agrees with the median distance of the OB association. And several small clouds (see IDs 3 and 5 in Table \ref{tab:tab1}) toward this direction are at $\sim$1 kpc.  

Toward the Cygnus OB8, we find an interesting $\rm ^{13}CO$ cloud G078.12+3.63 that is associated with a maser G078.12+03.63 at a distance of $\sim$1.55 kpc \citep{2019ApJ...885..131R}. 
The median distance of Cygnus OB8 from EDR3 \citep{2021MNRAS.508.2370Q} is a bit farther than the maser's distance. However, considering the distance uncertainty, they are probably at the same place. 
Based on our models, another two MCs (IDs 33 and 35 in Table \ref{tab:tab1}) are located in $\sim$1.2 kpc in this direction, indicating that the extended 1.3 kpc gas layer extends in front of the OB8 association (see Figure \ref{fig:fig17}). 
%There is a maser G078.12+03.63 (see m4 in Table 2) in Cygnus OB8, we find interesting $\rm ^{13} CO$ velocity structures are associated with it. Cloud G078.12+3.63 matches it best, while multiple components with similar velocities are displayed. Median distance of Cygnus OB8 is a bit further, thus the correlation between stars association and star-forming regions in this area is unknown. We also have two molecular clouds measured in 1.3 kpc layer in this direction, and it suggests the 1.3 kpc layer is much extensive, which can be seen in Figure 17.

$Other ~clouds ~associated ~with ~masers$.  
Two masers G074.03$-$1.71 and G080.80$-$1.92 (see m1 and m8 in Table \ref{tab:tab2}) are located at about 1.6 kpc \citep{42zz, 2013ApJ...769...15X}. We find two MCs (i.e., cloud IDs 16 and 116 in Table \ref{tab:tab1}) that are probably at $\sim$1.3 kpc and $\sim$1 kpc, respectively, based on Model A. However, due to lacking a convincing distance measurement from Model B, we suspect that the foreground molecular gas at 1-1.3 kpc is likely overlapped on the masers' cloud. This effect leads to where we just detect the nearer distance of the foreground gas. According to the mulitjump model (in preparation), jump features in 
the $A_{G}$--$Distance$ map can be discerned at $\sim$1.51 kpc and 1.73 kpc, respectively.

%Two regions of Cygnus X get our attention due to the $\sim$1.6 kpc masers, G74.03-1.71 and G80.80-1.92 (see m1 and m8 in Table 2). We find the molecular cloud the masers locate in (see cloud id 15 and 115 in Table 1), but distance measurements to both clouds are not favoured by model B, due to no convincing model B measurement, as a result foreground emission might have contamination to our measurement more than expected. 

We identify that an MC (see m5 in Table \ref{tab:tab2}) with an LSR velocity of $\rm -5~km~s^{-1}$ is associated with AFGL2591 at a distance of 3.3 kpc \citep{rygl_preliminary_2011}.
We find that the m5 cloud is overlapped with L889 filament (see Figure \ref{fig:fig11}d). The cloud has a high bright temperature and large line widths, indicating the massive star formation within it. 
%In spite of the proximity in CO emission, it locates neither in 1.5 kpc star formation region nor in foreground Cygnus Rift, but should be away in Perseus Arm, refer to the trend of Local Arm suggested by \citet{2019ApJ...885..131R}. 
%Notice the identified cloud associated with AFGL2591 is larger than other HMSFRs, and the big distance, leading to a large mass and small virial parameter.
Another small cloud (see m2 in Table \ref{tab:tab2}) with a medium bright temperature and column density is associated with a 2.7 kpc maser toward the direction of Cygnus OB1. 
Finally, for the two masers more than 3 kpc G075.76+0.33 
\citep{2013ApJ...769...15X} and G075.78+0.34 \citep{2011PASJ...63...45A}, we find intense CO emission at $\sim$800 pc located just in front of the two masers. As a result, we are unable to identify the MCs associated with them. 
 
The difficulty increases when looking for more distant CO emission in our field of view (FOV).  
For the more distant clouds, smaller angular sizes, together with large extinction and parallax uncertainty of stars from Gaia DR3, limit us from determining their distances. These more distant clouds also suffer from extinction of foreground layers (i.e. the 800 pc, 1 kpc, and 1.3 kpc gas layer) identified in this paper. 
%especially when the two more distance masers have close $\rm v_{LSR}$ to foreground emission, we suppose the associated clouds are compact with small angular sizes, the contributed emission could be merged into foreground large angular size clouds, as a result we fail to identify the molecular clouds associated with the two further masers. 

We measured at least seven clouds (classes I and II) with distances larger than 1.4 kpc. All of them are in a large radius (i.e., at least larger than $5'$ based on $\rm ^{13}CO$ emission). Most of them are located in relatively high latitude regions and/or simple extinction environments, leading to less contamination of foreground emission. The above two facts enable us to determine their distances after background elimination. 
%latitudes, due to majority of samples in this layer are suffered from large extinction of foreground emission. However clouds in higher latitudes and more clean environment are easier to take distance measurements, and this would not be the focus of this work. 

\begin{longrotatetable}
\begin{deluxetable*}{ccccccccccccc}
\setlength{\tabcolsep}{3 mm}
\tablecaption{Parameters of 207 MC structures with distance measurement \label{chartable}}
\tablewidth{700pt}
\tabletypesize{\scriptsize}
\tablehead{
\colhead{(1)} & \colhead{(2)} & \colhead{(3))} & \colhead{(4)} & \colhead{(5)} &
\colhead{(6)} & 
\colhead{(7)} & \colhead{(8)} & \colhead{(9)} &
\colhead{(10)} & \colhead{(11)} & \colhead{(12)} & 
\colhead{(13)} \\
\colhead{ID} & \colhead{$l$} & \colhead{$b$} & 
\colhead{$v_{\rm LSR}$} & \colhead{$\sigma_{v}$} & \colhead{$R$} & 
\colhead{$d^{A}$} & \colhead{$d^{B}$} & 
\colhead{$N_{\rm mean}$} & \colhead{$\rm Mass^{\rm LTE}$} & 
\colhead{$\rm Mass^{\rm Xfactor}$} & \colhead{$\rm \alpha$} &
\colhead{Adopted Model} \\ 
\colhead{} & \colhead{(deg)} & \colhead{(deg)} & \colhead{($\rm km~{s^{-1}}$)} & \colhead{($\rm km~{s^{-1}}$)} &
\colhead{(pc)} & 
\colhead{(pc)} & \colhead{(pc)} & \colhead{($\rm cm^{-2}$)} &
\colhead{($m_{\odot}$)} & \colhead{($m_{\odot}$)} & \colhead{ } & 
\colhead{ } \\
}
\startdata
  1 &  72.29 &   2.31 &   -2.2 &    1.2 &   5.2 &   $1784^{+27}_{-20}\pm89$ &    $1745^{+20}_{-26}\pm87$ &  5.98e+21 &  10,920 &         12,170 &         0.8 &     A \\
  2 &  72.43 &   0.38 &    4.1 &    1.4 &   4.1 &  $2478^{+76}_{-42}\pm123$ &                            &  1.50e+21 &   1740 &          3000 &         5.5 &     A \\
  3 &  72.62 &   0.76 &    5.6 &    0.7 &   1.5 &    $959^{+18}_{-19}\pm47$ &     $964^{+22}_{-22}\pm48$ &  7.21e+20 &    110 &           210 &         8.3 &     A \\
  4 &  72.68 &  -0.74 &    5.0 &    1.2 &   2.5 &   $1072^{+30}_{-28}\pm53$ &                            &  1.39e+21 &    570 &           870 &         6.9 &     A \\
  5 &  73.23 &   0.58 &    5.4 &    0.6 &   1.4 &   $1001^{+37}_{-33}\pm50$ &     $994^{+39}_{-36}\pm49$ &  1.03e+21 &    140 &           240 &         3.9 &     A \\
  6 &  73.23 &  -0.69 &    7.1 &    0.9 &   1.9 &   $1095^{+24}_{-34}\pm54$ &    $1052^{+36}_{-35}\pm52$ &  1.68e+21 &    400 &           360 &         4.1 &     A \\
  7 &  73.31 &  -0.06 &    3.1 &    1.2 &   6.5 &   $1671^{+32}_{-31}\pm83$ &                            &  2.39e+21 &   6750 &         10,240 &         1.5 &     A \\
  8 &  73.44 &  -0.69 &    3.7 &    1.3 &   2.5 &   $1060^{+22}_{-30}\pm53$ &    $1038^{+25}_{-26}\pm51$ &  1.67e+21 &    710 &           940 &         6.4 &     A \\
  9 &  73.99 &   3.56 &   -4.5 &    1.2 &   3.6 &   $1909^{+74}_{-63}\pm95$ &    $1872^{+50}_{-74}\pm93$ &  4.62e+21 &   3930 &          4610 &         1.6 &     A \\
 10 &  74.08 &   0.38 &    1.9 &    1.7 &   4.4 &   $1108^{+22}_{-20}\pm55$ &                            &  2.70e+21 &   3460 &          5350 &         4.2 &     A \\
 11 &  74.09 &  -0.50 &    6.2 &    0.9 &   2.1 &   $1000^{+33}_{-41}\pm50$ &                            &  1.91e+21 &    570 &           790 &         3.6 &     A \\
 12 &  74.18 &  -4.79 &   10.1 &    1.0 &   1.6 &    $959^{+22}_{-23}\pm47$ &     $969^{+24}_{-24}\pm48$ &  8.91e+20 &    160 &           240 &        12.5 &     A \\
 13 &  74.31 &   0.16 &    2.3 &    1.3 &   2.8 &   $1014^{+36}_{-34}\pm50$ &                            &  2.60e+21 &   1390 &          2180 &         4.3 &     A \\
 14 &  74.35 &  -2.47 &    3.1 &    1.2 &   1.7 &                           &    $1258^{+42}_{-57}\pm62$ &  2.33e+21 &    430 &           820 &         6.9 &     B \\
 15 &  74.88 &  -4.67 &   11.9 &    0.5 &   1.8 &    $908^{+20}_{-20}\pm45$ &     $899^{+35}_{-27}\pm44$ &  1.01e+21 &    220 &           290 &         2.8 &     A \\
\enddata 
\tablecomments{
%%A catalog of measured distances and physical properties of the MC structures in the Cygnus region. Only 10 lines are presented in here, see the online version for a full machine-readable version of this table. \\
\\
(1) ID: cloud ID in this work of distance measured molecular clouds, sorted by longitude.\\
(2) $l$: longitude of cloud center weighted by intensity; see Equation (3).\\
(3) $b$: latitude of cloud center weighted by intensity; see Equation (4).\\
(4) $v_{\rm LSR}$: velocity of cloud center weighted by intensity; see Equation (8).\\
(5) $\sigma_{v}$: intensity weighted velocity dispersion of clouds; see Equation (9).\\
(6) $R$: physical size of molecular cloud, in unit of pc; see Equation (17).\\
(7) $d^{A}$: median value of cloud distance measured using Model A. The uncertainties represent 16th/84th value in MCMC samples of cloud distance measured using Model A (lower/upper value of $1\sigma$ confidence interval).\\
(8) $d^{B}$: median value of cloud distance measured using Model B. The uncertainties represent 16th/84th value in MCMC samples of cloud distance measured using Model B (lower/upper value of $1\sigma$ confidence interval).\\
(9) $N_{\rm mean}$: the mean column density of each cloud based on LTE, which is calculated by Equation (13) and (16). \\ 
(10) $\rm Mass^{\rm LTE}$: LTE mass derived from $\rm ^{13} CO$ emission with determined distance, which is calculated by column density of $\rm ^{13} CO$ emission; see Equation (13) and (18).\\
(11) $\rm Mass^{\rm Xfactor}$: mass calculated by $\rm ^{12} CO$ emission with determined distance, in a $\rm ^{13} CO$ traced cloud boundary. The CO-to-$\rm H_{2}$ conversion factor adopts $X_{\rm CO} = 2 \times 10^{20} \rm cm^{-2} (K\cdot km\cdot s^{-1})^{-1}$. \\
(12) Virial parameter of $\rm ^{13}CO$ traced cloud, which is calculated by Equation (21). \\ 
(13) Adopted Model: determine which model to give a more accurate distance. \\
(This table is available in its entirety in machine-readable form.) Only a portion of this table is shown here to demonstrate its form and content. A machine-readable version of the full table is available.  
}
\label{tab:tab1}
\end{deluxetable*}
\end{longrotatetable}

\begin{deluxetable*}{ccccccccccccc}
\tablenum{2}
% \resizebox{1 pt}{!}{}
%\resizebox{0.5\textwidth}{!}{}
\tablewidth{0pt}
\tablecaption{Physical Properties of molecular clouds matched with masers\label{tab:masers}}
\tablewidth{1pt}
\tablehead{
\colhead{ID} & \colhead{$l$} & \colhead{$b$} & \colhead{Parallax\tablenotemark{\tiny$a$}} & \colhead{Parallax Error\tablenotemark{\tiny$a$}} &
\colhead{$v_{\rm maser}$\tablenotemark{\tiny$a$}} & \colhead{$v_{\rm LSR}$} & \colhead{$\sigma_{v}$}  & \colhead{$T_{\rm peak}$} & \colhead{Column Density} & \colhead{$R_{\rm eff}$}  & 
\colhead{Mass} & \colhead{$\alpha_{\rm vir}$} \\
\colhead{ } & \colhead{(deg)} & \colhead{(deg)} & \colhead{(mas)} & \colhead{(mas)} &
\colhead{($\rm km~s^{-1}$)} & \colhead{($\rm km~s^{-1}$)} & \colhead{($\rm km~s^{-1}$)} & \colhead{(K)} & \colhead{($\rm cm^{-2}$)} & \colhead{(pc)} & \colhead{($M_{\odot}$)} & \colhead{ }
}
%\decimalcolnumbers
\startdata
m1\tablenotemark{\tiny$\rm (1)$}  &  73.93 & -1.86 &     0.629 &         0.017 &   5.0 &    5.6 &    1.3 &    6.9 &  3.4e+21 &    5.8 &   7580 &         1.4 \\
m2\tablenotemark{\tiny$\rm (2)$}  &  74.56 &  0.79 &     0.367 &         0.083 &  -1.0 &   -0.6 &    1.0 &    4.6 &  2.8e+21 &    2.7 &   1420 &         2.3 \\
m3\tablenotemark{\tiny$\rm (1)$}  &  76.39 & -0.66 &     0.770 &         0.053 &  -2.0 &   -2.1 &    1.3 &   10.6 &  9.6e+21 &    1.1 &    810 &         2.8 \\
m4\tablenotemark{\tiny$\rm (3)$}  &  78.12 &  3.63 &     0.645 &         0.030 &  -6.0 &   -3.3 &    1.3 &    7.9 &  9.8e+21 &    0.6 &    250 &         4.6 \\
m5\tablenotemark{\tiny$\rm (4)$}  &  78.60 &  0.89 &     0.300 &         0.010 &  -5.7 &   -5.3 &    1.7 &   10.8 &  3.1e+21 &   14.1 &  41,530 &         1.2 \\
m6\tablenotemark{\tiny$\rm (4)$}  &  79.74 &  1.01 &     0.737 &         0.062 &  -3.0 &   -1.6 &    1.0 &    4.5 &  3.3e+21 &    1.9 &    840 &         2.9 \\
m7\tablenotemark{\tiny$\rm (1)$}  &  79.87 &  1.19 &     0.620 &         0.027 &  -5.0 &   -4.2 &    1.2 &    8.0 &  9.9e+21 &    0.6 &    290 &         3.6 \\
m8\tablenotemark{\tiny$\rm (6)$}  &  80.39 & -2.33 &     0.620 &         0.047 &  -3.0 &   -4.1 &    1.1 &    8.4 &  2.9e+21 &    5.5 &   5770 &         1.3 \\
m9\tablenotemark{\tiny$\rm (4)$}  &  80.88 &  0.36 &     0.687 &         0.038 &  -3.0 &   -2.1 &    1.4 &   11.7 &  7.0e+21 &    3.1 &   4510 &         1.7 \\
m10\tablenotemark{\tiny$\rm (4)$} &  80.91 & -0.28 &     0.666 &         0.035 &  -3.0 &   -3.4 &    1.0 &   10.9 &  9.5e+21 &    1.1 &    770 &         1.7 \\
m10\tablenotemark{\tiny$\rm (7)$} &  80.98 & -0.13 &     0.666 &         0.035 &  -3.0 &   -3.1 &    0.8 &   10.6 &  1.0e+22 &    1.3 &   1180 &         0.8 \\
m10\tablenotemark{\tiny$\rm (7)$} &  80.99 & -0.09 &     0.666 &         0.035 &  -3.0 &   -2.4 &    0.9 &   10.0 &  6.9e+21 &    0.8 &    300 &         2.7 \\
m10\tablenotemark{\tiny$\rm (7)$} &  81.08 & -0.19 &     0.666 &         0.035 &  -3.0 &   -4.9 &    1.0 &   14.1 &  7.8e+21 &    2.2 &   2620 &         0.9 \\
m10\tablenotemark{\tiny$\rm (7)$} &  81.22 &  0.88 &     0.666 &         0.035 &  -3.0 &   14.6 &    1.6 &   10.9 &  1.1e+22 &    2.7 &   5790 &         1.4 \\
m10\tablenotemark{\tiny$\rm (7)$} &  81.35 & -0.05 &     0.666 &         0.035 &  -3.0 &   -5.0 &    1.2 &   16.3 &  1.1e+22 &    3.6 &   9440 &         0.7 \\
m10\tablenotemark{\tiny$\rm (7)$} &  81.54 &  0.09 &     0.666 &         0.035 &  -3.0 &   -6.0 &    1.0 &   11.0 &  7.5e+21 &    2.3 &   2740 &         0.9 \\
m10\tablenotemark{\tiny$\rm (7)$} &  81.57 &  0.71 &     0.666 &         0.035 &  -3.0 &   -1.6 &    1.0 &   14.4 &  8.0e+21 &    3.8 &   7700 &         0.5 \\
m10\tablenotemark{\tiny$\rm (7)$} &  81.61 & -0.03 &     0.666 &         0.035 &  -3.0 &    9.6 &    1.0 &   10.1 &  5.7e+21 &    2.3 &   2110 &         1.4 \\
m10\tablenotemark{\tiny$\rm (7)$} &  81.64 &  0.49 &     0.666 &         0.035 &  -3.0 &   -2.0 &    0.7 &   10.1 &  1.4e+22 &    0.5 &    230 &         1.1 \\
m10\tablenotemark{\tiny$\rm (7)$} &  81.64 &  0.75 &     0.666 &         0.035 &  -3.0 &    8.5 &    0.8 &   10.4 &  5.9e+21 &    1.4 &    770 &         1.2 \\
m10\tablenotemark{\tiny$\rm (7)$} &  81.74 &  0.58 &     0.666 &         0.035 &  -3.0 &   -4.1 &    1.7 &   17.3 &  2.2e+22 &    0.9 &   1390 &         2.4 \\
m10\tablenotemark{\tiny$\rm (7)$} &  81.83 &  0.97 &     0.666 &         0.035 &  -3.0 &   -0.7 &    1.1 &   14.6 &  7.5e+21 &    2.9 &   4140 &         1.0 \\
m10\tablenotemark{\tiny$\rm (7)$} &  81.83 &  1.26 &     0.666 &         0.035 &  -3.0 &   11.3 &    1.3 &   13.3 &  1.0e+22 &    2.6 &   4630 &         1.1 \\
m10\tablenotemark{\tiny$\rm (7)$} &  81.86 &  0.88 &     0.666 &         0.035 &  -3.0 &   -2.1 &    1.1 &   15.9 &  1.2e+22 &    2.3 &   4160 &         0.7 \\
m11\tablenotemark{\tiny$\rm (4, 5)$} &  81.95 &  0.71 &     0.772 &         0.042 &   7.0 &   11.2 &    1.7 &   11.6 &  6.4e+21 &    2.6 &   2990 &         3.1 \\
\enddata
% \tablecomments{All masers in Cygnus region cross matched with our identified molecular clouds traced by $\rm ^{13} CO$ emission in velocity interval [-40, 50] $\rm km \cdot s^{-1}$. }
\tablenotetext{Note. }
{ \\
    (1): cloud parameters of m3 \citep{2013ApJ...769...15X} are alternatively from S106 cloud or G076.39$-$0.66; here is the latter one (see Section \ref{subsec:s106}).   \\
    (2): \citet{2014PASJ...66..102B}.  \\
    (3): \citet{66MM}, \citet{66NN}.  \\  
    (4): in order, AFGL2591, IRAS 20290+4052, DR20, DR21, and W 75N from \citet{2012Rygl}.  \\
    (5): \citet{rygl_preliminary_2011}.  \\
    (6): \citet{42zz}.   \\
    (7): clouds associated with DR21 filament \citep{2012Rygl}.   \\
    $^{a}$ Parameters derived from masers' measurement. 
}
\label{tab:tab2}
\end{deluxetable*}

\begin{deluxetable*}{cccccccccc}
\tablenum{3}
% \resizebox{1 pt}{!}{}
%\resizebox{\textwidth}{2mm}{}
\renewcommand\arraystretch{1.2}
\tablecaption{Physical Properties of MCs matched with PDR interfaces \label{tab:irdc}}
\tablewidth{1pt}
\tablehead{
\colhead{ID} & \colhead{$l$} & \colhead{$b$} & 
\colhead{ $v_{\rm LSR}$ } & \colhead{ $~\sigma_{v}~$ }  & \colhead{$T_{\rm peak}$} & \colhead{Column Density} & \colhead{$R_{\rm eff}$}  & 
\colhead{Mass} & \colhead{$\alpha_{\rm vir}$} \\
\colhead{ } & \colhead{(deg)} & \colhead{(deg)} & 
\colhead{($\rm km~s^{-1}$)} & \colhead{($\rm km~s^{-1}$)} & \colhead{(K)} & \colhead{($\rm cm^{-2}$)} & \colhead{(pc)} & \colhead{($M_{\odot}$)} & \colhead{ }
}
%\decimalcolnumbers
\startdata
p1\tablenotemark{\tiny$a$}  &  76.94 &   2.22 &    1.4 &    1.1 &   17.2 &  6.0e+21 &    4.3 &   7440 &         0.8 \\
p2\tablenotemark{\tiny$b$}  &  77.22 &   1.52 &    0.0 &    0.8 &    7.8 &  3.3e+21 &    0.8 &    140 &         4.0 \\
p3\tablenotemark{\tiny$a$}  &  77.47 &   1.85 &    1.4 &    1.4 &   13.3 &  1.1e+22 &    4.6 &  16,170 &         0.6 \\
p4\tablenotemark{\tiny$c$}  &  77.97 &   0.03 &   -2.5 &    1.2 &    6.5 &  3.6e+21 &    2.7 &   1830 &         2.6 \\
p5\tablenotemark{\tiny$a$}  &  77.98 &   1.77 &   -1.9 &    1.0 &   11.7 &  8.4e+21 &    3.4 &   6490 &         0.6 \\
p6\tablenotemark{\tiny$c$}  &  78.00 &   0.56 &   -2.5 &    2.0 &    7.3 &  6.3e+21 &    2.0 &   1610 &         5.5 \\
p7\tablenotemark{\tiny$c$}  &  78.16 &  -0.31 &   -0.5 &    2.1 &    7.1 &  6.4e+21 &    2.7 &   3080 &         4.3 \\
p8\tablenotemark{\tiny$b$}  &  78.46 &   2.67 &    0.5 &    0.8 &   15.2 &  1.4e+22 &    1.0 &    980 &         0.9 \\
p9\tablenotemark{\tiny$a$}  &  79.18 &   2.21 &   -2.2 &    0.8 &   11.1 &  8.3e+21 &    1.4 &   1120 &         1.0 \\
p10\tablenotemark{\tiny$d$} &  79.26 &   0.25 &    3.4 &    1.0 &   10.0 &  6.2e+21 &    1.0 &    440 &         2.8 \\
p11\tablenotemark{\tiny$d$} &  79.87 &   1.19 &   -4.2 &    1.2 &    8.0 &  9.9e+21 &    0.7 &    330 &         3.4 \\
p12\tablenotemark{\tiny$d$} &  79.98 &   0.84 &  -10.8 &    1.2 &   12.5 &  9.6e+21 &    1.3 &   1150 &         1.8 \\
p13\tablenotemark{\tiny$d$} &  80.38 &   0.44 &    8.3 &    0.9 &   16.8 &  1.3e+22 &    1.0 &    890 &         1.0 \\
p14\tablenotemark{\tiny$d$} &  80.84 &   0.56 &   11.6 &    1.0 &    8.8 &  8.4e+21 &    1.0 &    600 &         2.1 \\
p15\tablenotemark{\tiny$e$} &  80.91 &  -0.28 &   -3.4 &    1.0 &   10.9 &  9.5e+21 &    1.2 &    980 &         1.5 \\
p16\tablenotemark{\tiny$e$} &  80.98 &  -0.13 &   -3.1 &    0.8 &   10.6 &  1.0e+22 &    1.5 &   1510 &         0.7 \\
p17\tablenotemark{\tiny$e$} &  81.08 &  -0.19 &   -4.9 &    1.0 &   14.1 &  7.8e+21 &    2.5 &   3360 &         0.8 \\
p18\tablenotemark{\tiny$a$} &  81.27 &   1.05 &   15.5 &    0.7 &    7.9 &  7.4e+21 &    0.8 &    350 &         1.3 \\
p19\tablenotemark{\tiny$a$} &  81.50 &   0.48 &    7.0 &    0.8 &   11.6 &  7.7e+21 &    1.5 &   1120 &         1.1 \\
\enddata
% \begin{tablenotes}  
    % \footnotesize  
\tablenotetext{Note. }
{\\
    $^{a}$ Cometary clouds are likely shaped by Cygnus OB associations.     \\
    $^{b}$ Irregular clouds.          \\
    $^{c}$ Oval-shaped HII regions identified by \citet{1966ApJ...144..937D}, in order: DR9, DR6, and DR13.   \\
    $^{d}$ Globules near the center of far-FUV sources.   \\
    $^{e}$ Clouds match a hub-like HII region in Cygnus X North; see panel (e) of Figure \ref{fig:fig8}.
}
% \end{tablenotes}
%\tablecomments{.}
\label{tab:tab3}
\end{deluxetable*}

%\usepackage{booktabs}

%\multicolumn{}{}{} 
\begin{table}
\tablenum{4}
\caption{\textbf{Mass Evaluation}}
\centering
\renewcommand\arraystretch{1}
\begin{tabular}{ccccc}
\toprule
\textbf{Region}&\textbf{Comments}&\textbf{LTE}&\textbf{X-factor}&by $\rm ^{13}CO$ (2--1)\tablenotemark{a} \\ %[0.3cm]
  &  &($M_{\odot}$)&$(M_{\odot}$)&($M_{\odot}$)  \\
\midrule
\multirow{3}{*}{\textbf{Cygnus}}
&by Table 1&$2.4\times10^{5}$&$3.5\times10^{5}$&  \\
&by Table 1--3&$4\times10^{5}$&$5.3\times10^{5}$&  \\ 
&whole&$1.1\times10^{6}$&$2.7\times10^{6}$&  \\ 
\midrule
\multirow{1}{*}{\textbf{Cygnus within 1 kpc}}
&by Table 1&$2.6\times10^{5}$&$6.2\times10^{5}$&  \\
\midrule
\multirow{5}{*}{\textbf{Cygnus X}}
&Cygnus X North (0$\sim$3 kpc)&$2.3\times10^{5}$&$4\times10^{5}$& \\ 
&Cygnus X North (1.3$\sim$1.7 kpc)&$1.7\times10^{5}$&$3.2\times10^{5}$& $2.8\times10^{5}$ \\
&Cygnus X South (0$\sim$3 kpc)&$4.1\times10^{5}$&$8.3\times10^{5}$& \\
&Cygnus X South (1.3$\sim$1.7 kpc)&$3.2\times10^{5}$&$7\times10^{5}$& $4.5\times10^{5}$ \\
\bottomrule
\end{tabular}
\caption{The mass estimated in different subregions/distance intervals. }
{$^{a}$ Estimated by \citet{2006A&A...458..855S} based on $\rm ^{13}CO$ (2--1) and the LTE method.   \\
}
\label{tab:tab4}
\end{table}

\section{Discussion} \label{subsec:discuss}
\subsection{Big picture} 
\subsubsection{The Cygnus Rift}
%Another problem is large extinction from the foreground Cygnus Rift. 
Cygnus Rift is a foreground molecular structure in Cygnus complex \citep{1956ApJ...124..367H, 1985ApJ...297..751D}, which makes heavy obscuration in optical images. \citet{1985ApJ...297..751D} assumed that the Rift is confused together with Cygnus X above $l$=$74^\circ$. However, the exact spatial and distance distribution is still up for debate, especially in the part overlapping with the background Cygnus X star-forming region. 
What is the real extension of Cygnus Rift in true 3D space? What are the accurate distances for the different gas distribution for the whole Cygnus region? What is the difference between the molecular gas in Cygnus X star-forming region and those in Cygnus Rift? 
%It's time to investigate the whole region based on the new CO survey and improved distance measurement methods.

The overall distributions of MCs toward the Cygnus region (including both the Cygnus X star-forming region and the foreground Cygnus Rift) show multiple gas layers based on our new measurements (see Section \ref{subsec:subregions}). 
%at the same distances with the major emission of at Cygnus OB7 , probably belongs to the known Gould’s Belt. 
We reveal that the clouds in the 800 pc gas layer mainly concentrate along the 800 pc loop, mainly including the foreground emission of the Cygnus X North and NAP region therein (see Sections \ref{subsec:cygn} and \ref{subsec:nap}). 
Based on the MC distance measurements toward the whole region, the 800 pc gas layer also extends to the  Cygnus X South region. The 800 pc gas layer extends to at least $\sim$135 pc along the longitude and a thickness of $\sim$80 pc (i.e., the FWHM of the Gaussian fitting from the distances distribution of MCs). The average value and variance of the column density in this layer (700 pc $\leqslant$ $d$ $\leqslant$ 900 pc) are $\rm 3.1\times 10^{21}~cm^{-2}$ and $\rm 2.3\times 10^{21}~cm^{-2}$, respectively.

The 1 kpc gas layer is also widespread in both of the Cygnus X North and Cygnus X South regions. We find that these clouds have similar velocity distributions, peak temperature, and mean column density. Based on our results, we attribute them as a whole gas layer at $\sim$1 kpc; nevertheless, the clouds in the Cygnus X North region are slightly more distant. Similarly, we estimate the length and depth of the 1 kpc gas layer with $\sim$145 pc along the longitude and a thickness of $\sim$70 pc, respectively. The average value and variance of the column density in this layer (900 pc $\leqslant$ $d$ $\leqslant$ 1100 pc) are $\rm 2.5\times 10^{21}~cm^{-2}$ and $\rm 1.2\times 10^{21}~cm^{-2}$, respectively.    %supporting the  are in good agreement, so they should be linked together. 

Combining the 800 pc and 1 kpc gas layers, the Cygnus Rift is distributed across a large spatial extent of at least in $l$ = [$72^{\circ}$, $87^{\circ}$] and $b$ = [$-5^{\circ}$, $4^{\circ}$]. The distance of Cygnus Rift ranges from the nearer NAP region ($\gtrsim$ 700 pc) to the farther gas layers (e.g., Cygnus X North, Cyg-NW at 1 kpc) according to MCs’ distances. These dark clouds toward the region, composing multiple gas layers with slightly different distances. The accumulation of molecular gas along these layers makes large extinctions in some directions \citep[i.e. the ``extinction wall" effect,][]{1993BaltA...2..171S}. Therefore, it is hard to determine the distances for the distant MCs behind the Cygnus Rift. For example, the contamination from foreground emission at the Cygnus Rift leads to a biased distance measurement in previous studies of S106. The average value and variance of the column density of combined layers are $\rm 2.8\times 10^{21}~cm^{-2}$ and $\rm 1.9\times 10^{21}~cm^{-2}$, respectively.

As is presented above, the molecular gas in Cygnus Rift is better identified and measured. It plays an important role in revealing the 3D view along the LOS. Studying molecular gases (e.g., distributions and physical properties) in the foreground is also helpful to reveal the MCs behind them. For example, we may reveal the detailed gas structures associated with the high-energy emission from Cygnus Cocoon \citep{2024SciBu..69..449L} We need more advanced methods (e.g., multijump) to properly separate the gas at different distances. Finally, a big picture of the whole region will be well delineated.

\subsubsection{Molecular Gas behind the Cygnus Rift}
The distribution of clouds in the 1.3 kpc gas layer is different between the Cygnus X North and South regions. The majority of clouds in this layer are located in the south region based on our distance measurements (see Section \ref{subsec:cygs}; also see the special cases measured only by Model B in Appendix \ref{app:appd}). The result also agrees with the previous studies \citep{2020A&A...633A..51Z, 2022A&A...658A.166D}. However, it does not mean that there is no molecular gas at 1.3 kpc gas layer in the Cygnus X North region. We propose that clouds (see m11 in Table \ref{tab:tab2} and Figure \ref{fig:fig10}a) associated with W 75N \citep[$\sim$1.3 kpc,][]{2012Rygl} in the Cygnus X North are also in this layer. 
%as its column density is similar to clouds in 1.3 kpc, but more bright in $\rm ^{13}CO$ emission. 
These clouds' distances are relatively smaller than that of the Cygnus North Filament \citep[$\sim$1.5 kpc for DR20 and DR21 in][]{2012Rygl}. We notice clouds in this layer have a slightly larger brightness temperature and mean column densities than the clouds within 1 kpc (see Figure \ref{fig:fig12}). The clouds are probably heated by nearby star-forming regions, i.e. substructures of Cygnus OB2 identified by \citet{2019MNRAS.484.1838B}, \citet{2021MNRAS.502.6080O}. 

We note 1.3 kpc layer is also distributed extensively (see middle layer in Figure \ref{fig:fig17}) based on the MWISP data and our results. It is interesting that some clouds associated with HII regions (DR6, DR9, DR13, etc.) have similar physical properties as compared to those clouds in 1.3 kpc layer (see Figure \ref{fig:fig12}). Furthermore, DR13 cloud (see ID 48 in Table \ref{tab:tab1}) has been measured in $1320^{+182}_{-157}$ pc, supporting the association between the heated molecular gas and PDR interfaces. Therefore, some clouds with intense infrared emission are probably located in the 1.3 kpc layer, although we temporarily put the clouds matching with infrared structures (Table \ref{tab:tab3}) at $\sim$1.7 kpc. 

On the contrary, few clouds in the 1.3 kpc gas layer toward the Cygnus X North are successfully measured in our method. We suspect that the majority of molecular gas behind the 800 pc and 1 kpc gas layers are likely located at $\sim$1.5 kpc. For example, the massive Cygnus North Filament (see Figure \ref{fig:fig10}a) contributes to the majority of CO emission in this region, which also agrees with the presence of the densest region toward Cygnus X North at $\sim$1.5 kpc \citep{2012Rygl, 2022A&A...658A.166D}.

The clouds associated with background PDR interfaces in both regions are likely related to the Cygnus OB2, while cometary clouds in Cygnus X South (labeled in A, B, C ,and S106; see Figures \ref{fig:fig11} and \ref{fig:fig16}) could also be influenced by star members in Cygnus OB1 \citep[$\rm \sim$1.7 kpc from][]{2021MNRAS.508.2370Q, 2023AstBu..78..119R}. 
%which are in the same distance with the major OB2 stars group \citep[][$\rm \sim$1.7 kpc]{2021MNRAS.508.2370Q, 2023AstBu..78..119R}. 
In addition, the molecular gas is likely concentrated in 1.5 kpc toward the Cygnus X North region (see blue histogram in right panel in Figure \ref{fig:fig7}). On the other hand, the gas in the Cygnus X South region is mainly at $\sim$1.3 and $\sim$1.7 kpc. 
Actually, only two MCs (IDs 19 and 67 in Table \ref{tab:tab1}) in the Cygnus X South region are at 1.5 kpc with Model B.  
%in Cygnus South, dense gases at 1.5 kpc are not as prominent as those along Cygnus North filament, 
%while MCs associated with masers in foreground layer (see m6 at 1.3 kpc in Table 2) and more distant layer (see m7 at 1.6 kpc in Table 2) are confirmed. 
%Tens of MCs with distance measurement and maser compose 1.3 kpc layer, 
%while FUV radiation shaped clouds is directly associated with OB clusters. 
%These indicate the significant difference of star formation activities and physical properties between Cygnus X North and South regions. 
These indicate the different molecular gas distribution between the Cygnus X North and South regions. 
%The CO-IR discrepancies denote different evolutionary stages of gases with evolved phase of star formation. In spite of the displacement of infrared emission and $\rm ^{13}CO$ gases (see Figure 8e), or small proportion of $\rm ^{13}CO$ emission might merge into foreground L889 (see Figure 11e). There are a bulk of molecular gases are not coexisting with ionized gases. 
%These dense gases which are infrared dark are in an earlier stage while those bright mid-IR emission without correspondent clouds are in an end of gas dissociation and ionizing. 
We suggest that the Cygnus X star-forming region is not an integral structure located in the same distances based on maser measurements and our new MC distance measurements from Gaia DR3. 

\subsubsection{Comparison with the distribution of YSOCs}
Furthermore, we use selected YSO candidates (see Appendix \ref{app:appf}) to trace the spatial distribution of young stars toward the Cygnus region. We find that the majority of YSO candidates spatially coexist very well with the $\rm ^{13}CO$ structures, especially for molecular gas within 1 kpc (see Figure \ref{fig:fig17}). Along the 800 pc gas loop (see 800 pc layer in Section \ref{subsec:cygn}), the significant overdensity of YSO candidates can be found toward the NAP region and the superposed subregion toward the Cygnus X North (see Figure \ref{fig:fig10}b), while the coincidence between the YSOs and molecular gas is also discerned in L914 and molecular gas at [$l\sim83^{\circ}$, $b\sim2^{\circ}$]. 
Additionally, the overdensity of YSO candidates also coincides with the molecular gas in the Cygnus X South region. Different from the Cygnus X North region, the distribution of YSO candidates extends to farther gas layers. It again indicates the 1.3 kpc middle gas layer is mainly distributed toward the Cygnus X South region. 

Behind the Cygnus Rift, YSOCs still roughly follow the distribution of molecular gas. 
Interestingly, the spatial distribution of YSO candidates is much more extended toward the star-forming regions at $\gtrsim$ 1.4 kpc. We find that the overdensities of YSOs are roughly associated with identified molecular gas and OB associations in those regions. 

For the multiple gas layers, the results reveal that a significant overdensity of potential young stars occurs in the vicinity of molecular gas traced by $\rm ^{13}CO$, indicating the connection between them. 

\subsection{Molecular gas mass toward the Cygnus Region} 
A summary of the derived masses is listed in Table \ref{tab:tab4}.
Here, we first discuss the total mass of the molecular gas based on $\rm ^{13}CO$ emission in LTE (see equation 14 and 18). Obviously, the fractional detection rate ($FDR$) of clouds with distance has a major influence on the total mass estimation. $FDR$ can be estimated by the ratio of flux between the distance measured clouds (see Tables 1--3) and raw data. %Generally, closer structures are easier to measure distances, with relatively large angular sizes and enough background on source stars. On the contrary, bright clouds without measured distances tend to be in distant positions. It's obvious that $\rm CR_{Cygnus~Rift} > CR_{1.3kpc~layer} > CR_{bakground}$. Massive clouds in the background are ignored when we compute mass based on our results, $\rm M_{measured}\sim 2.4\times10^{5}~M_{\odot}$. It would lower the underestimate rate when we add the mass of background massive clouds evaluated from masers and OB associations measurements. 
%Based on Table 1 (column 11 of class I and II clouds), $\rm M_{Table 1}\sim 2.4\times10^{5}~M_{\odot}$. 

First we defined $FDR_{1}$ (here, 48.6\% for the whole region) as the flux ratio between MCs with distance and the total identified $\rm ^{13}CO$ clouds. Second, $FDR_{2}$, the flux ratio between the total identified $\rm ^{13}CO$ clouds and masked raw data, is 72.6\% (see Section \ref{subsec:clustering}). 
Thus, we obtain $FDR=FDR_{1}\times FDR_{2}$ = 35.3\%. 
The total mass of clouds with known distances in Cygnus traced by $\rm ^{13}CO$ structures, including clouds within 1 kpc and Cygnus X MSFRs, is about $M_{\rm Table 1-3}\sim 4\times10^{5}~M_{\odot}$, leading to the total mass 
%The underestimation rate to all identified clouds (overall CR of different layers) $UR_{1}=51.4\%$, and the underestimation rate to raw data $UR_{2}=27.6\%$ (see Section 3.1). 
$M_{\rm overall}\sim 1.1\times10^{6}~M_{\odot}$ for the whole region. 

We note that $FDR_{1}$ in different layers are not the same, which has a dominant influence on the total mass estimation in Cygnus traced by $\rm ^{13}CO$. 
% Even thoudh we include clouds in star-forming regions validated by masers and OB stars, the completion rate in background is still smaller than foreground Cygnus Rift with distance measurement. 
In addition, the beam filling factor \citep{2021ApJS..256...32S, 2021ApJ...910..109Y} and the smaller coverage of $\rm ^{13}CO$ emission would lead to the underestimation of the molecular gas mass. 
So the total mass estimated above is still a lower limit toward the whole Cygnus region traced by $\rm ^{13}CO$ emission.

Alternatively, we further use $\rm ^{12} CO$ emission to estimate the total mass of molecular gas toward the whole Cygnus region. 
For each identified $\rm ^{13}CO$ structure, we give an X-factor mass from corresponding $\rm ^{12}CO$ emission in the same PPV space. 
%Each cloud gives an X-factor mass (see column 11 in Table 1),
%The flux of $\rm ^{12} CO$ is integrated in the velocities of measured $\rm ^{13} CO$ clouds. 
Similarly, the molecular mass can be estimated to be $M_{\rm Table 1-3}^{\rm Xfactor}\sim 5.3\times10^{5}~M_{\odot}$ for the corresponding $\rm ^{13}CO$ structure by adopting $X_{\rm CO} = 2 \times 10^{20} \rm cm^{-2} (K\cdot km\cdot s^{-1})^{-1}$ \citep{2013ARA&A}. 
%It's the gas mass traced by $\rm ^{12} CO$ in identified $\rm ^{13} CO$ mask. 
The $\rm ^{12} CO$ flux ratio between the identified structures with distance and the raw data is about 19.5\%, leading to the total molecular mass of $M_{\rm overall}^{\rm Xfactor}\sim 2.7\times10^{6}~M_{\odot}$ in the whole Cygnus region. 
%The underestimation rate UR is evaluated directly by comparing with raw data, which mitigate the effect brought by optical depth in $\rm ^{13} CO$ traced dense region. Considering the effect of above two hypotheses, 
Considering the underestimation of $FDR_{1}$ in the above calculation, the estimated mass of $M_{\rm overall}^{\rm Xfactor}\sim 2.7\times10^{6}~M_{\odot}$ is still a lower limit. 

%It suffers from the following problems: 
%$\rm ^{12} CO$ line profile deviates from Gaussian since it suffers from opacity broadening and line blending effects \citep{2016A&A...591A.104H}, physical quantities derived from $\rm ^{12} CO$ line are supposed to be oriented from $\rm ^{13} CO$ results. Ideally, we assume $\rm ^{13} CO$ clouds are one-to-one matching to $\rm ^{12} CO$ clouds. For both spatially isolated one, and those with hierarchical structures. Their centroid velocities, line widths have similar distributions. But with the complicated interstellar environment, it's hard to check the associations between $\rm ^{12} CO$ cloud and $\rm ^{13} CO$ cloud spatially.
%As we discussed, $\rm ^{12} CO$ traced emission present cloud complexes in spatially distribution, as it's more extended and its line width is broader. Giant $\rm ^{12} CO$ clouds often contains several $\rm ^{13} CO$ components in PPV space \citep{2022ApJS..261...37Y, 2023ApJ...944...91Y}. 

In Table \ref{tab:tab4}, we make further efforts to estimate the mass of molecular gas within 1 kpc (mainly from Cygnus Rift). Similar to the above analysis, we use $FDR_{2}$ =72.6\% and $FDR_{1}$ $\sim$1 to evaluate the total mass within 1 kpc. 
According to Table \ref{tab:tab1}, the total molecular mass within 1 kpc is estimated to be $M_{\rm 1 kpc}^{\rm LTE}\sim 2.6\times10^{5}~M_{\odot}$. Assuming the same ratio from $M_{\rm overall}^{\rm Xfactor}$/$M_{\rm overall}^{\rm LTE}$ for the gas layers within 1 kpc, the low limit of $M_{\rm 1 kpc}^{\rm Xfactor}$ is about $6.2\times10^{5}~M_{\odot}$ because the emissions of those small structures are ignored. 
%considering the estimated masses in Table 4, the results are reasonable to reveal the mass distribution of the molecular gas within 3 kpc. 
%the underestimation ratio can be partially reflected by subtracting background emission. 

For comparison with previous work \citep[e.g.,][]{2006A&A...458..855S}, we evaluate the mass of the Cygnus X region. 
For the Cygnus X North region ($FDR_{1}=59.3\%$, $FDR_{2}=83.9\%$), 
%$\rm M^{Cygnus North}\sim 6.6\times10^{4}~M_{\odot}$, 
the total mass (0--3 kpc) of molecular gas is $M_{\rm Cyg North}^{\rm LTE} \sim 2.3\times10^{5}~M_{\odot}$ and $M_{\rm Cyg North}^{\rm Xfactor} \sim 4\times10^{5}~M_{\odot}$, respectively. Assuming $FDR_{1}$ $\approx$ 1 for foreground molecular gas within 1 kpc, we estimated that molecular mass from background emission (mainly from 1.3 to 1.7 kpc) is $M_{\rm Cyg North}^{\rm LTE} \sim 1.7\times10^{5}~M_{\odot}$ and $M_{\rm Cyg North}^{\rm Xfactor} \sim 3.2\times10^{5}~M_{\odot}$ by subtracting the foreground mass. 
Similarly, for the Cygnus X South region ($FDR_{1}=48.9\%$, $FDR_{2}=74.6\%$), the total molecular mass is $M_{\rm Cyg South}^{\rm LTE} \sim 4.1\times10^{5}~M_{\odot}$ and $M_{\rm Cyg South}^{\rm Xfactor} \sim 8.3\times10^{5}~M_{\odot}$. 
Additionally, the molecular masses associated with the nearby star-forming region are $M_{\rm Cyg South}^{\rm LTE} \sim 3.2\times10^{5}~M_{\odot}$ and $M_{\rm Cyg South}^{\rm Xfactor} \sim 7\times10^{5}~M_{\odot}$ for 1.3$\sim$1.7 kpc. 

The total mass toward Cygnus X in 1.3$\sim$1.7 kpc is $M_{\rm Cygnus X}^{\rm LTE} \sim 4.9\times10^{5}~M_{\odot}$, $M_{\rm Cygnus X}^{\rm Xfactor} \sim 1\times10^{6}~M_{\odot}$, respectively. We note $\sim$20\% flux toward the direction is from the molecular gas within 1 kpc. 
%Although we subtract the foreground mass within 1 kpc, our results from $\rm ^{13}CO$ emission are comparable to 
We also attached the mass evaluated by $\rm ^{13}CO$ (2--1) \citep{2006A&A...458..855S} based on LTE method. 
Our new results are supposed to be smaller than those of \citet{2006A&A...458..855S}, knowing that we put part of the MCs in a closer place (e.g., the gas layers within 1 kpc). On the other hand, the difference of estimated molecular mass between the two works is likely from the different transitions from $\rm ^{13}CO$. 
%two works use different tracers, $\rm ^{13}CO$ (1-0) trace a larger coverage than $\rm ^{13}CO$ (2-1), thus trace more molecular gases. 

%Obviously, the estimated mass of molecular gas from X-factor method is at least 2 times larger than LTE method. 
%We suggest using $\rm ^{12}CO$ emission based on X-factor is closer to MCs masses. 

\subsection{From PPV to PPP} \label{subsec:ppvp}
It is sometimes assumed that the CO emission in PPV space corresponds to coherent structures in position--position--position (PPP) space. However, in the simulation results in \citet{2013ApJ...777..173B}, there are two possible problems in cloud decomposition from PPV to PPP. The first one is that the gas emission with different velocity components (thus probably with different distances) are combined into the same spatial structure. And the other is that clouds belonging to a coherent structure in real 3D space exhibit two or more velocity components that are separated in PPV space. Due to the prevalent feedback from star-forming activities, great caution must be exercised when discussing the properties of identified features in PPV space \citep{2018MNRAS.479.1722C}. Many structures with a single feature in PPV space would not be directly identified as coherent features in PPP space. For example, Perseus Arm in longitude--velocity diagram is not a continuous structure in true spatial space \citep{2022ApJ...925..201P} based on the 3D dust map \citep{2019ApJ...887...93G}. We also find that some identified gas structures with different velocities are located at the same distance (e.g., see Figures \ref{fig:fig9}, \ref{fig:fig12}, and \ref{fig:fig14}). 
%show clouds that are very near each other in PPV often have very different distances.

Without a distance measurement, cloud clustering methods can actually introduce two problems. One is overdecomposition to the whole MC complex linked by many substructures with different velocity features, while the other is a mistaken aggregation of unrelated structures in PPP space because of close velocities. 
In our work, 
%we identified velocity coherent structures \citep{2013A&A...554A..55H} traced by $^{13}CO$, 
%combined with our results, the clustering based on Gaussian decomposition tend to identify substructures. 
%\citep{2013A&A...554A..55H}
we confirmed that the clouds in NAP region are from a whole MC complex, although the algorithm decomposes the complex into different velocity structures (see Section \ref{subsec:nap}, Figures \ref{fig:fig13} and \ref{fig:fig14}). On the contrary, it is necessary to avoid a wrong aggregation of different spatial structures, considering the complicated velocity structure toward Cygnus region.   
%to get more accurate distances. 
For the first case, with the distance measurement, substructures from the overdecomposition can be linked together to form larger cloud complexes (see the 800 pc loop with a large extension of $\sim4^{\circ}\times4^{\circ}$ in Figure \ref{fig:fig17}). 
%CO spectral lines overlap on the wing in many areas. With spatially coherent features, they are presented as multiple velocity components. So it's necessary to avoid wrong aggregation to get more accurate distances, and with these solid distances, substructures are linked together forming large cloud complexes. 
Finally, many identified MCs with close velocities are indeed located in different distances based on our cloud clustering and distance measurements (see Figures \ref{fig:fig8} and \ref{fig:fig11}).  
%For the identified substructures in different distances, they are separated although close velocities are presented. 
Therefore, accurate distance measurements, together with appropriate cloud clustering, allow us to better describe the true distribution of molecular gas in the Milky Way.

\citet{1992ApJ...384...95A} indicate that the size--line width relationship from modeling galactic disk is not a reliable indicator of the physical nature of cloud complexes. We also caution against using the results of clustering MCs with various methods and size--line width relation from PPV space \citep{2010ApJ...712.1049S, 2015MNRAS.453.3082P} without an accurate distance measurement. 
%The existence of a size-line-width relationship is not a reliable indicator of the physical nature of cloud complexes \citep{2010ApJ...712.1049S, pan_what_2015}. 
Our results clearly demonstrate a very wide range of distances for clouds with close velocities toward the Cygnus region. And conversely, the clouds located in the same layers might have very different velocity features (see Figures \ref{fig:fig9} and \ref{fig:fig12}). 
From this work, we propose that the studies of MCs within 3 kpc should be revisited in details at Gaia's age of precise astronomy \citep{2023ASPC..534...43Z}. The true 3D distribution of MCs is essential to construct a large-scale structure of our Galaxy.

\section{Summary} \label{subsec:sum}
%We study the properties of MCs toward Cygnus region based on the unbiased MWISP CO survey and the measured distances from Gaia DR3. We identify 3829 structures from $\rm ^{13} CO$ emission based on coherent spatial distribution and velocities, as well as line widths. About 70\% fluxes are recovered.  
%%decompose $\rm ^{13} CO$ emission with Gauss fitting to unbiased CO survey, an uniform clouds catalogue is derived by a hierarchical clustering algorithm. $\rm ^{13} CO$ structures are checked to be coherent in spatial distribution and velocities as well as line widths. 
We study the properties and distribution of MCs toward the Cygnus region ($\rm \sim150~deg^{2}$) from the MWISP CO survey. Here, we summarize the main conclusions in this work: 

(1) We identified 3829 structures based on coherent spatial and velocity structures of $\rm ^{13} CO$ emission. About 72.6\% fluxes are recovered after Gaussian decomposition and clustering from GaussPy+ and ACORNS. 

(2) Combining the identified cloud structures and data from Gaia DR3, we design two models (A and B) to measure distances of molecular gas in the Cygnus region. Among the identified $\rm ^{13} CO$ structures, we obtain distances for over 200 large clouds (i.e., $\rm \geqslant 60~arcmin^{2}$). 120 clouds are measured by both Models A and B (class I). 22 clouds are only measured by Model B (class II), and over 60 clouds are alone measured by Model A (class III). The flux of MCs with distance (classes I and II in Table \ref{tab:tab1}) contributes to about 31.2\% total flux of the identified $\rm ^{13} CO$ structures.

(3) About 20 clouds are coincidental with bright mid-IR emission (see panel e in Figures \ref{fig:fig8} and \ref{fig:fig11}; also see Table \ref{tab:tab3}). The association between the MCs and surrounding star-forming regions is also supported by MC properties (cometary morphology, high peak temperature, and intense emission of the gas, etc.). Moreover, based on our models, some clouds with oval-shaped and irregular morphology (e.g., DR13) are indeed measured to be $\sim$1.3 kpc. These clouds associated with PDR interfaces are probably related to OB subgroups at $\sim$1.3 kpc. 

(4) Additionally, we find that tens of MC structures are associated well with masers in Table \ref{tab:tab2}. The distances of these MCs can be obtained based on the association between masers and the corresponding molecular gas. We also find that our independent measurement of a cloud (cloud 94 in Table \ref{tab:tab1}) is consistent with the corresponding maser's distance of 1.36 kpc \citep{2012Rygl}. 

(5) The spatial distribution of YSOs' candidates coincides well with $\rm ^{13}CO$ structures within 1 kpc, indicating the tight connection between them. These cases in (2), (3), (4), and (5) show that our models are effective for distance measurement to MCs in velocity crowding regions. 
 
%Most of these clouds are failed to do distance measurements using our model due to the large extinction from foreground emission. 
 
%Evidences (cometary structure, high peak temperature and intense emission, etc.) show that those clouds identified in Table 2 and 3 are likely associated with star-forming regions. 

(6) Our distance measurements of MCs, combined with the additional information from molecular gas associated with masers and nearby OB associations, show that there are multiple layers of gas structure toward Cygnus region: (I) the gas in 800 pc and 1 kpc layer composing Cygnus Rift, (II) the 1.3 kpc layer majorly in Cygnus X South, and (III) the Cygnus North Filament and the adjacent dense gas at 1.5 kpc, as well as many cometary MCs directly shaped by Cygnus OB associations at $\sim$1.7 kpc (see Figure \ref{fig:fig17}). The results reveal the complex distribution of molecular gas toward Cygnus region, both in spatial and velocity distribution. The total masses of molecular gas of the whole Cygnus region are $\sim 1.1\times10^{6}~M_{\odot}$ by LTE, and $\sim 2.7\times10^{6}~M_{\odot}$ by X-factor (see Table \ref{tab:tab4}). 

(7) Our work determines the large spatial extent of Cygnus Rift at least in $l=[72^{\circ}, 87^{\circ}]$ and $b=[-5^{\circ}, 4^{\circ}]$. 
The distances of the MCs are well determined in the range of 700 pc to 1 kpc, revealing the multilayer nature toward the Cygnus Rift (see Figures \ref{fig:fig7} and \ref{fig:figd1}). For example, the foreground gas toward the Cygnus X North region is composed by 800 pc and 1 kpc layers. The superposition of gas structures toward L889 in the Cygnus X South also exists. The large extinction of the foreground gas in these directions causes the failure of the distance measurement for MCs at larger distances (i.e., ``extinction wall'' effect). The molecular mass of the foreground Cygnus Rift (within 1 kpc) contains $\sim$ 25\% of the whole region. 

(8) We propose that the molecular gas associated with the Cygnus X star-forming region does not come from an integral structure. Actually, there are different molecular structures at different distances, such as  $\sim$1.3 kpc molecular gas that are likely associated with subgroups of Cygnus OB2, the dense gas in Cygnus North Filament at $\sim$1.5 kpc, and the cometary MCs shaped by Cygnus OB associations at $\sim$1.7 kpc (see histogram in right panel of Figure \ref{fig:fig7}). Toward the Cygnus X SFR, the molecular gas within 1.3$\sim$1.7 kpc is about $\sim 4.9\times10^{5}~M_{\odot}$ by LTE, and $\sim 1\times10^{6}~M_{\odot}$ by X-factor.

We find that abnormal jumps and/or multijump features of $A_{G}$--$Distance$ map are common toward the Cygnus region. Besides the gas layers discussed above, there are likely other gas structures at different distances in the whole region. For example, some MCs at $\sim$550 pc (i.e. clouds 198 to 203 and 205 in Table \ref{tab:tab1}; also see light red contours in Figure \ref{fig:fig17}) are in front of Cygnus Rift in the region $l$ $\geqslant85.^{\circ}$3. These clouds have relatively small sizes and column density, thus contribute to the small proportion of mass within 1 kpc (see histograms in Figure \ref{fig:fig7}). These clouds at 500--600 pc are probably related to the MC complex toward the Cygnus OB7. We indeed detect that two jump features are exactly at $\sim$550 pc and $\sim$760 pc in NAP region, corresponding to the different gas layer therein. 
%The abnormal velocity deviation of the ``twin clusters'' mentioned in Section 4.3.1. All 
We propose that the multilayer nature of molecular gas is ubiquitous. We will develop a multiple jumps detection model to further reveal the 3D molecular gas distribution of Cygnus region in a forthcoming paper.

\begin{acknowledgments}
Acknowledgments. We would like to thank the anonymous referee for the valuable comments and suggestions that have largely improved the manuscript.  This research made use of the data from the Milky Way Imaging Scroll Painting (MWISP) project, which is a multiline survey in $\rm ^{12}CO$/$\rm ^{13}CO$/$\rm C^{18}O$ along the northern Galactic plane with the PMO 13.7 m telescope. We are grateful to all the members of the MWISP working group, particularly the staff members at the PMO 13.7m telescope, for their long-term support. MWISP was sponsored by the National Key R\&D Program of China with grants 2023YFA1608000, 2017YFA0402701, and the CAS Key Research Program of Frontier Sciences with grant QYZDJ-SSW-SLH047. This work is supported by NSFC grants 12173090, 12073079. X.C. acknowledges the support  from the Tianchi Talent Program of Xinjiang Uygur Autonomous Region and the support by the CAS International Cooperation Program (grant No. 114332KYSB20190009). This work also made use of data from the European Space Agency (ESA) mission Gaia (https://www.cosmos.esa.int/gaia), processed by the Gaia Data Processing and Analysis Consortium (DPAC, https://www.cosmos.esa.int/web/gaia/dpac/consortium). Funding for the DPAC has been provided by national institutions, in particular the institutions participating in the Gaia Multilateral Agreement. 
%For the space telescope, Gaia provides unprecedented accuracy for parallax measurements to stars. The number density of Gaia DR3 stars is comparable to the spatial resolution of MWISP data, thus they work well together on distance measurements to MCs near Galactic plane.
\end{acknowledgments}

%% Following the acknowledgments Section, use the following syntax and the

\vspace{5mm}
\facilities{PMO: DLH, Gaia}


\software{astropy \citep{2013A&A...558A..33A, 2018AJ....156..123A, 2022ApJ...935..167A},
scikit-learn \citep{2011JMLR...12.2825P},
emcee \citep{2013PASP..125..306F},
Pymc3 \citep{2016ascl.soft10016S}, 
Matplotlib \citep{2007CSE.....9...90H}.
          }

%% subSections with the same markup you use in the main body of the paper.


%\begin{figure}
%\plotone{cygnus_R_clean_rgb_final.pdf}

\begin{figure}[!htb]
\centering
%\hspace{-0.12\linewidth}
\includegraphics[width=1\linewidth]{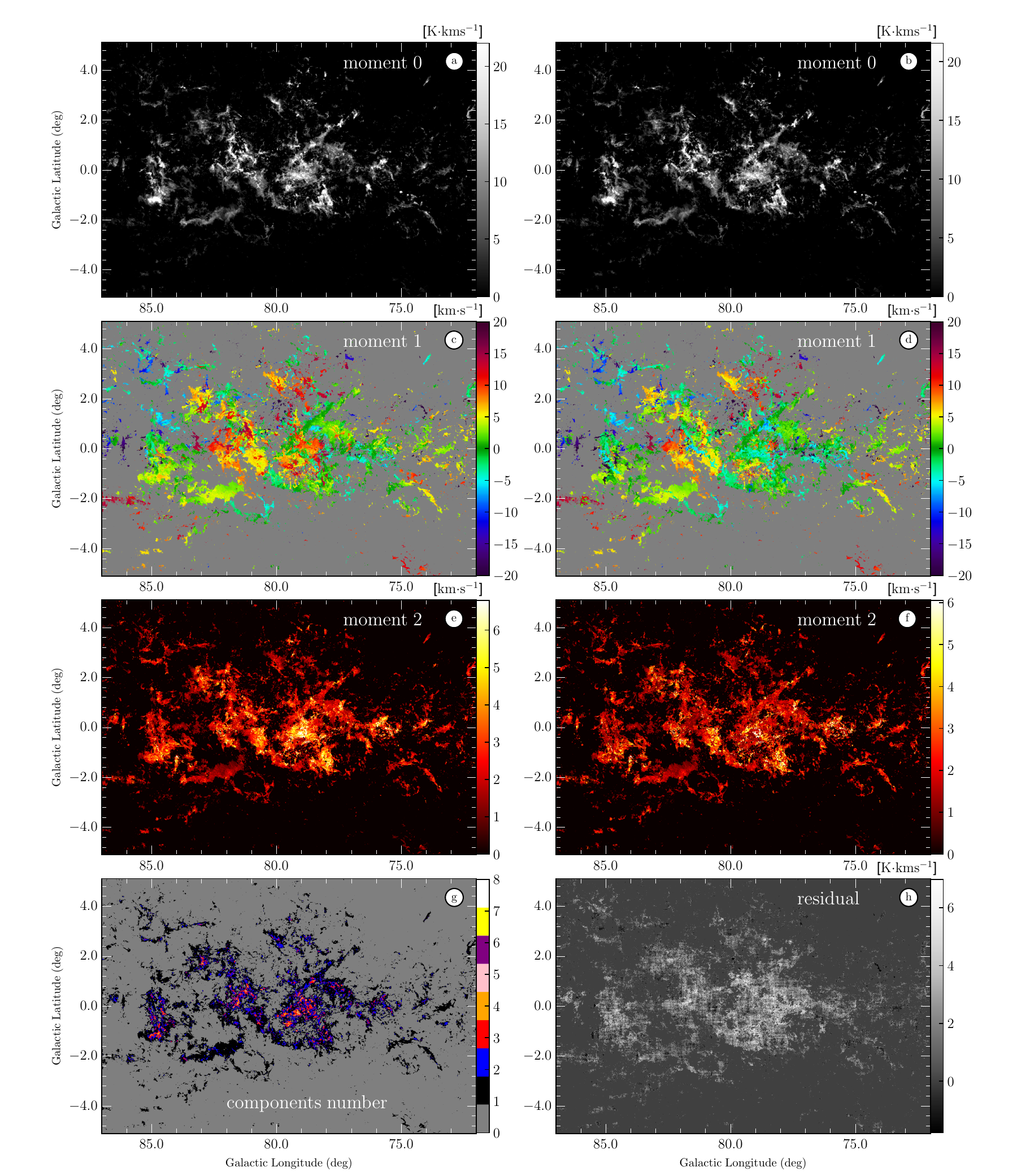}
\caption{(a) Reconstruction of moment 0 map of $\rm ^{13} CO$ emission; the values in gray scale are fluxes summed up by all Gaussian components in each sight line. (b) Similar to (a), but only for the emission with Gaussian components after clustering. (c) Reconstruction of moment 1 map, showing the centroid velocity of Gaussian components. We choose most positive velocities on the top (top positive), but not a weighted average. (d) The same with (c), but smallest velocities on the top (top minus). (e) The moment 2 map, showing the line width of most positive velocities on the top. (f) The same with (e), but smallest velocities on the top. (g) The distribution of the number of Gaussian components in the whole region. And (h) the residuals between integrated emission of raw data and reconstructed image by Gaussian decomposition.
\label{fig:fig2}}
\end{figure}

\begin{figure}
\plotone{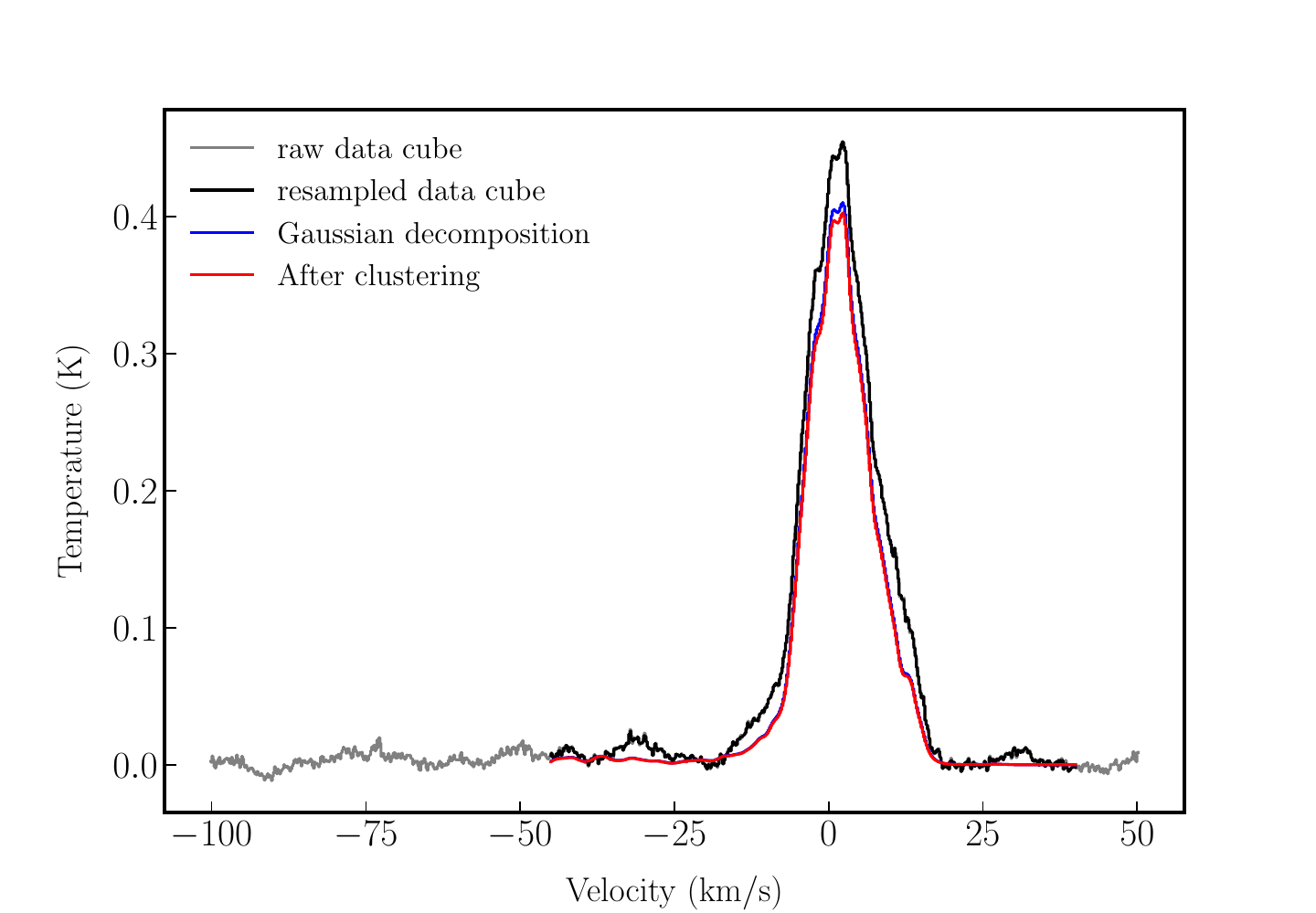}
\caption{The average spectrum of $\rm ^{13}CO$ toward the Cygnus region.
%Data cube is averaged by all pixels successfully fitted by GaussPy+.  We did resample to all three CO lines in a $\rm [-45,40] km \cdot s^{-1}$ intervals, within which the emission is significant. Then $\rm ^{12} CO$, $\rm ^{13} CO$, $\rm C^{18}O$ are all regrided to $\rm 0.2 km \cdot s^{-1}$ for convenience in following studies. 
Gray line shows the raw data truncated from $\rm -100$ to $\rm 50~ km~s^{-1}$, while black line denotes the average spectrum after resampling. Blue curve indicates the average of all signals restored by Gaussian decomposition, while red line shows the average spectrum after clustering based on ACORNS \citep{2019MNRAS.485.2457H}. Toward the Cygnus region, we note that a collective cloud emission overlapped together within a narrow velocity range ($\rm -20$ to $\rm 20~ km~s^{-1}$). The flux reconstructed by Gaussian decomposition and clustering are 81.5\% and 79.5\% (see Section \ref{subsec:check}), respectively. 
\label{fig:fig3}}
\end{figure}

\begin{figure*}
\gridline{\fig{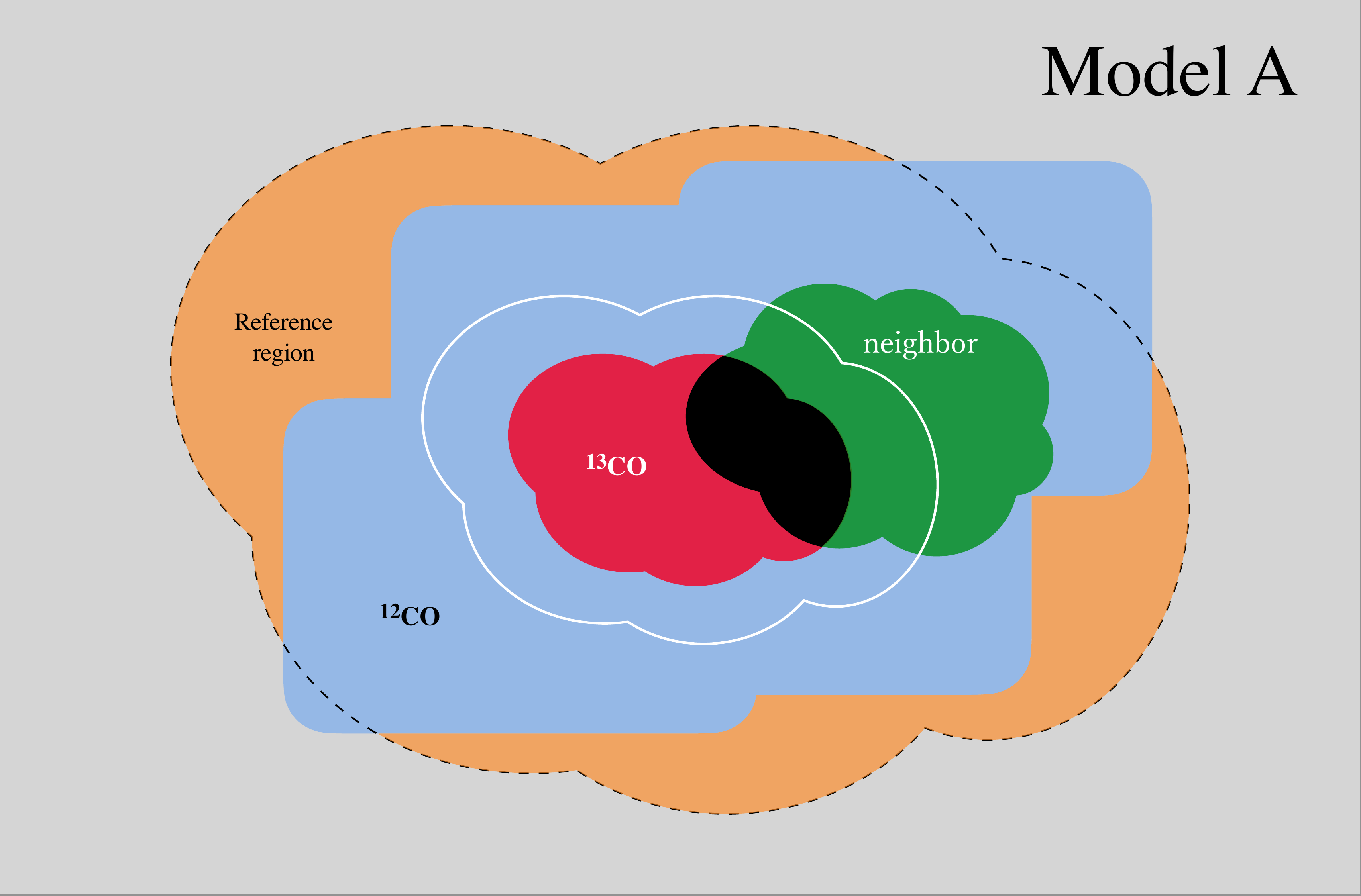}{0.5\textwidth}{(a)}
          \fig{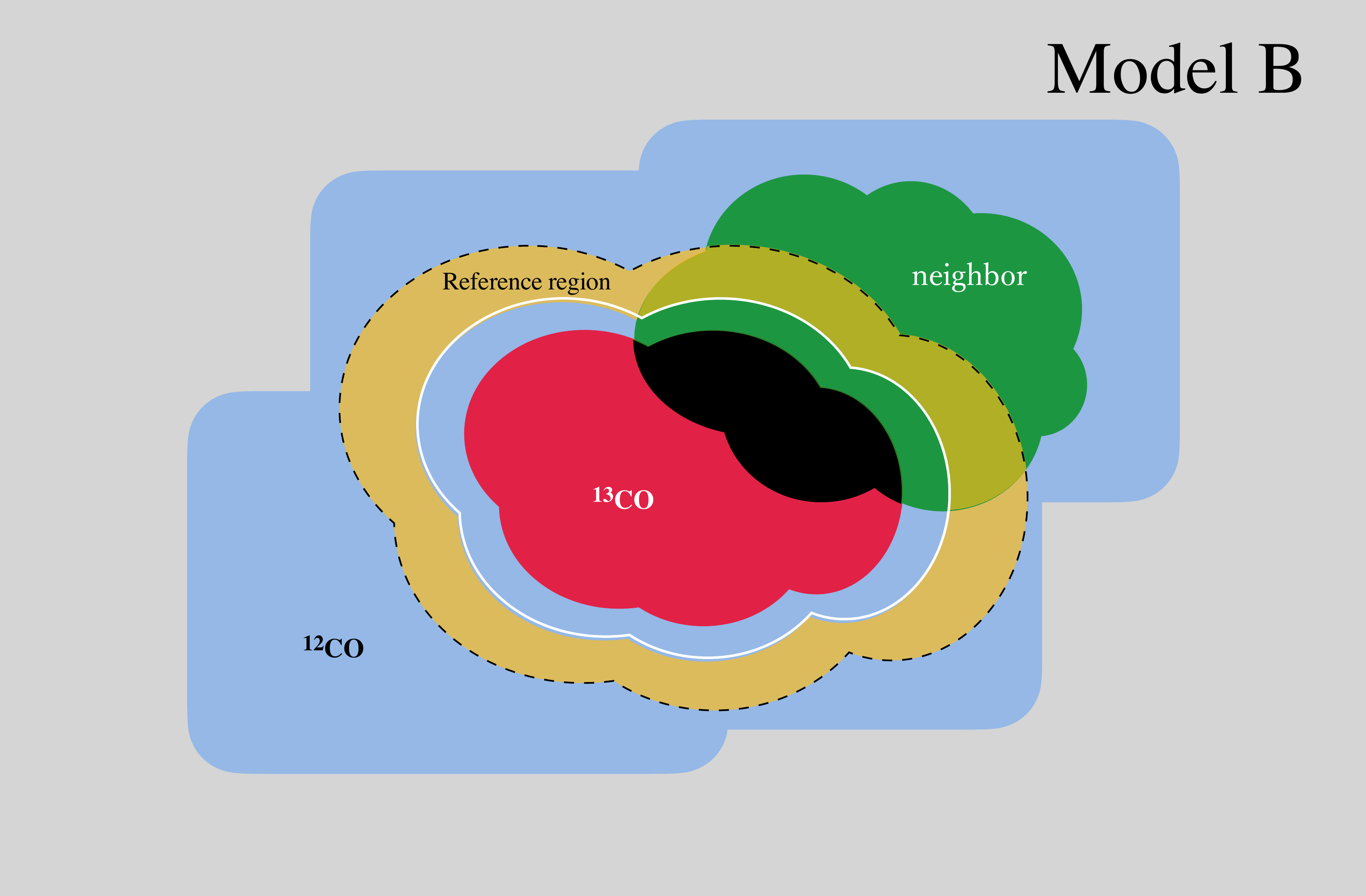}
{0.5\textwidth}{(b)}
          }
\caption{The sketches illuminate our method based on BEEP \citep{yan_molecular_2019}. On-sources are selected within the $\rm ^{13} CO$ structure. 
(a) Use $\rm ^{12} CO$ emission to choose field stars. (b) Use $\rm ^{13} CO$ from its own morphology to select field stars in its periphery. 
%A simple model based on CO emission, represent typical molecular clouds in velocity crowded region traced by CO emission. 
Red region is the identified $\rm ^{13} CO$ cloud, which is partly overlapped by a neighboring cloud in green. The black part is the region with superposition, which include more than one Gaussian component. White solid contour and black dashed contour denote different dilation of cloud boundary. Orange part represents the reference region. 
%we select referenced stars in the ring like field between two curves(in orange). For the smaller white contour lower the contamination caused by the extensive emission(the extinction to star lights), and wider dashed contour include enough field stars to do baseline fitting as well as avoid contamination of other line of sights far away from our target cloud. 
Blue area represents the widespread $\rm ^{12} CO$ emission. Obviously, Model A considers the $\rm ^{12} CO$ free region as the reference region, while Model B refers to $\rm ^{13} CO$ morphology to select the reference stars (see Section \ref{subsec:selectf}). 
}
\label{fig:fig4}
\end{figure*}

\begin{figure*}
\gridline{\fig{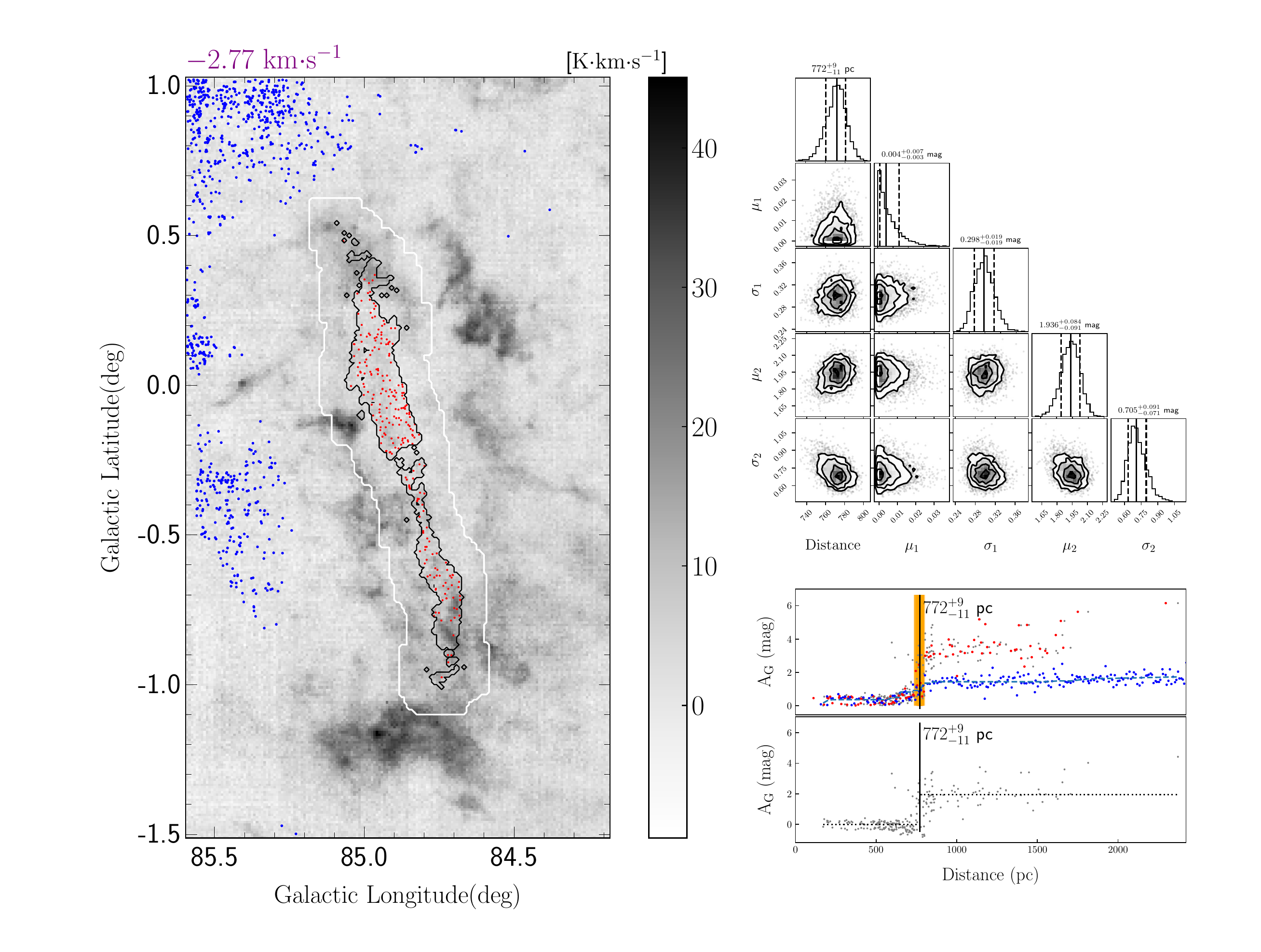}{0.5\textwidth}{(a)}
          \fig{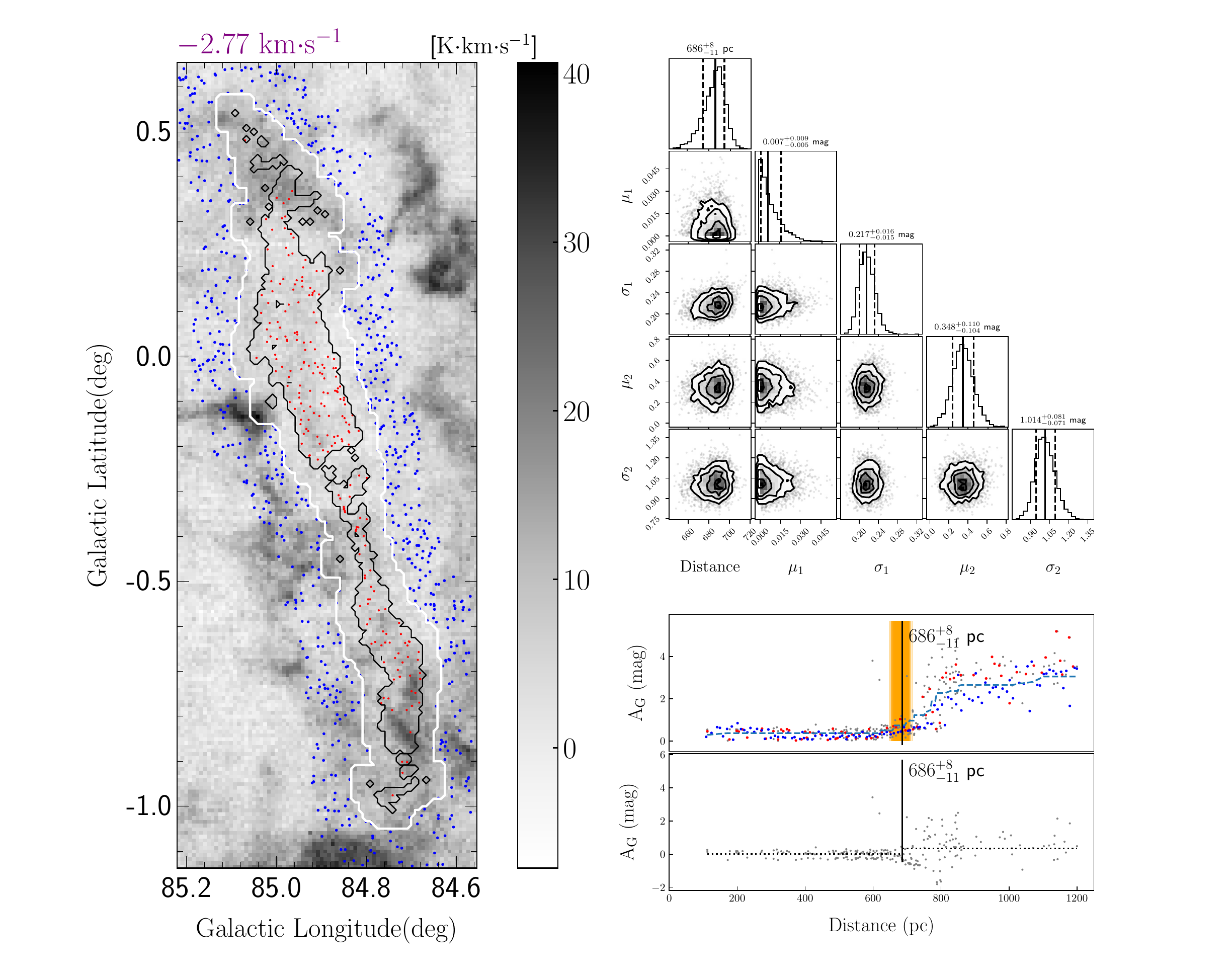}{0.5\textwidth}{(b)}
          }
\caption{The distance measurement of cloud G084.88$-$0.20 (ID 185 in Table \ref{tab:tab1}) from Models A and B. 
%Model A give a significant jump, and it matches better with distances in literature of NAP region. Most of differences in our samples between A and B are within an allowed error limits. 
The Gray-scale map in the left side presents the $\rm ^{13} CO$ integrated intensity of the cloud and its neighbors. Black contour plots the MC structure identified with ACORNS based on the Gaussian decomposition, while white contour expands the MC structure boundary with $5'$. All on-cloud stars located within the cloud are marked as red dots, while referenced stars (blue dots) are chosen out of white contour; see details in Figure \ref{fig:fig4}. Corner plot in the upper right gives the MCMC sampling results from emcee. We set three parameters in our model, nwalker=50, the number of burn-in samples for Markov chain with 1000, and steps with 2000. All binary relations among parameters are presented as 2D Gaussian distributions. We compute median value (solid line) of all samples for result, and $1\sigma$ confidence interval in dashed lines (16th and 84th percentile). Subplot in the lower right presents an $\ A_{G}$--$Distance$ relation for on-cloud stars. The upper one gives $A_{G}$ values before baseline elimination, while the lower one gives the result by subtracting a monotonous fitting of reference stars (see Section \ref{subsec:undertaintyb}). Gray dots are raw data from Gaia DR3, while red and blue dots are binned values in a 10 pc interval for on-cloud and off-cloud stars, respectively. Orange vertical lines denote all the jump points in MCMC sampling. 
Model A give a significant jump, and it matches better with distances in literature of NAP region. 
% We use baseline eliminated $A_{\rm G}$ values for Bayes analysis.
\label{fig:fig5}}
\end{figure*}

\begin{figure*}
\gridline{\fig{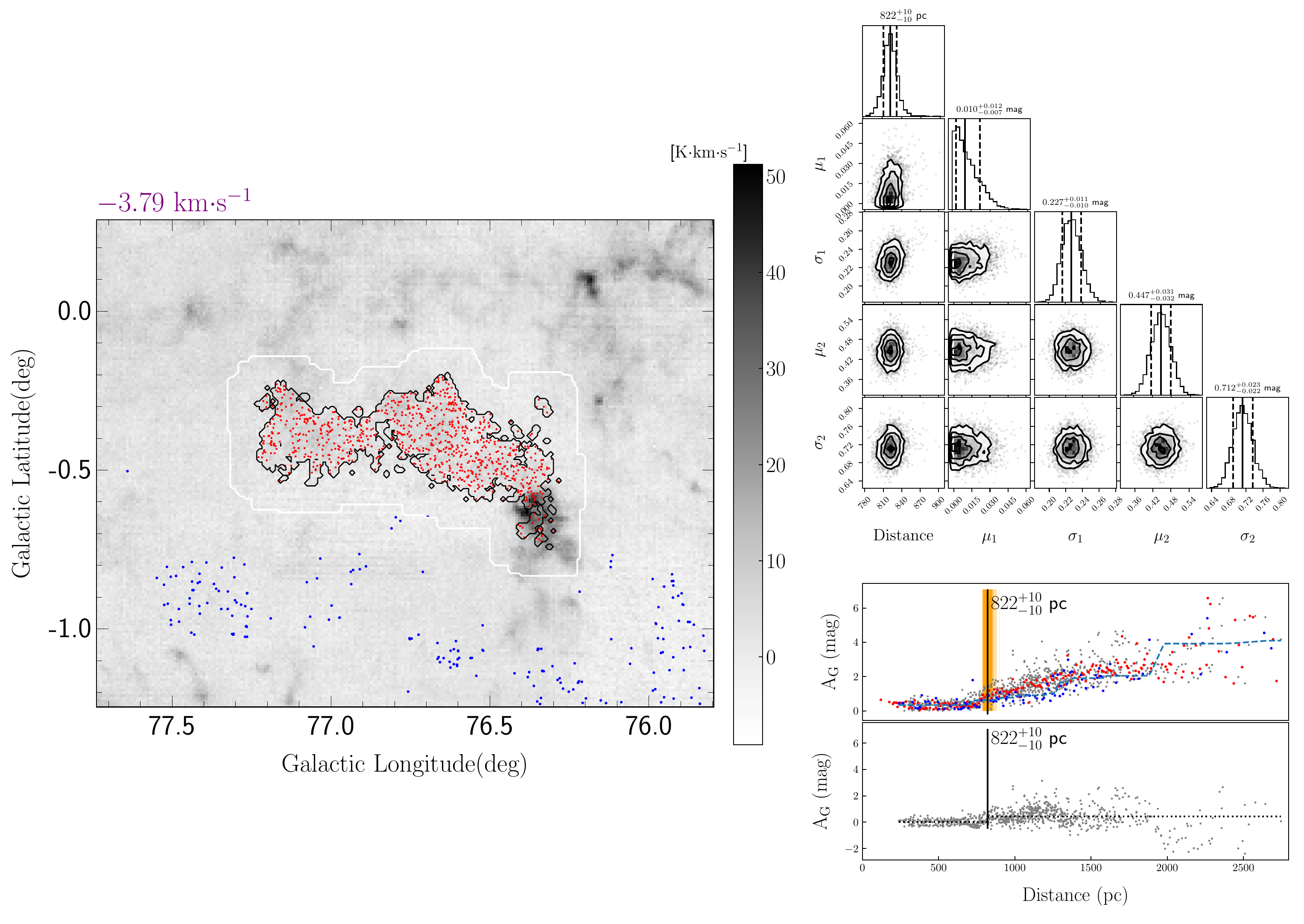}{0.5\textwidth}{(a)}
          \fig{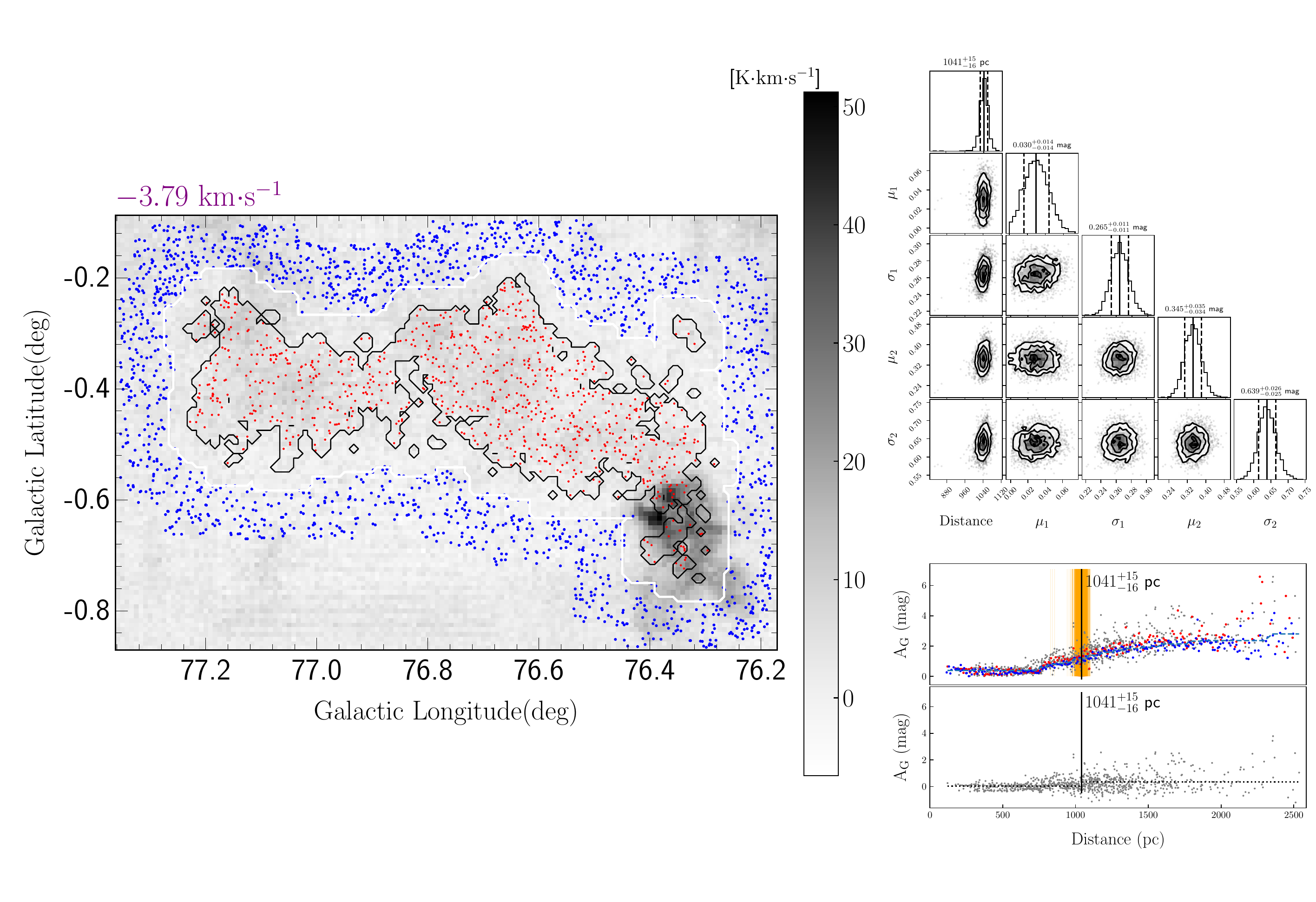}{0.5\textwidth}{(b)}
          }
\caption{The same as Figure \ref{fig:fig5}, but for cloud G76.74-0.43 (ID 30 in Table \ref{tab:tab1}). The fluctuation of $A_{G}$--$Distance$ relation for on-cloud stars for Model A denotes worse baseline fitting. We thus choose the result from Model B (see Section \ref{subsec:undertaintyb}). 
\label{fig:fig6}}
\end{figure*}

\begin{figure}[!htb]
\centering
\hspace{0\linewidth}
\includegraphics[width=1.\linewidth]{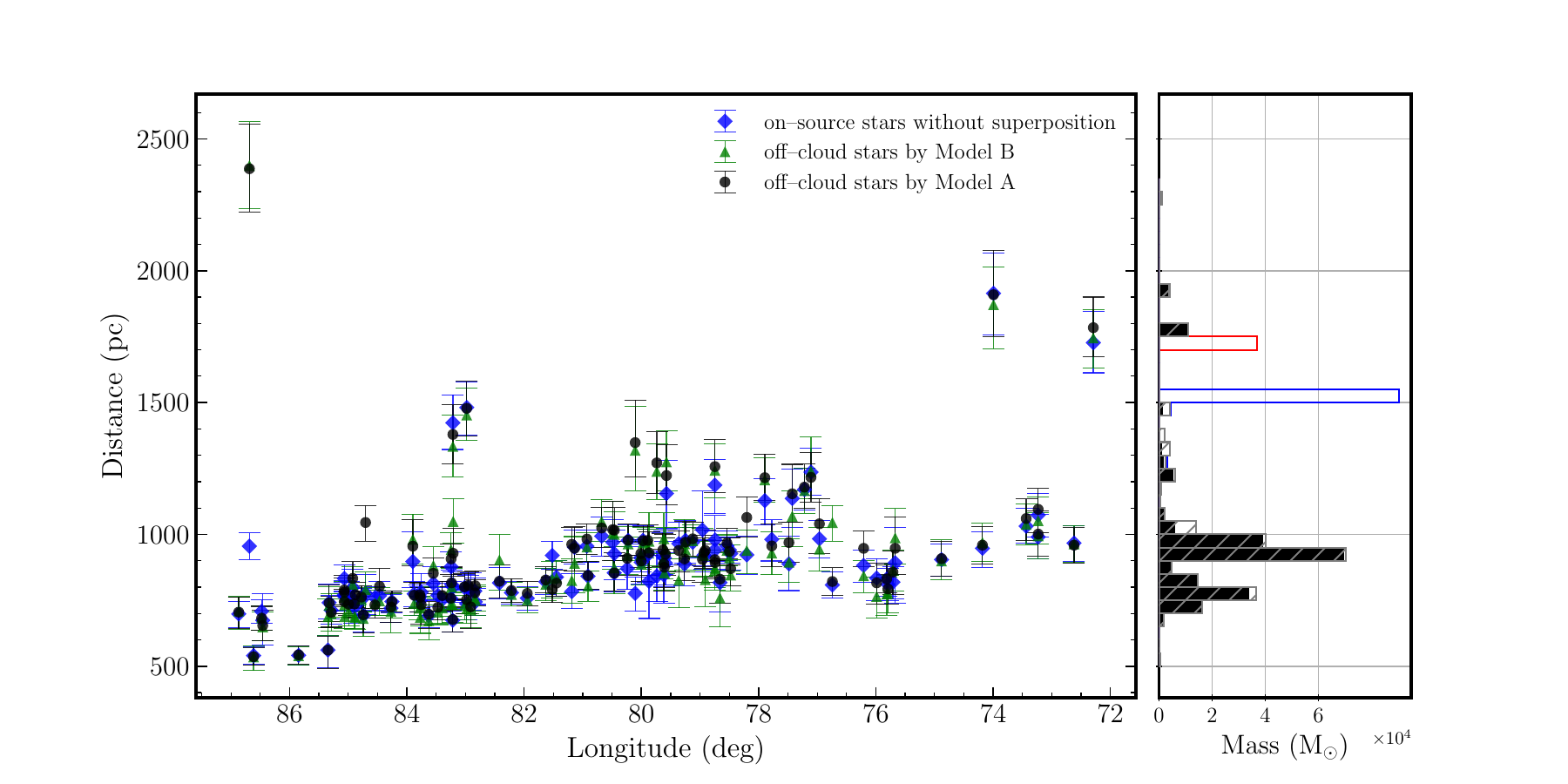}
\caption{Left panel: Our results for 120 robust distance measurements both in Models A and B (class I, see details in Section \ref{subsec:mcp}). We find that all differences between Model A and Model B are smaller than 150 pc, except cloud G76.74-0.43 (ID 30 in Table \ref{tab:tab1}). 
Right panel: LTE mass statistics based on our cloud catalog, i.e. black histogram from samples of class I in Table \ref{tab:tab1}, gray histogram with slashes included class II results from Table \ref{tab:tab1}, while blue and red histograms from Tables \ref{tab:tab2} and \ref{tab:tab3}. 
%Solid grey bins are clouds associated with masers, and the dotted grey bins are those dense clouds along filamentary structure in Cygnus North, and we propose they are in the same distance with DR21 and DR20. 
%While the red bins are clouds matching mid-IR emission, we put them in 1.7 kpc.
\label{fig:fig7}}
\end{figure}

\begin{figure}[htbp]
%\plotone{c311.pdf}
\centering
\includegraphics[width=18cm,angle=0]{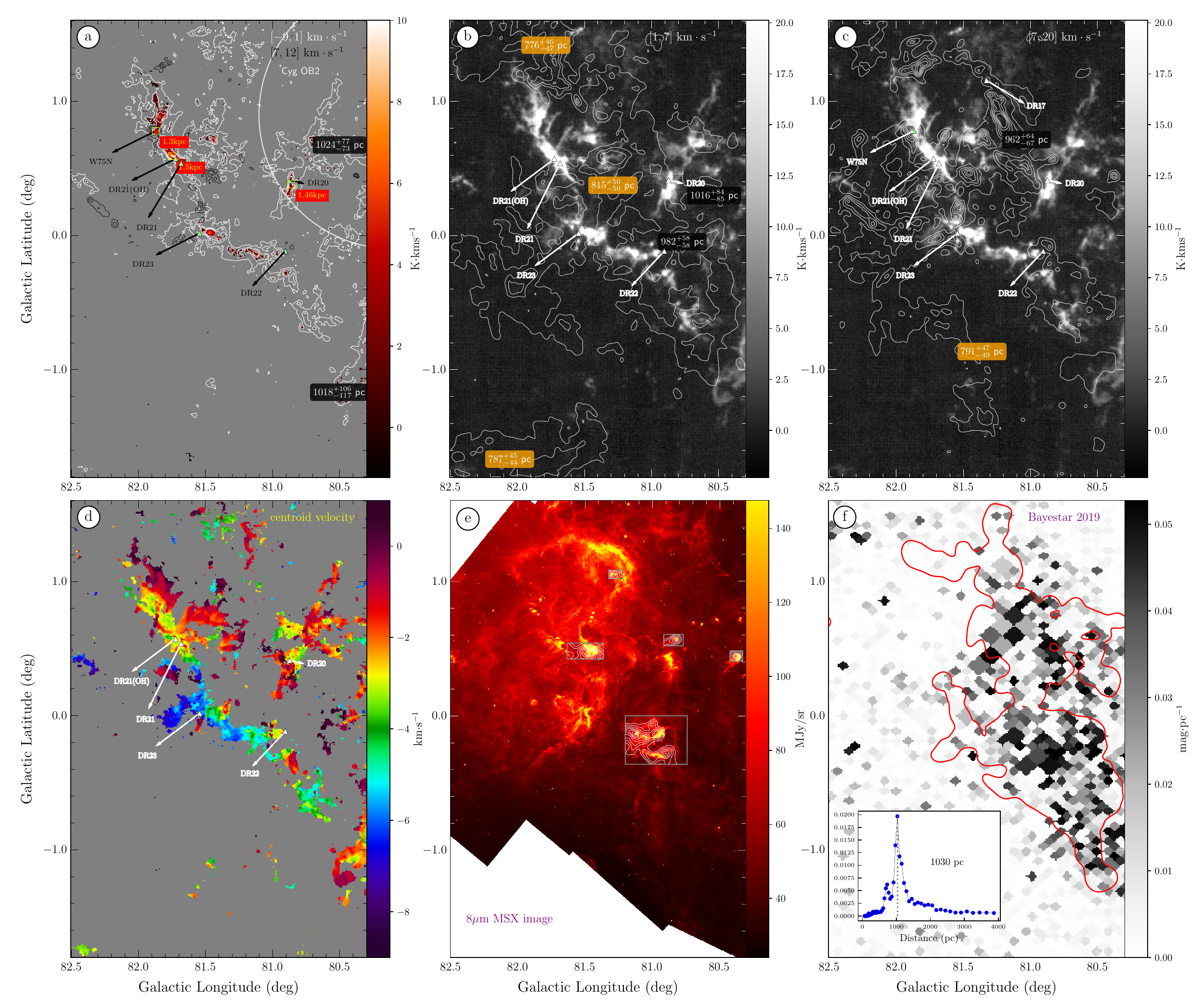}
\caption{Subplots for the Cygnus X North region. Panel (a) shows $\rm ^{13}CO$ (white contours in [5, 20, 35, 50] $\times \sigma$, $\sigma$ is the noise level of intensity map) and $\rm C^{18}O$ (colormap) emission in [$-9$, 1] $\rm km~ s^{-1}$. Massive star-forming regions are sited along the dense filamentary structure, i.e. DR21, DR23, and DR22. 
%we keep the mask where all pixels have three continuous channels bigger than $3\sigma$, here $\sigma$ is the noise level of $\rm C^{18}O$, $\sigma \sim 0.3K$, 
Black contours also denote the dense molecular gas in [7, 12] $\rm km~s^{-1}$ in same levels, e.g. W 75N is overlapped on the DR21 filament. Some HII regions are marked with triangles. The extent of Cygnus OB2 is plotted in a white circle. Distances with 5\% system error of our measured clouds and those from masers are presented; different gas layers are also labeled with different colors. Panel (b): the gray-scale map is $\rm ^{13}CO$ emission from velocity interval of [$-9$, 1] $\rm km ~s^{-1}$, overlaid with smoothed data from [1, 7] $\rm km~ s^{-1}$ (white contours in [2, 6, 10, 14, 18, 22] $\rm K~ km~s^{-1}$). Panel (c): all the same with panel (b), but contours for velocities in [7, 20] $\rm km ~s^{-1}$. Panel (d): A moment 1 map of centroid velocities extracted from Gaussian decomposition for clouds in [$-9$, 1] $\rm km ~s^{-1}$. Obviously, prominent velocity gradient along filamentary structure can be seen. Panel (e): An 8$\mu$m Midcourse Space Experiment (MSX) image toward the Cygnus X North region. Gray edged rectangles mark the identified molecular cloud with the associated mid-IR features. The extent of these rectangles is derived from the $\pm3\sigma$ of v, l, b relative to cloud center in Table \ref{tab:tab3}. Panel (f): A 1030 pc slice of 3D dust map from Bayestar 2019 \citep{2019ApJ...887...93G}, overlaid with the 1 kpc MC layer (red contours) from our samples (class I in Table \ref{tab:tab1}) with well-determined distances. Note that our results are in good agreement with \citet{2019ApJ...887...93G}. \label{fig:fig8}}
\end{figure}

\begin{figure}[htbp]
%\plotone{c311.pdf}
\centering
\includegraphics[width=18cm,angle=0]{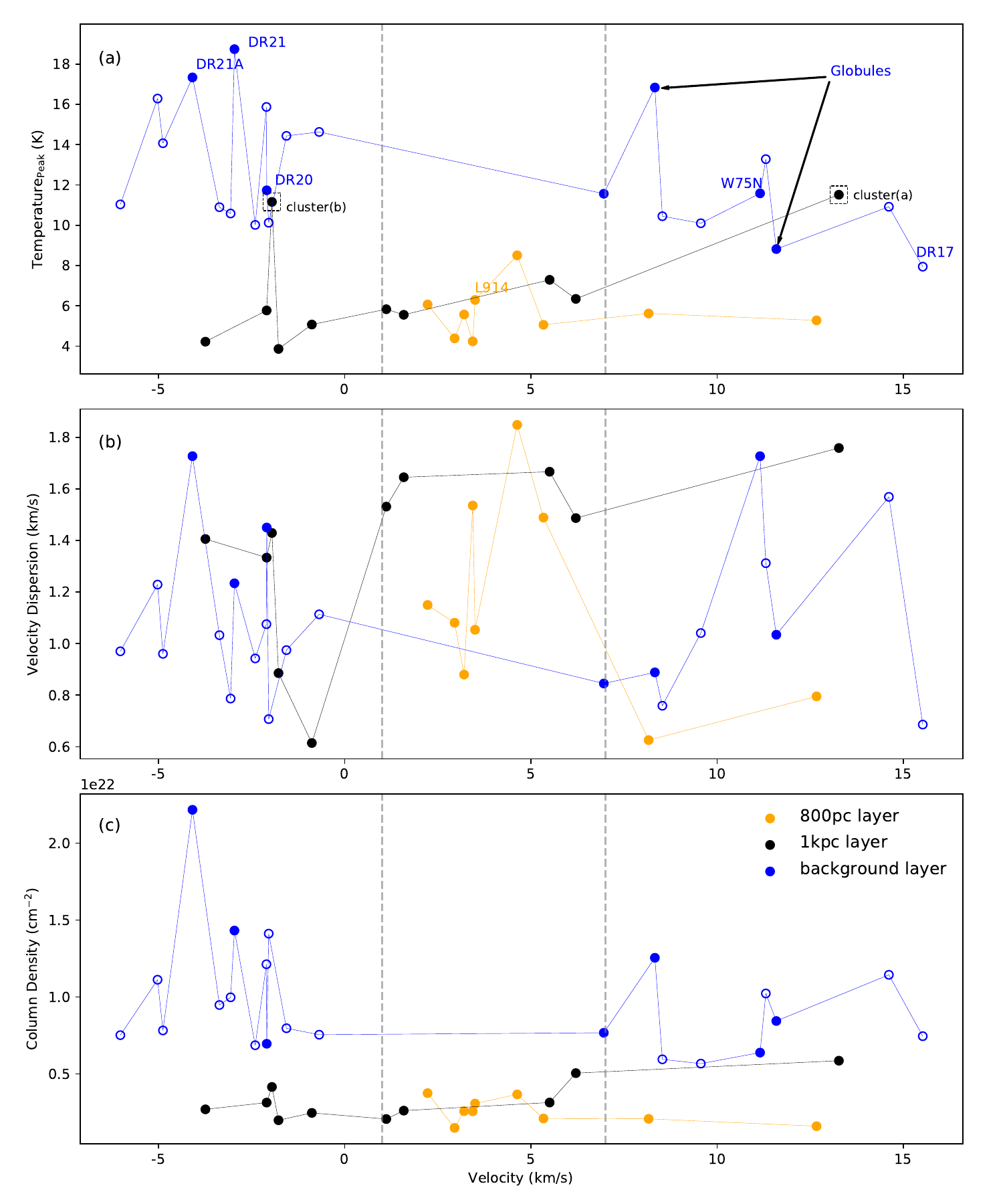}
\caption{ Physical properties of clouds in the Cygnus X North region. Panel (a): $\rm ^{13}CO$ peak temperature of MCs. Different colors are marked for clouds in different layers.  Clouds in background layer are associated with masers or bright cometary mid-IR features (filled blue dots) and the nearby dense gas structures (empty dots) along the filamentary structure of Cygnus X North star-forming region. Panels (b) and (c), the same with panel (a), but for velocity dispersion and column densities of the MCs. \label{fig:fig9}}
\end{figure}

\begin{figure*}
\gridline{\fig{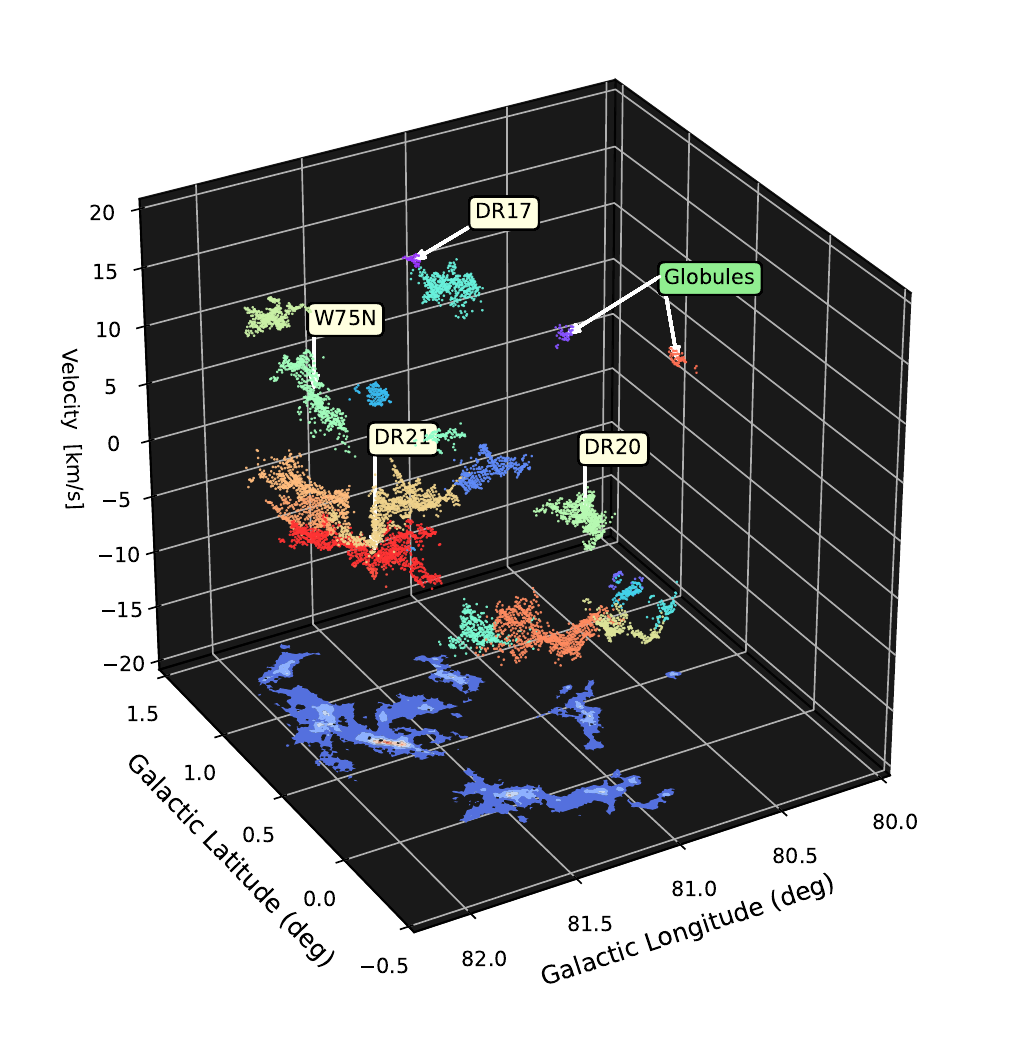}{0.48\textwidth}{(a)}
          \fig{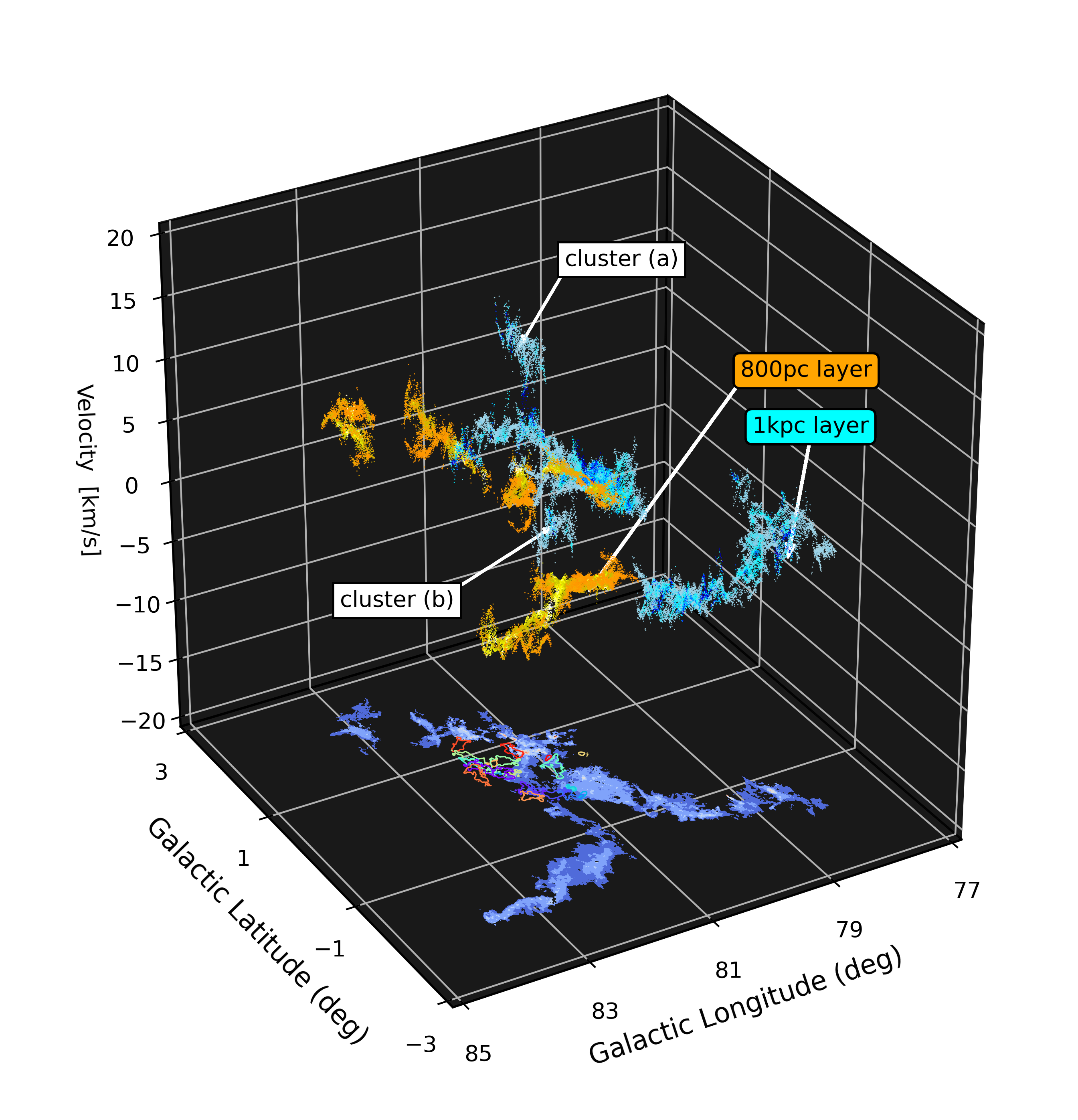}{0.48\textwidth}{(b)}
          }
\caption{3D plot of different gas layers toward the Cygnus X North region. Panel (a): Scatter plot of clouds in background layer (i.e., $1.3\sim1.7$ kpc). Colors mark different $\rm ^{13}CO$ structures identified by ACORNS, and each dot presents pixel-by-pixel centroid velocities. Their summed intensities are projected at the bottom. Panel (b): Scatter plot of clouds in 800 and 1 kpc layers. Clouds in 800 pc are in orange, while clouds in 1 kpc are in cyan. Both of the two layers display large filamentary complexes overlapped together. Clouds in $1.3\sim1.7$ kpc layer are also plotted as colored contours on the bottom, which shows the spatial relation of different clouds on projection. 
\label{fig:fig10}}
\end{figure*}

\begin{figure}[htbp]
%\plotone{c311.pdf}
\centering
\includegraphics[width=18cm,angle=0]{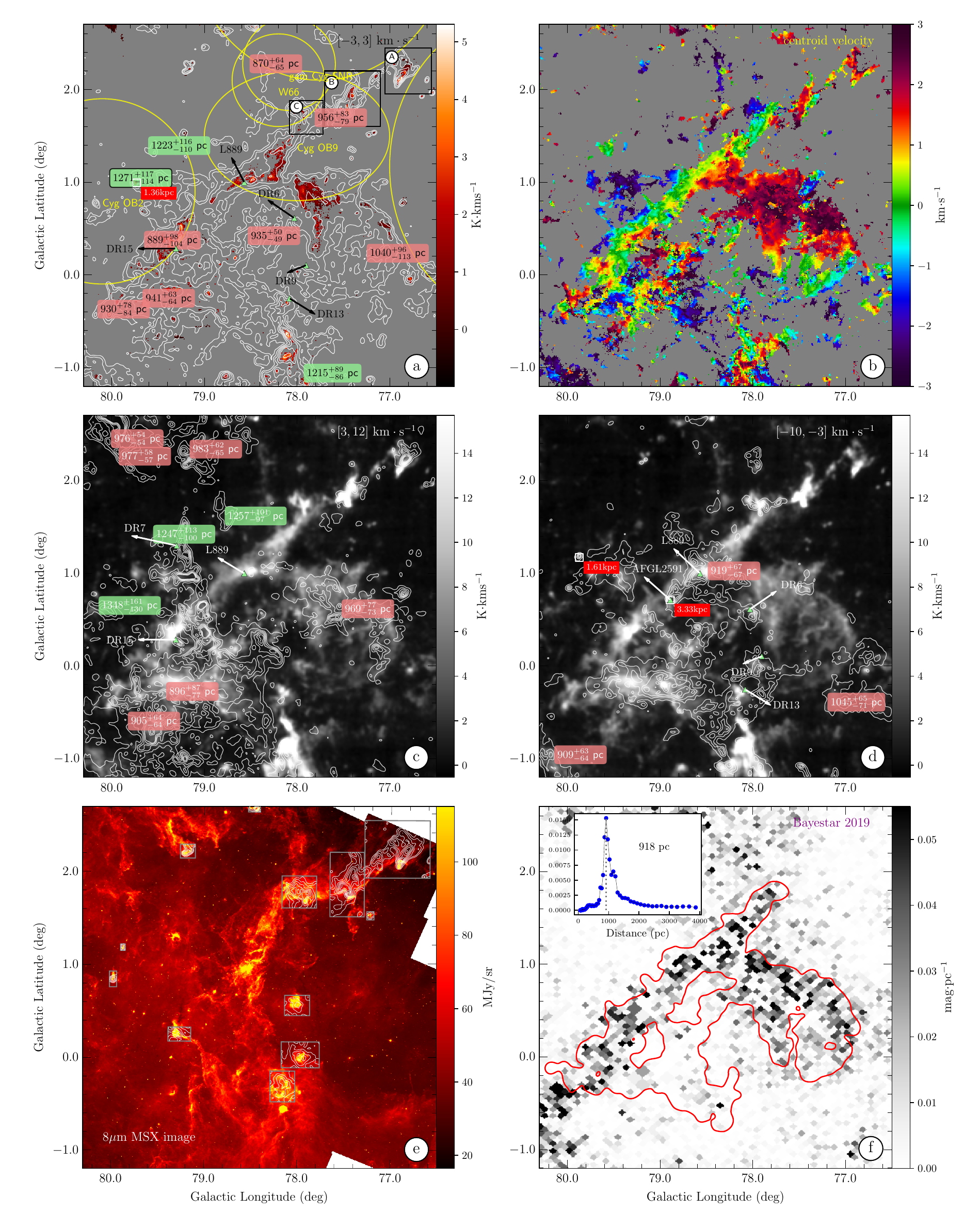}
\caption{Subplots toward the Cygnus X South region. Panel (a): $\rm ^{13}CO$ (white contours in [5, 10, 20, 35, 50, 70] $\times \sigma$, $\sigma$ is the noise level of intensity map) and $\rm C^{18}O$ (colormap) emission in [$-3$, 3] $\rm km~ s^{-1}$. These HII regions and dense cores of L889 are marked with triangles. The extent of OB associations and SNRs are plotted in yellow ellipses. Distances (with 5\% system error) of our measured clouds and those from masers are also presented. Panel (b): moment 1 map in [$-3$, 3] $\rm km~ s^{-1}$. Panel (c): the grey scale map is $\rm ^{13}CO$ emission from [$-3$, 3] $\rm km ~s^{-1}$, overlaid with smoothed data from [3, 12] $\rm km~ s^{-1}$ with levels [10, 25, 40, 55, 70, 85] $\times \sigma$. Panel (d): all the same with panel (c), but contours for velocities in [$-10$, $-3$] $\rm km~ s^{-1}$. Panel (e): same with Figure \ref{fig:fig8}e, but for Cygnus X South region. Panel (f): a 918 pc slice of 3D dust map from Bayestar 2019 \citep{2019ApJ...887...93G}, overlaid with L889 cloud at $\sim 935$ pc (red contour) from our samples (see cloud 57 in Table \ref{tab:tab1}).  \label{fig:fig11}}
\end{figure}

\begin{figure}[htbp]
%\plotone{c311.pdf}
\centering
\includegraphics[width=18cm,angle=0]{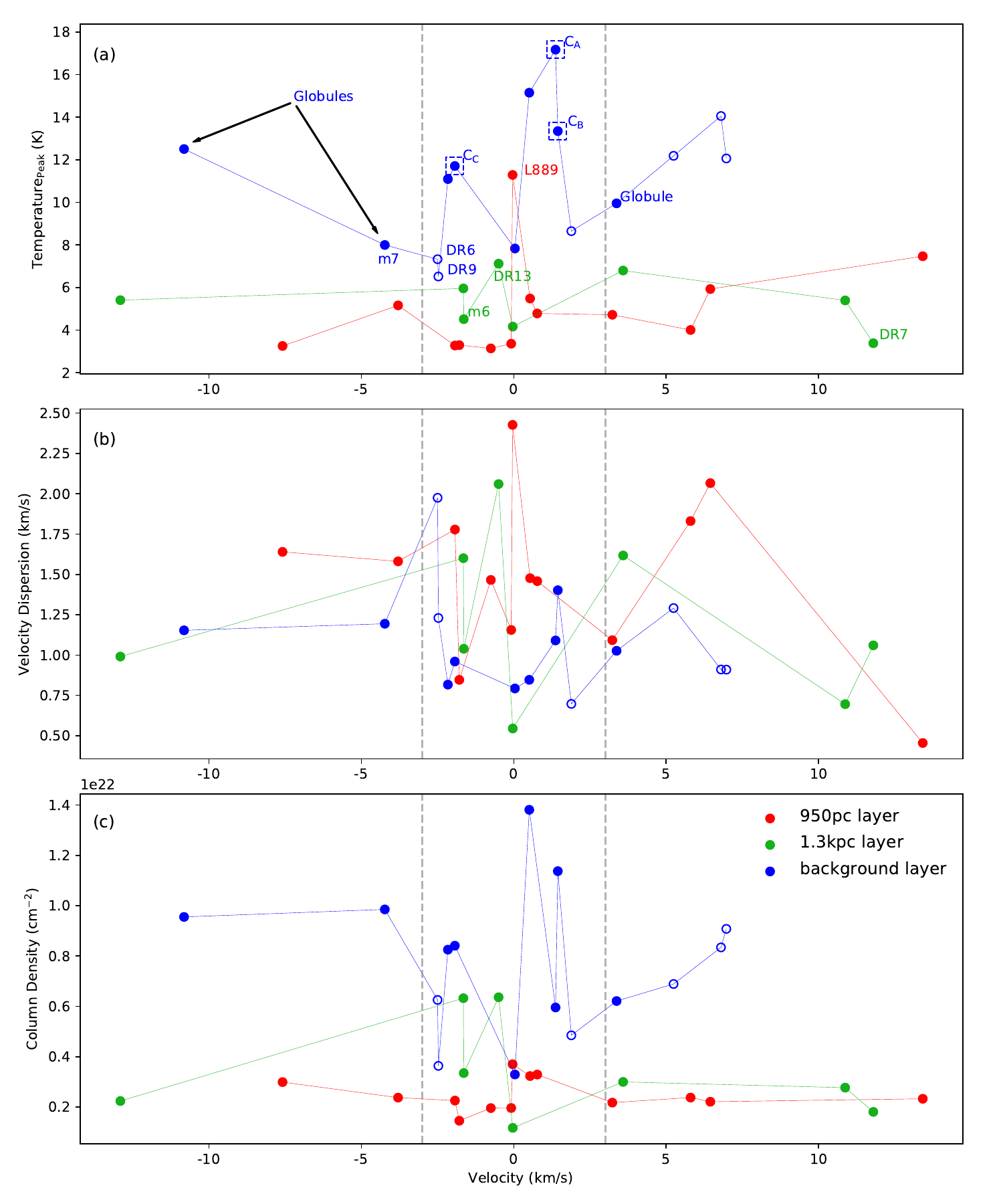}
\caption{Physical properties of clouds in the Cygnus X South region. Panel (a): $\rm ^{13}CO$ peak temperature of MCs. Different colors are marked for clouds in different gas layers.  Clouds in background layer (blue dots) are all associated with bright mid-IR features. Clouds in filled dots have cometary or globular morphology, which is a strong evidence that has been shaped by massive OB star groups, e.g., cloud A, B marked by \citet{2006A&A...458..855S}, while clouds in empty dots are oval shaped or irregular, e.g. DR6 and DR9. Panels (b) and (c): the same with panel (a), but show velocity dispersion and column densities of the clouds. \label{fig:fig12}}
\end{figure}

\begin{figure*}
\gridline{\fig{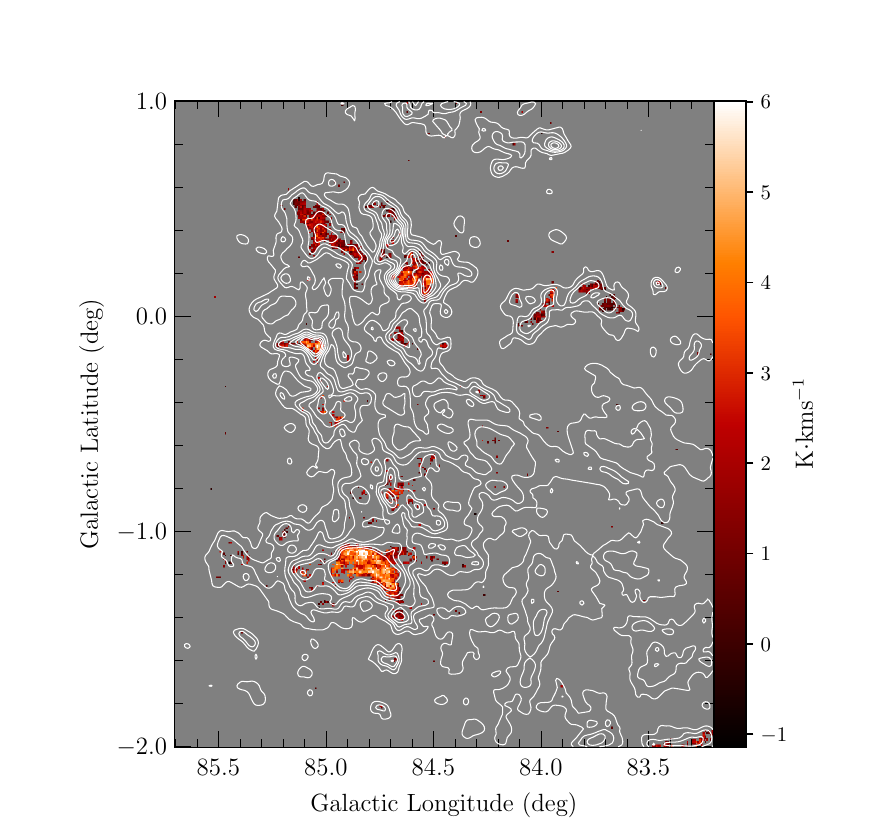}{0.48\textwidth}{(a)}
          \fig{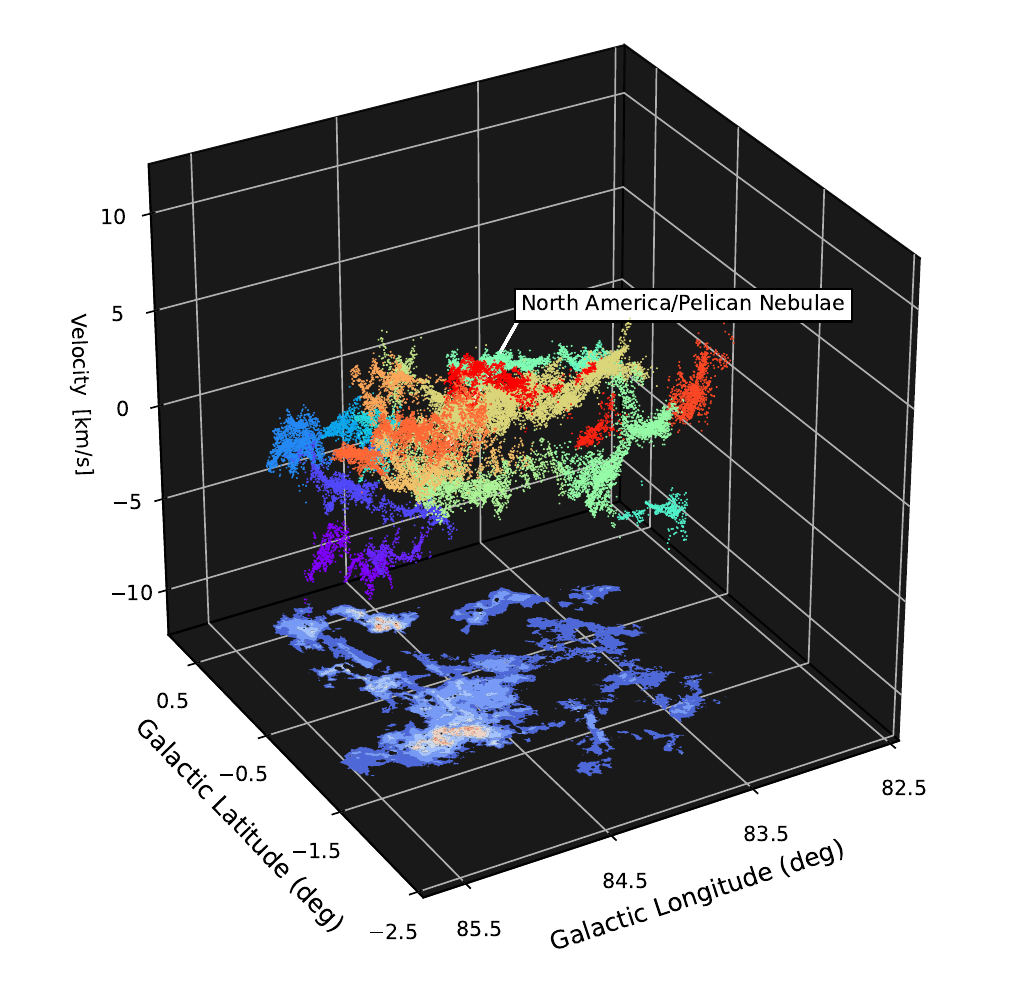}{0.48\textwidth}{(b)}
          }
\caption{Panel (a): $\rm ^{13}CO$ (white contours in [10, 20, 30, 45, 60, 75, 90] $\times \sigma$, $\sigma$ is the noise level of intensity map) and $\rm C^{18}O$ (colormap) emission in [$-10$, 10] $\rm km~ s^{-1}$ toward the NAP region. Panel (b): a 3D plot for all $\rm ^{13} CO$ structures with measured distances in the region.
\label{fig:fig13}}
\end{figure*}

\begin{figure*}
\gridline{\fig{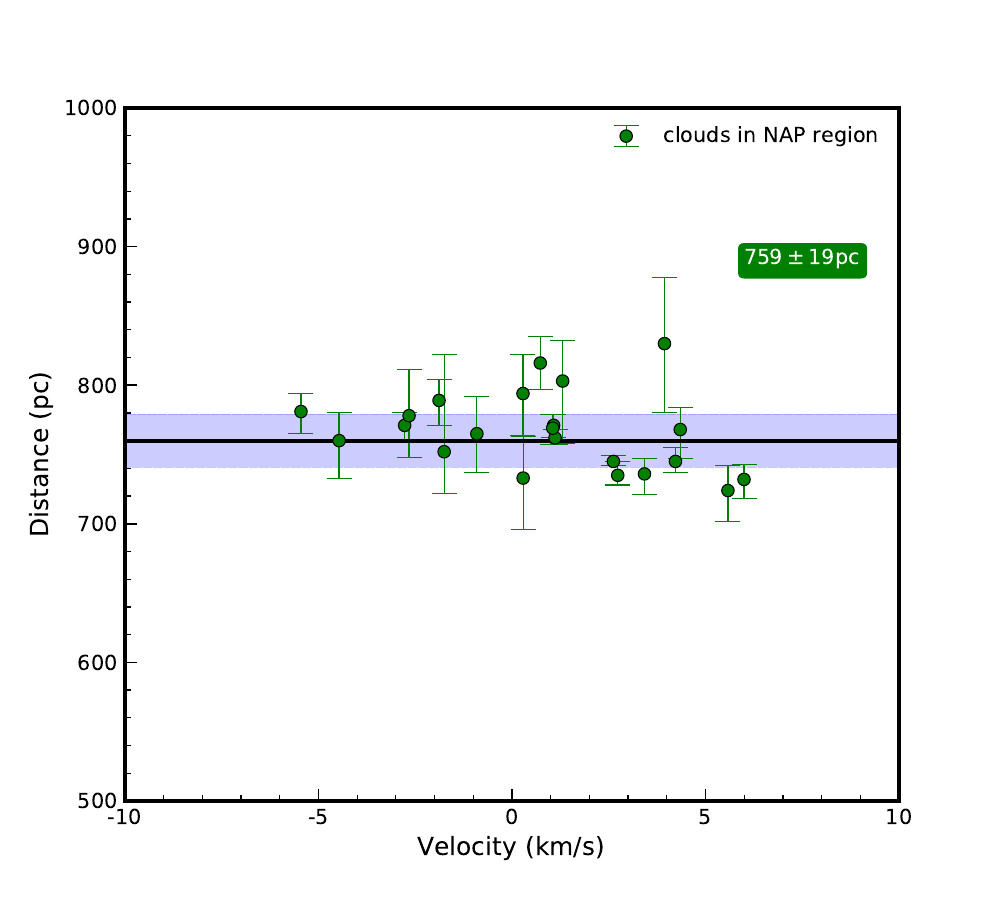}{0.48\textwidth}{(a)}
          \fig{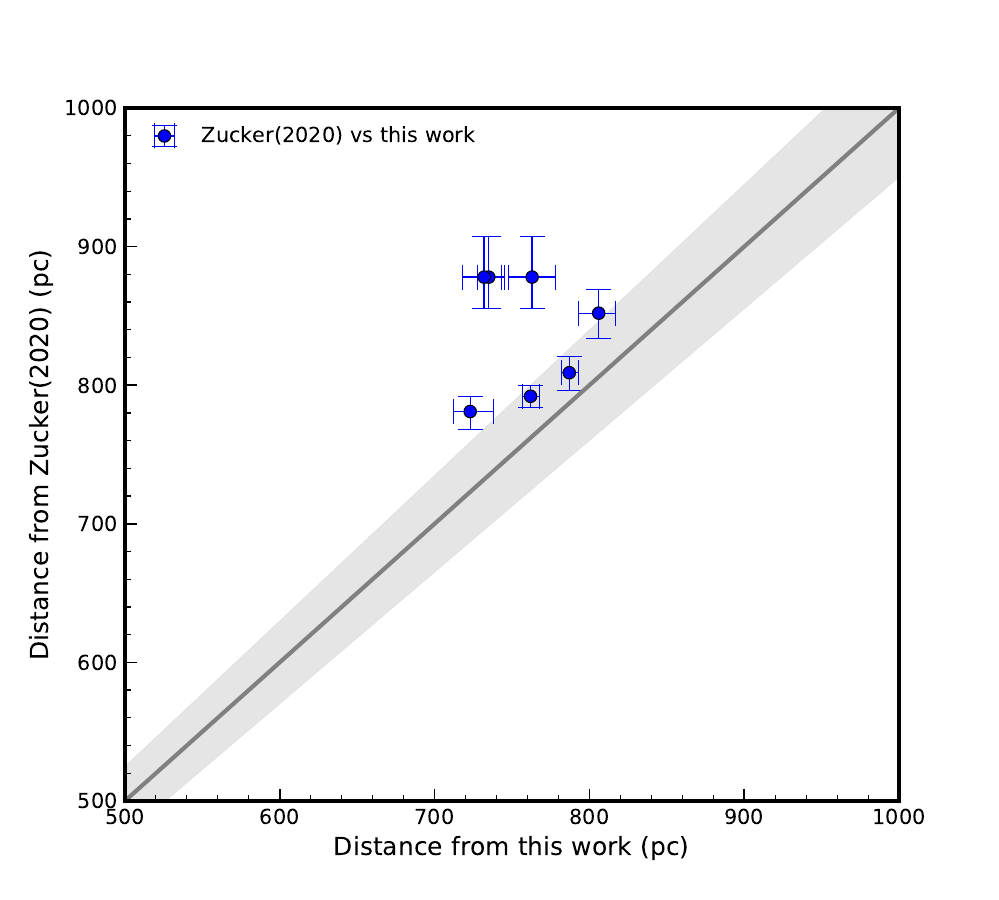}{0.48\textwidth}{(b)}
          }
\caption{Panel (a): Measured distances of different $\rm ^{13}CO$ structures toward the NAP region. Black line and blue shadow show the averaged distance and dispersion (without the 5\% system error) weighted by clouds' masses. Panel (b): distances compared to \citet{2020A&A...633A..51Z} with labeled 5\% system error (gray shadow) relative to distance; a systematic deviation can be found (see details in Section \ref{subsec:nap}). 
\label{fig:fig14}}
\end{figure*}

\begin{figure}[htbp]
%\plotone{c311.pdf}
\centering
\includegraphics[width=18cm,angle=0]{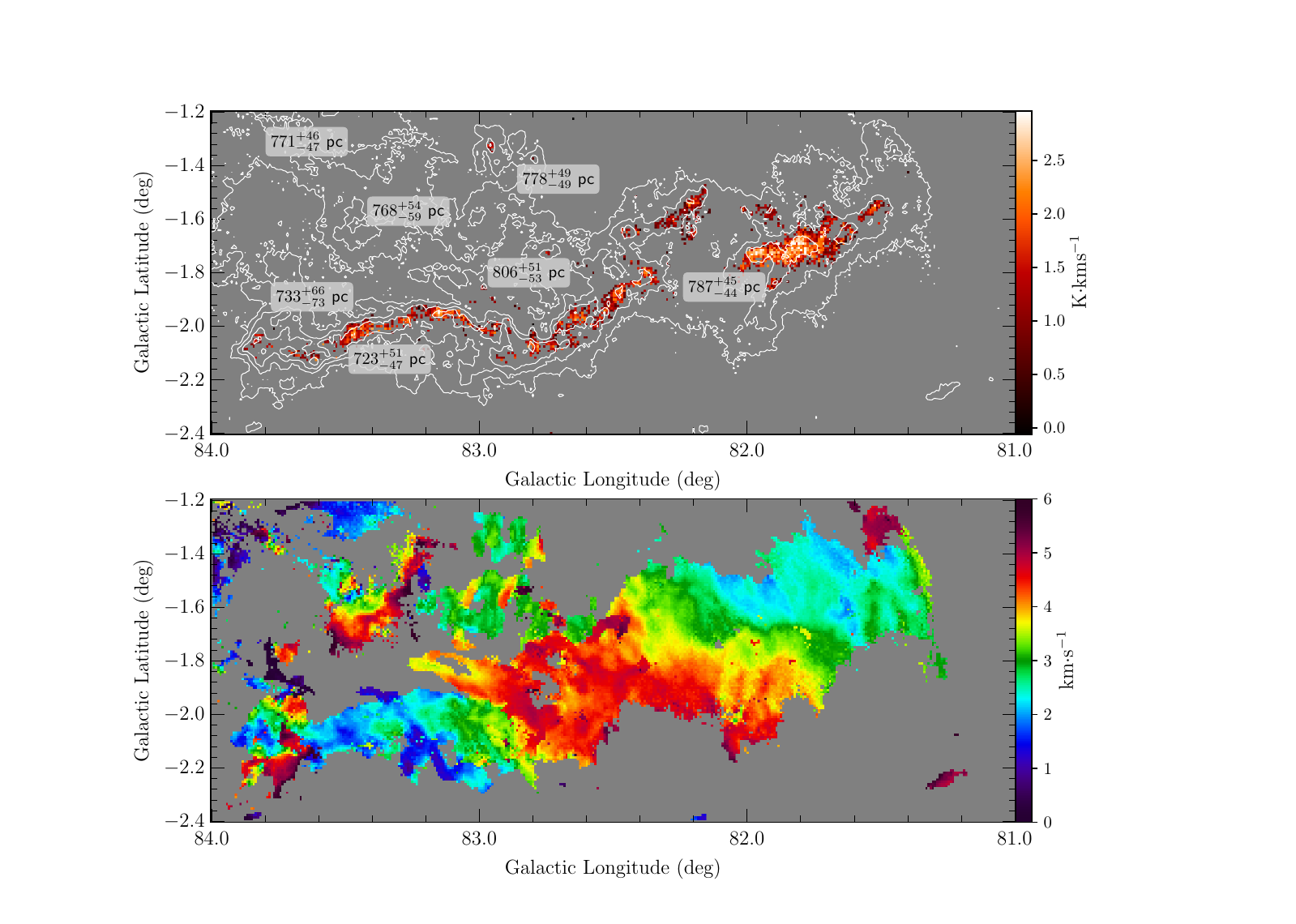}
\caption{Upper panel: $\rm ^{13}CO$ (white contours in [5, 15, 25, 35, 45, 55] $\times \sigma$, $\sigma$ is the noise level of intensity map) and $\rm C^{18}O$ emission (colormap) in [0, 6] $\rm km~ s^{-1}$ toward L914. 5\% system error of the distances is labeled for identified $\rm ^{13}CO$ structures. Lower panel: moment 1 map of centroid velocities extracted from Gaussian decomposition for the above velocity interval.  \label{fig:fig15}}
\end{figure}

\begin{figure}[htbp]
%\plotone{c311.pdf}
\centering
\includegraphics[width=18cm,angle=0]{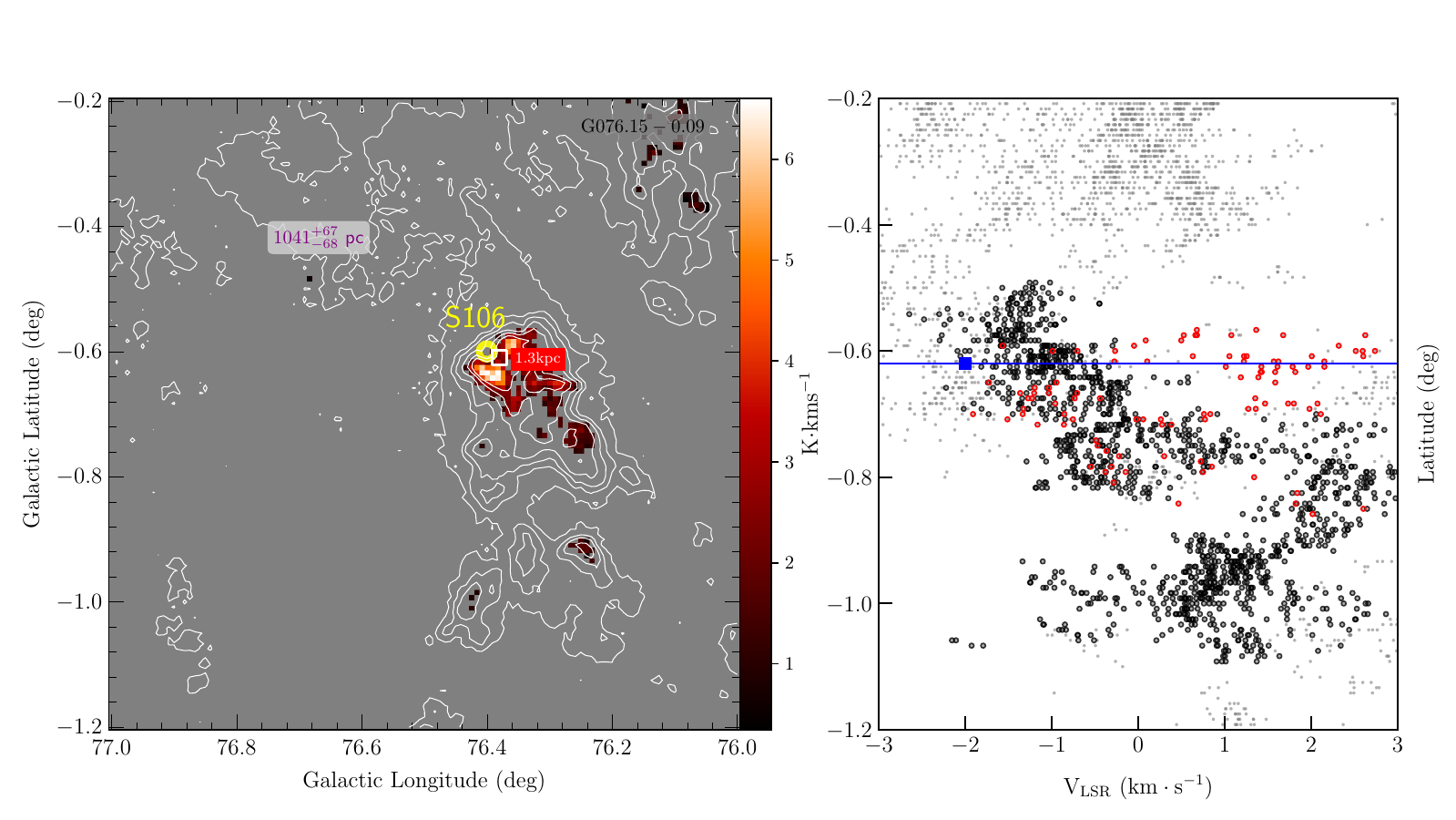}
\caption{Left panel: $\rm ^{13}CO$ (white contours in [5, 15, 25, 35, 45, 55] $\times \sigma$, $\sigma$ is the noise level of intensity map) and $\rm C^{18}O$ emission (colormap) in [$-3$, 3] $\rm km~s^{-1}$ toward S106. 5\% system error of the distances is labeled for identified $\rm ^{13}CO$ structures. The location of HII region (yellow circle) and maser (white box) are also labeled on the map. Right panel: %Moment 1 map of centroid velocities extracted from Gaussian decomposition. The insert panel show 
the different velocity structures labeled in black and red dots in the velocity--latitude coordinates from Gaussian decomposition (see details in Section \ref{subsec:s106}). The blue square with error represents the observed maser from \citet{2013ApJ...769...15X}.  \label{fig:fig16}}
\end{figure}

\begin{figure}[htbp]
%\plotone{c311.pdf}
\centering
\includegraphics[width=18cm,angle=0]{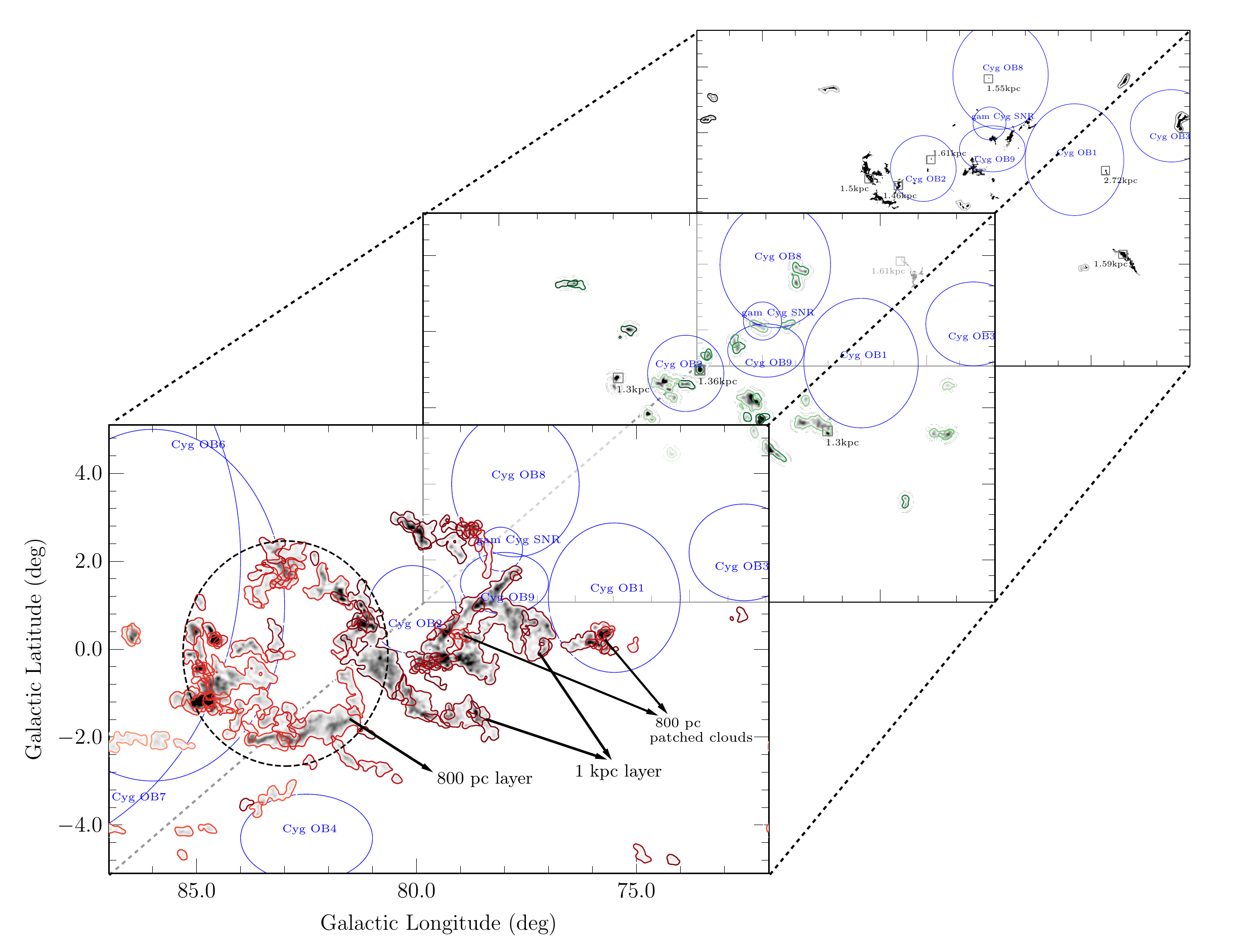}
\caption{Big picture of multiple gas layers toward the Cygnus region. We divide clouds with distance measurement (see Tables \ref{tab:tab1}, \ref{tab:tab2}, and \ref{tab:tab3}) into three intervals: [500, 1000] pc in red, [1000, 1400] pc in green, and [1400, 3000] pc in black. Lighter contours indicate nearer distance for each layer, while the darker shows the farther distances. Gray-scale maps are reconstructed by identified clouds from ACORNS, while their boundaries are delineated using a low-pass filtering method ($butterworth$ in Python). The dashed ellipse shows a large-scale molecular loop with diameter of $\sim56$ pc at a distance of 800 pc. Different layers are also annotated in the [500, 1000] pc map. Middle map includes two layers, the $\approx$ 1 kpc clouds in light green, and $\sim$ 1.3 kpc clouds in green. Clouds in $\geqslant$ 1.4 kpc layer without contours are MCs associated with masers and/or mid-IR bright features in Tables \ref{tab:tab2} and \ref{tab:tab3}. 7 clouds in Table \ref{tab:tab1} are successfully measured in this layer. Blue ellipses show OB associations and SNR, while boxes show the masers in Table \ref{tab:tab2}.  \label{fig:fig17}}
\end{figure}

\clearpage

\appendix

\section{Coherence of identified large structures} \label{app:appa}

The large-size MC structures traced by $\rm ^{13}CO$ usually display a hierarchical and spatially extended structure (see details in Section \ref{subsec:gaussian decomposition} and \ref{subsec:clustering}). See Figure \ref{fig:figa1}a for an example; this large cloud, at 982 pc (see cloud 118 in Table \ref{tab:tab1} and Figure \ref{fig:figa1}b) spanning $2^{\circ}$ along latitude, probably suffers a different extinction environment. 
%Here, the 0.25 deg width squares were chosen along the structure to cover various sub-regions as well as enough stars. We leverage the model A to estimate the sub-regions' distances, use the same field stars for reference, their distances and velocity are shown as blue squares on the plot;
%We another took the secondary trunks in the hierarchy provided by ACORNS, stars with the boundary of trunks were chosen to be on-cloud stars, then we implement model a the same way to derive their distances , which are shown in green triangles. 

We choose subregions along different LOS to measure their distances to check whether the substructures belong to a physical cloud.
Firstly, the 225 $\rm arcmin^{2}$ boxes were chosen along the structure to cover various subregions as well as enough on-cloud stars. 
Alternatively, we took the secondary trunks in the hierarchy provided by ACORNS to include on-cloud stars.
Using the same field stars for reference, we implement Model A the same way to derive their distances (blue squares for the boxes' samples and green triangles for the trunks in Figure \ref{fig:figa1}c).

We find that the measured distance of the whole cloud is consistent with the mean value of the distances of every small part (see Figures \ref{fig:figa1}bc). It indicates that the whole structure identified by our method does not suffer from significant contamination of unrelated emission at different distances. 
Similar results can also be found toward other large-scale structures, e.g. Figure \ref{fig:figa2}. 
We suggest the large-scale structures identified by the $\rm ^{13}CO$ emission are at similar distances. 

Some subregions probably tend to be contaminated by other different MC structures along LOS (see the left three samples in Figure \ref{fig:figa1}c). These small substructures in the east side are measured to be nearer (i.e. $\sim$800 pc). We find that these subregions are just located at the overlapping regions of 800 pc and 1 kpc gas layers of the Cygnus X North (see Figure \ref{fig:figa1}d and details in Section \ref{subsec:cygn}). Our further study with the multijump detection model (in preparation) indeed shows that two layers of molecular gas are overlapped in the direction, i.e. $\sim$800 pc and $\sim$1 kpc gas in the east side of the cloud. 

For another case of L889 cloud (Figure \ref{fig:figa2}), the deviation of distance measurements in individual regions likely indicates the contamination from the farther 1.3 kpc layer and foreground emission (see Figure \ref{fig:figa2}d and details in Section \ref{subsec:cygs}). The above discussions tell us that the overlapping between different identified structures may lead to deviations of distance measurements for subregions of a cloud. 
%the contamination from more extended dust extinction in other layers should be more severe.

\renewcommand\thefigure{\Alph{section}\arabic{figure}}
\setcounter{figure}{0}
\begin{figure}[!htb]
\centering
\hspace{0\linewidth}
\includegraphics[width=1.\linewidth]{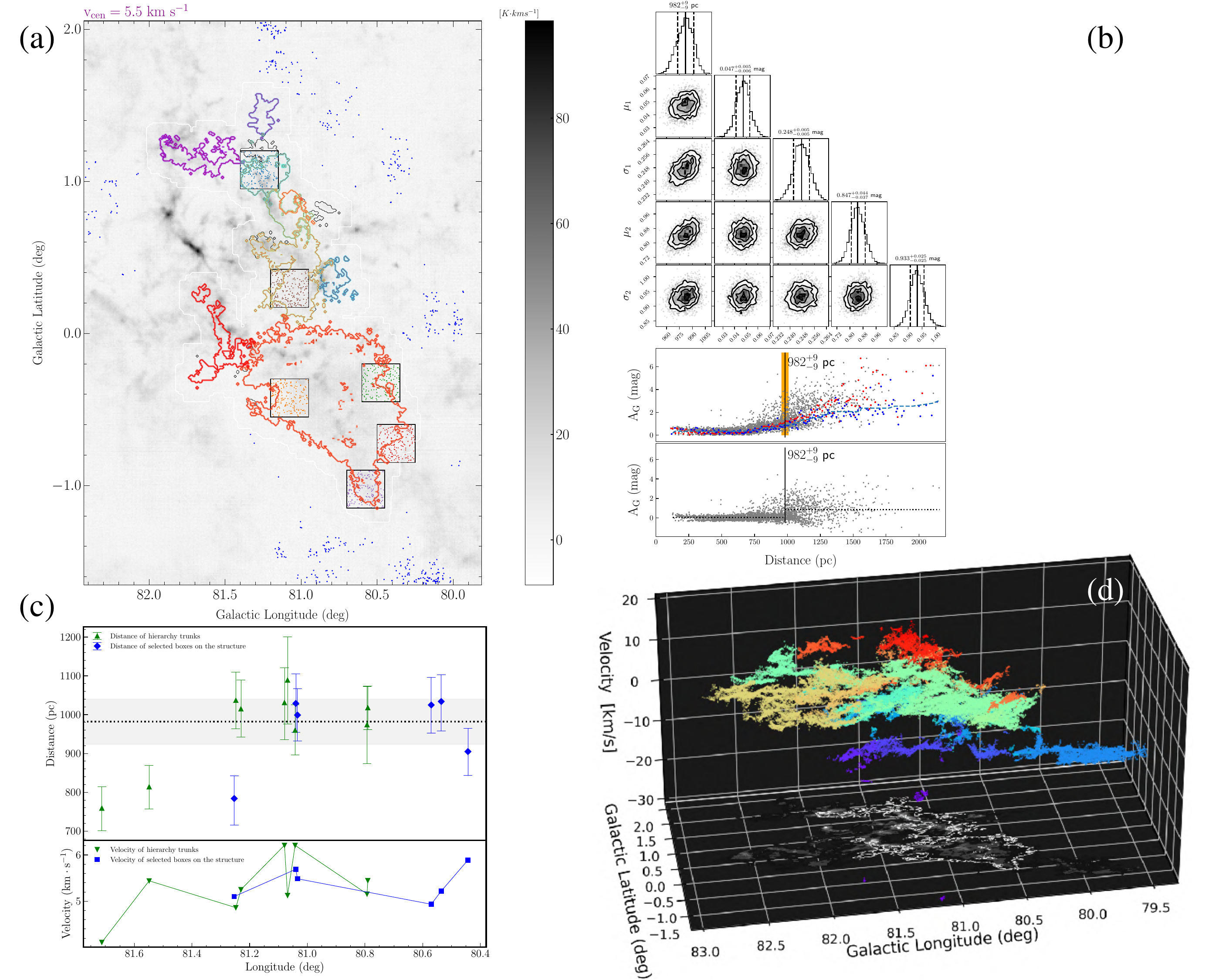}
\caption{A validation of coherent structures (ID 118 in Table \ref{tab:tab1}) identified by ACORNS based on Gaussian decomposition. We chose substructures of the large MC structure by manually marked boxes and hierarchical trunks (colored contours). Their distances are marked by blue squares and green triangles, respectively, while their confidence intervals (added 5\% system error of Gaia) are denoted by error bars. The $1\sigma$ confidence interval of distance is plotted in gray area of panel (c). In lower right panel, it is a 3D PPV scatter plot of centroid velocities of the cloud. All structures (other colors) overlaid with the cloud (in green) along LOS are also drawn, showing coherent velocity structures in PPV space.   
\label{fig:figa1}}
\end{figure}
\setcounter{figure}{0}

\renewcommand\thefigure{\Alph{section}\arabic{figure}}
\setcounter{figure}{1}
\begin{figure}[!htb]
\centering
\hspace{0\linewidth}
\includegraphics[width=1.\linewidth]{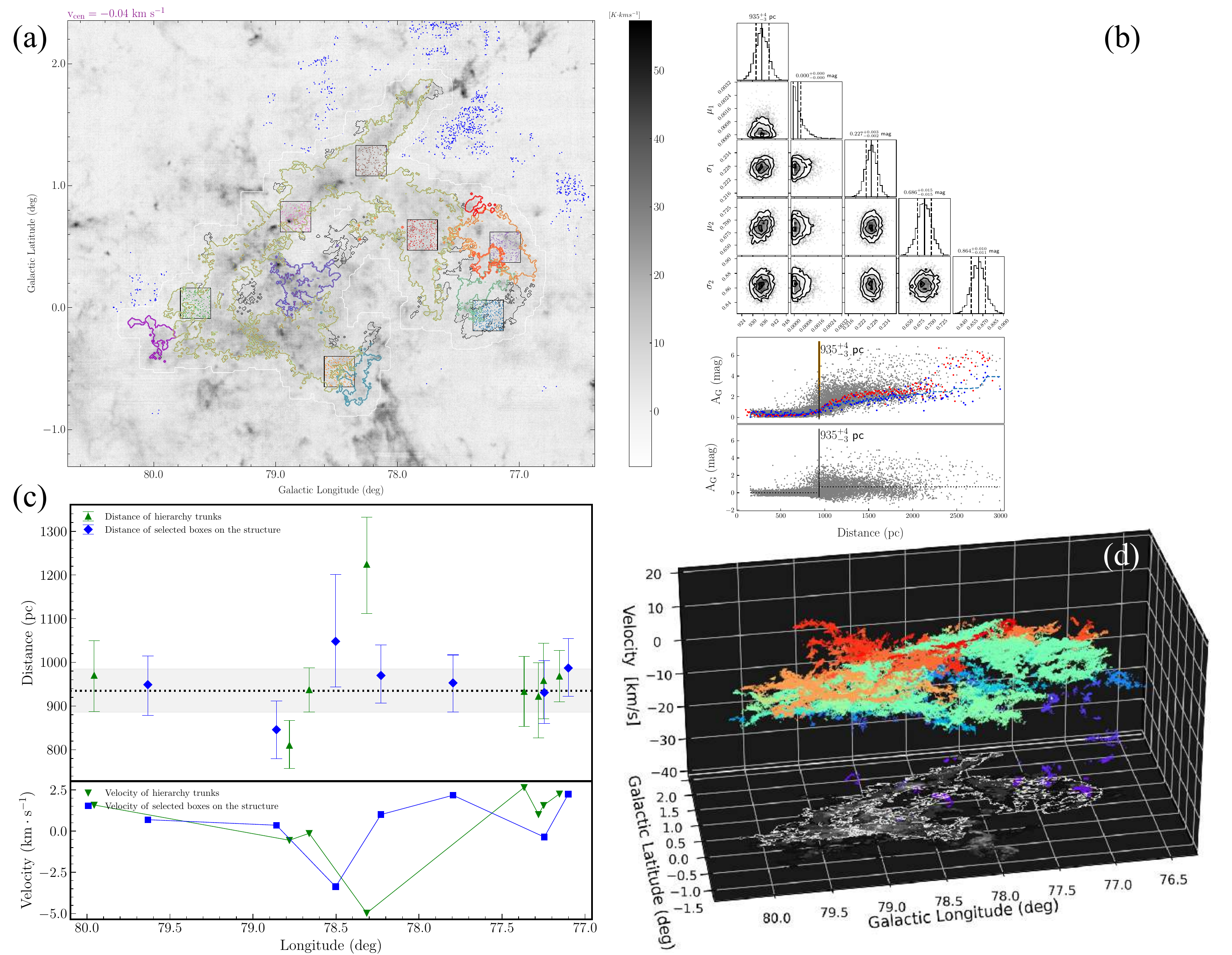}
\caption{The same as Figure \ref{fig:figa1}, but for MC L889 toward the Cygnus X South region. 
\label{fig:figa2}}
\end{figure}
\setcounter{figure}{1}

%\renewcommand\thefigure{\Alph{Section}\arabic{figure}}
%\setcounter{figure}{2}
%\begin{figure}[!htb]
%\centering
%\hspace{0\linewidth}
%\includegraphics[width=1.\linewidth]{figures/fio483857.pdf}
%\caption{The same as Figure A1, but for another structure.
%\label{four in one.}}
%\end{figure}
%\setcounter{figure}{2}

\section{Comparison for different on-cloud stars' selection and MCMC sampling modules} \label{app:appb}
$All$ $on$-$cloud$ $stars ~versus ~stars ~after ~removing ~overlapping ~regions$. 
We test the influence of different on-cloud star samples on the 
measured distances by Model A (see Figure \ref{fig:figb1}). 
In this work, we adopt all stars within the cloud boundary identified by $\rm ^{13}CO$ as on-cloud stars (see Section \ref{subsec:onstar}). Alternatively, we also exclude the stars in the overlapping region. That is, we reject the stars in the black area, and keep those in the red area in Figure \ref{fig:fig4}a. 
%are supposed to give more accurate result, with fewer contamination from neighbours. 

We find that both of on-cloud star samples generally give consistent measurements relative to their uncertainties. 
The distances from overlap-removed stars give overall larger uncertainties due to fewer on-source samples. Additionally, the samples based on overlap-removed stars will decrease the number of clouds with a measured distance (29 cloud structures, about 16\% in all Model A samples). \\

$Emcee ~working ~on ~the ~mock ~data$.  
Aiming to examine how the $emcee$ works on single-jump-point detection, the mock data of extinction and distance of stars from the Cygnus region were randomly generated.  
In the simulation, the preset jump point ($Distance$) is randomly distributed in [500, 2000] pc, while stars samples are uniformly produced in [100, 2500] pc in number density $\rm 0.2~pc^{-1}$. $\Delta{A_{G}}$ (i.e. $\mu_{2}-\mu_{1}$) follows a truncated exponential distribution from 0.2 mag and a rate parameter of 0.5 mag. To simplify the simulation, we fix $\sigma_{1}$ (0.2 mag) and $\sigma_{2}$ (0.4 mag). Based on star samples toward the Cygnus region, the median error over distance and dispersion of errors are set to 0.1 and 0.06, respectively. And the mean error of $A_{G}$ is set to 0.04 mag, and the dispersion of error of $A_{G}$ is set to 0.04 mag. 200 groups of stars sample (see two examples in Figure \ref{fig:figb2}) were produced.  

We present deviations and uncertainties of distances derived from $emcee$ in a relation with increasing $ \Delta A_{G}$ (see Figure \ref{fig:figb3}). 
Both the deviations from the preset distance and the  uncertainties from the posterior distribution decrease with the increasing $ \Delta A_{G}$. Since the fitted background extinction is likely to be larger for Model B after a background elimination, Model A gives a statistically better fit when they reveal the same jump. The deviations from most samples ($\gtrsim$ 90\%) are smaller than 150 pc, and the uncertainties and deviations are comparable at least in $ \Delta A_{G} \geqslant$ 0.2 mag. These explain why almost all the samples have distance differences within 150 pc between the two models (A and B) when they detect the same jump point (see Section \ref{subsec:undertaintyb}).
It proves that, in current uncertainties of Gaia DR3, the $emcee$ module provides robust results for the cloud distance measurement.

$Comparison ~with ~different ~sampling ~modules ~(emcee ~versus ~Pymc3)$. 
$Pymc3$ \citep{2016ascl.soft10016S} module implicitly builds up a $Theano$ function from the space of our parameters to their posterior probability density up to a constant factor. The probability distribution of $A_{G}$ can be treated as a piecewise function. We use $SWITCH$ function to describe the parameters in different intervals. In the sampling process, we use $Metropolis$ sampler for jump point, and $NUTS$ for the other four parameters (see Section \ref{subsec:distance result}).
After 1000 draws, the returned objects $TRACE$, including posterior predictive samples, are generated from MCMC sampling. The median values from the above sampling are chosen as our results. 

In Figure \ref{fig:figb4}a, we compare the distance measurement 
between $emcee$ and $Pymc3$ for clouds in Table \ref{tab:tab1}. The distances by $emcee$ are in good agreement with the posterior predictive samples produced by $Pymc3$. However, large discrepancies can be seen for some cases ($\sim$10\%). 
%To find out the reason and further assure our results, we design a pseudo data test based on stars parameters in Cygnus. 

Based on the 200 simulated samples (shown in Figure \ref{fig:figb4}b), we find that $Pymc3$ seems to work better than $emcee$ on mock data. 
But the results from $Pymc3$ match the mock data distances so closely with very small error bars, even when the jumps are not clear enough. This is actually caused by overfitting. Some results from $Pymc3$ are not robust due to deviating from the preset distance out of uncertainties from sampling (see a case in blue box in Figure \ref{fig:figb4}b). And a case detects multijump points (see red box in Figure \ref{fig:figb4}b). Samplings from $Pymc3$ are rather sensitive to small variations in $A_{G}$, and the detected jumps often deviate from the true value (by $emcee$ or human judge). We thus choose $emcee$ because it gives a more robust sampling result in application to real data by considering baseline fluctuations and $A_{G}$ dispersion uncertainties. 
%Knowing the random factors (baseline fluctuations, $A_{\rm G}$ dispersion changes) introduce more complicated situations, and emcee often gives a more robust sampling result, or an averaged one. 

%We notice the posterior samples from \textbf{Pymc3} are very sharp, those with larger uncertainties often show multiple jump points, probably be more sensitive to the number of jump points, in a way it's suitable for direct on-cloud stars' jump point detection, but not after background elimination. Some results are not robust, as it deviate from preset distance out of uncertainties from sampling. \textbf{emcee} give robust results with larger uncertainties, we choose this module because in application to real data, the random factors (baseline fluctuations, $A_{\rm G}$ dispersion changes) introduce more complicated situations, and emcee often gives a more robust sampling result, or an averaged one. 

\renewcommand\thefigure{\Alph{section}\arabic{figure}}
\setcounter{figure}{0}
\begin{figure}[!htb]
\centering
\hspace{0\linewidth}
\includegraphics[width=1.\linewidth]{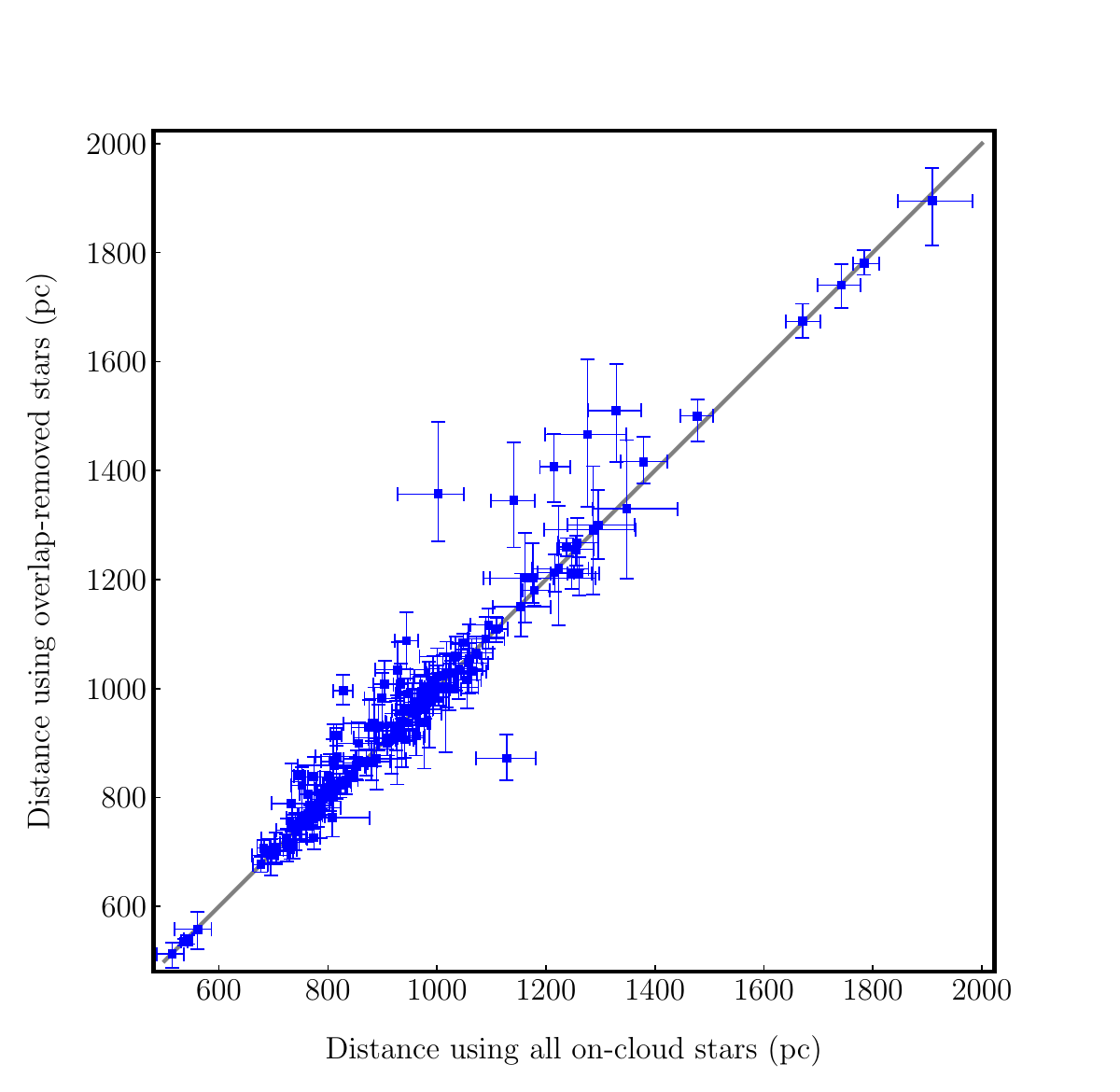}
\caption{All on-cloud stars (red+black region in Figure \ref{fig:fig4}) vs. overlap-removed on-cloud stars (red region in Figure \ref{fig:fig4}). 
\label{fig:figb1}}
\end{figure}
\setcounter{figure}{0}

\renewcommand\thefigure{\Alph{section}\arabic{figure}}
\setcounter{figure}{1}
\begin{figure*}
\gridline{\fig{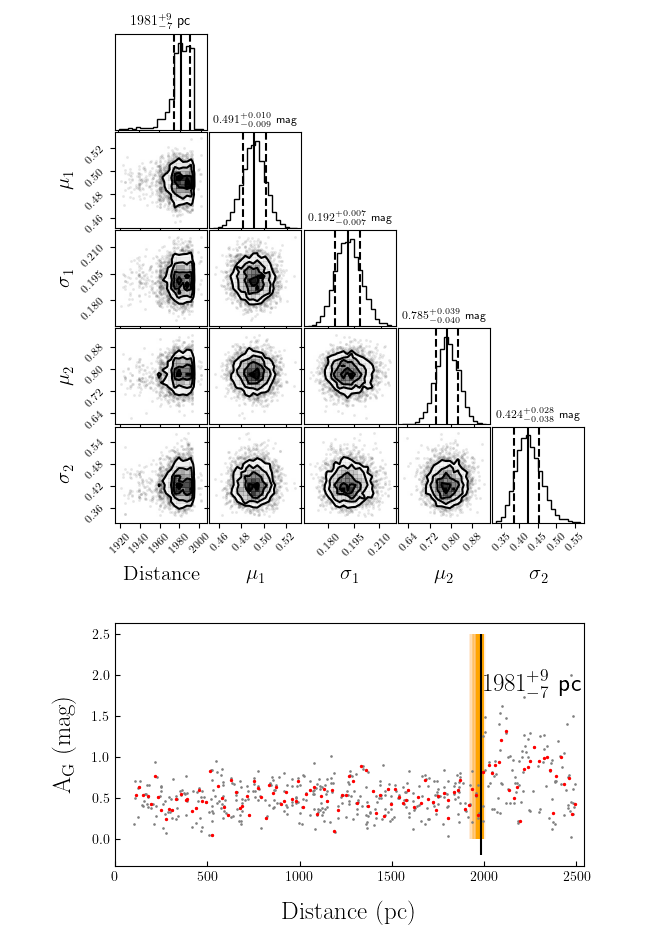}
{0.48\textwidth}{(a)}
          \fig{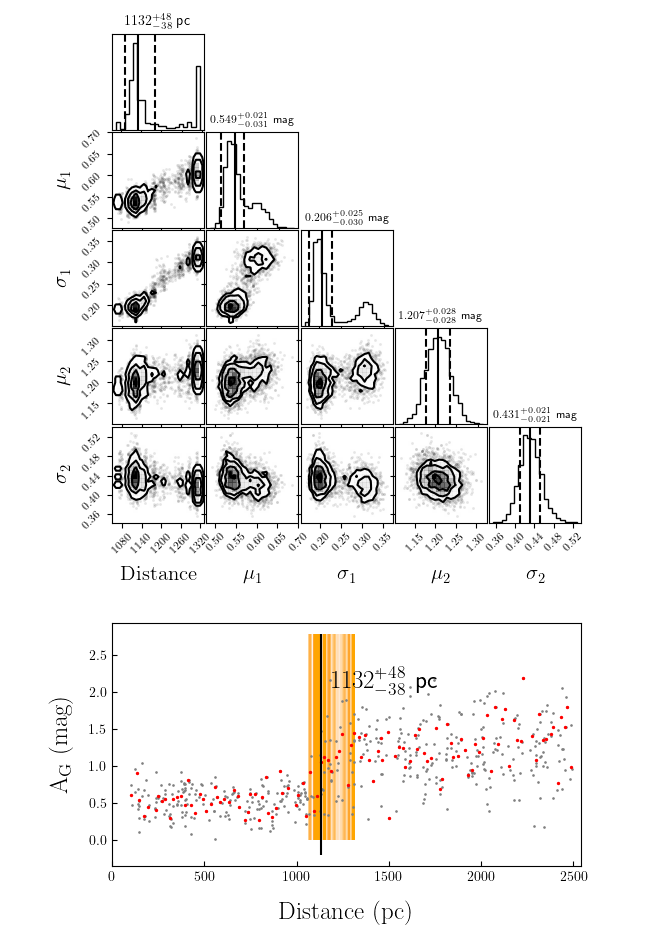}
          {0.48\textwidth}{(b)}
          }
\caption{Two examples of measuring distance with mock data based on the Bayes model. Details in the panels can be seen in the caption of Figure \ref{fig:fig4}, but points in lower panel are for the mock data. 
\label{fig:figb2}}
\end{figure*}
\setcounter{figure}{1}

\renewcommand\thefigure{\Alph{section}\arabic{figure}}
\setcounter{figure}{2}
\begin{figure*}
\gridline{\fig{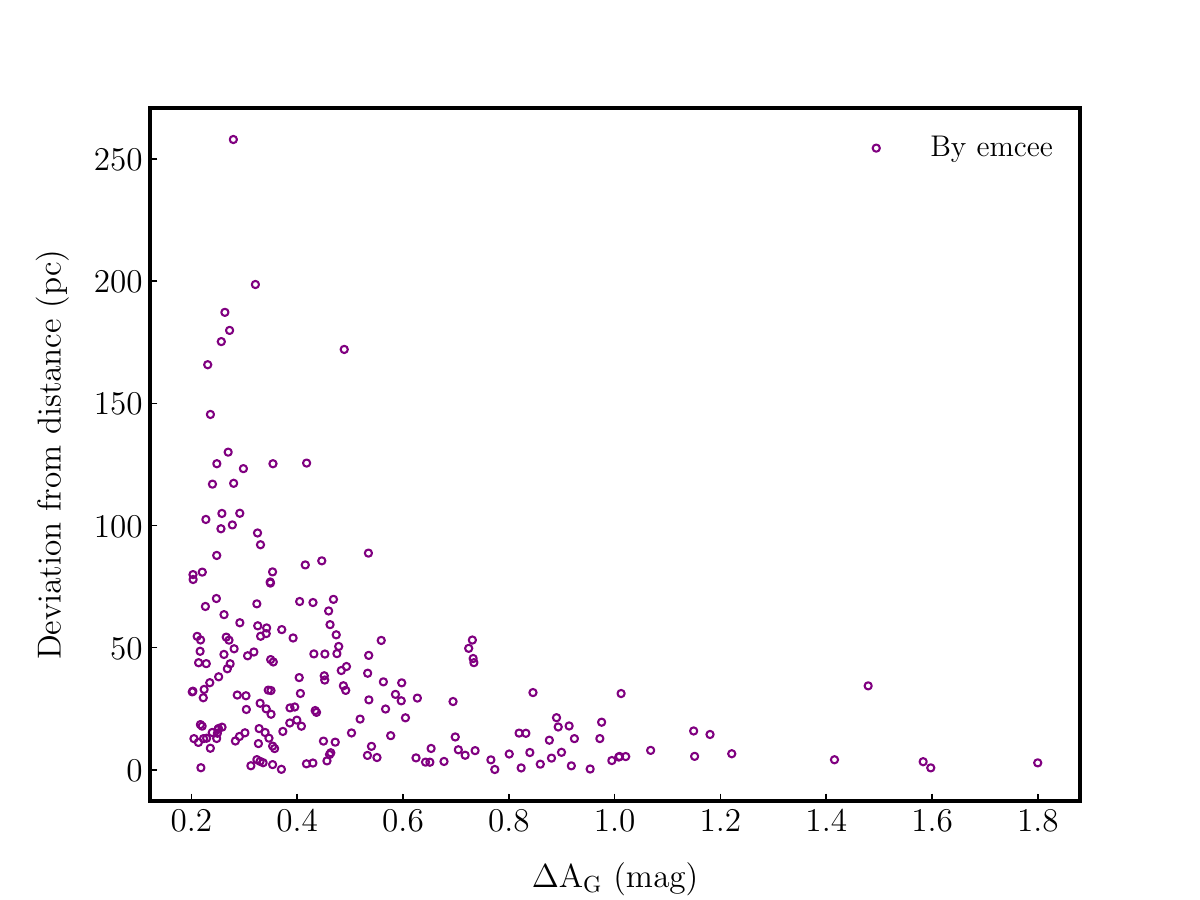}{0.48\textwidth}{(a)}
          \fig{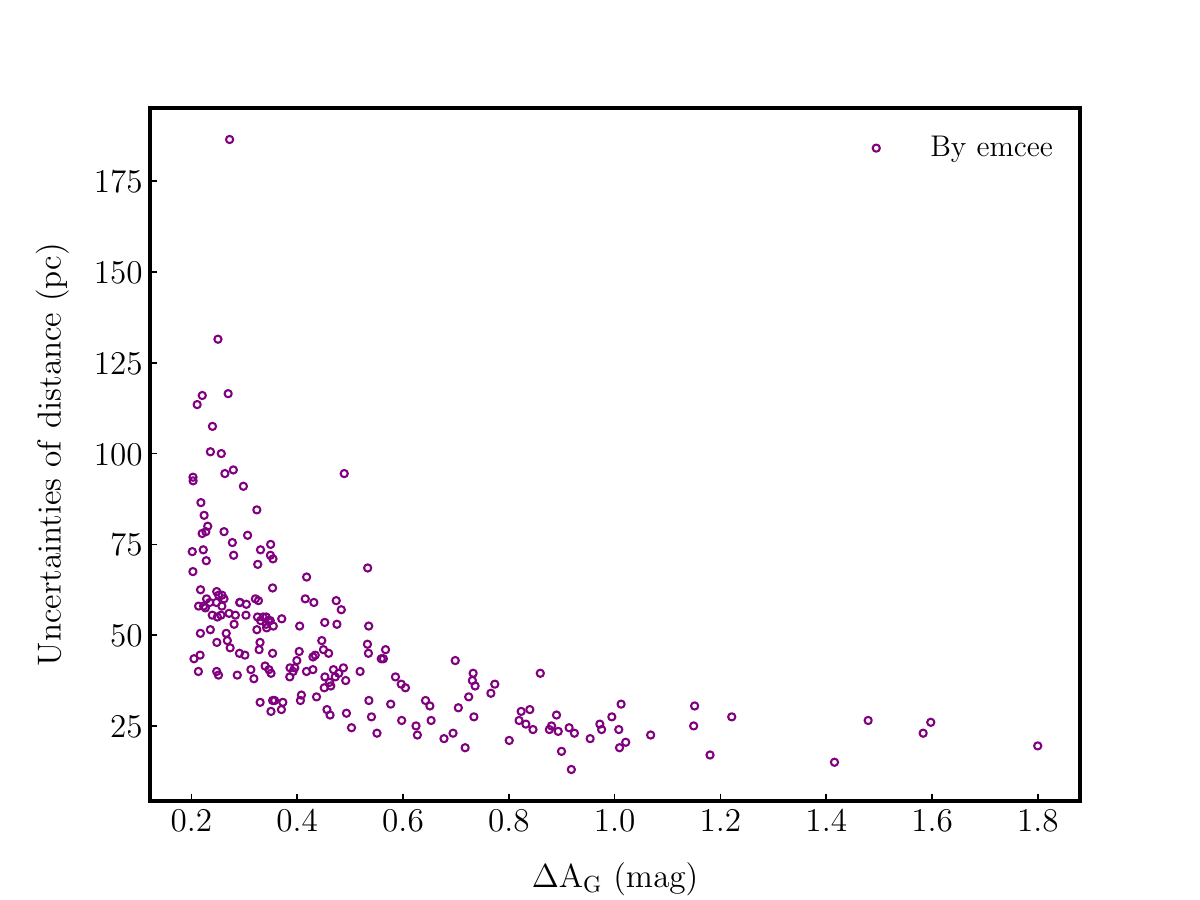}{0.48\textwidth}{(b)}
          }
\caption{  
(a) Deviation between distances derived from emcee and preset distances in a relation with increasing $ \Delta A_{G}$. 
(b) Uncertainties of distances derived from emcee in a relation with increasing $ \Delta A_{G}$. 
\label{fig:figb3}}
\end{figure*}
\setcounter{figure}{2} 

\renewcommand\thefigure{\Alph{section}\arabic{figure}}
\setcounter{figure}{3}
\begin{figure*}
\gridline{\fig{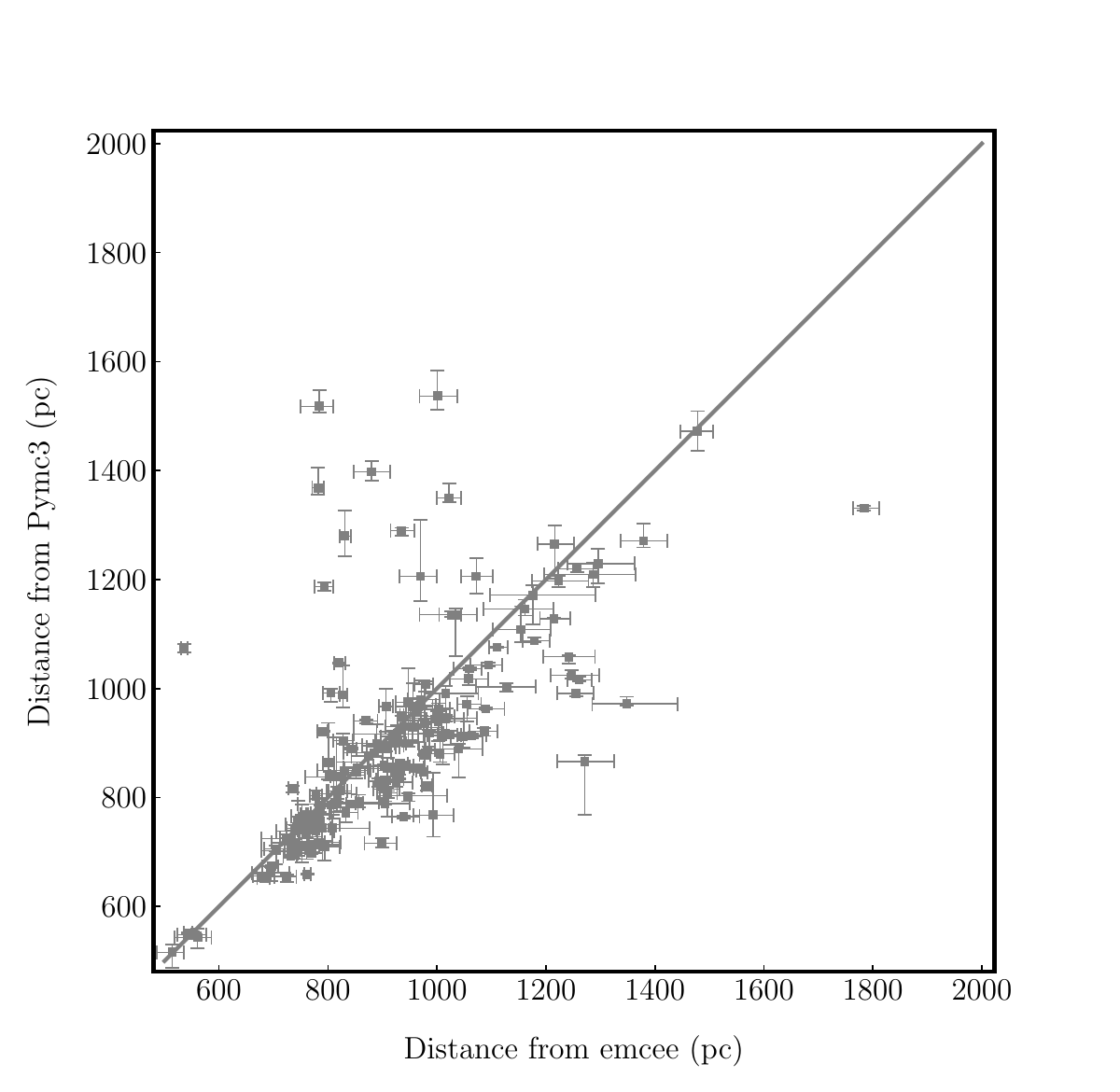}{0.48\textwidth}{(a)}
          \fig{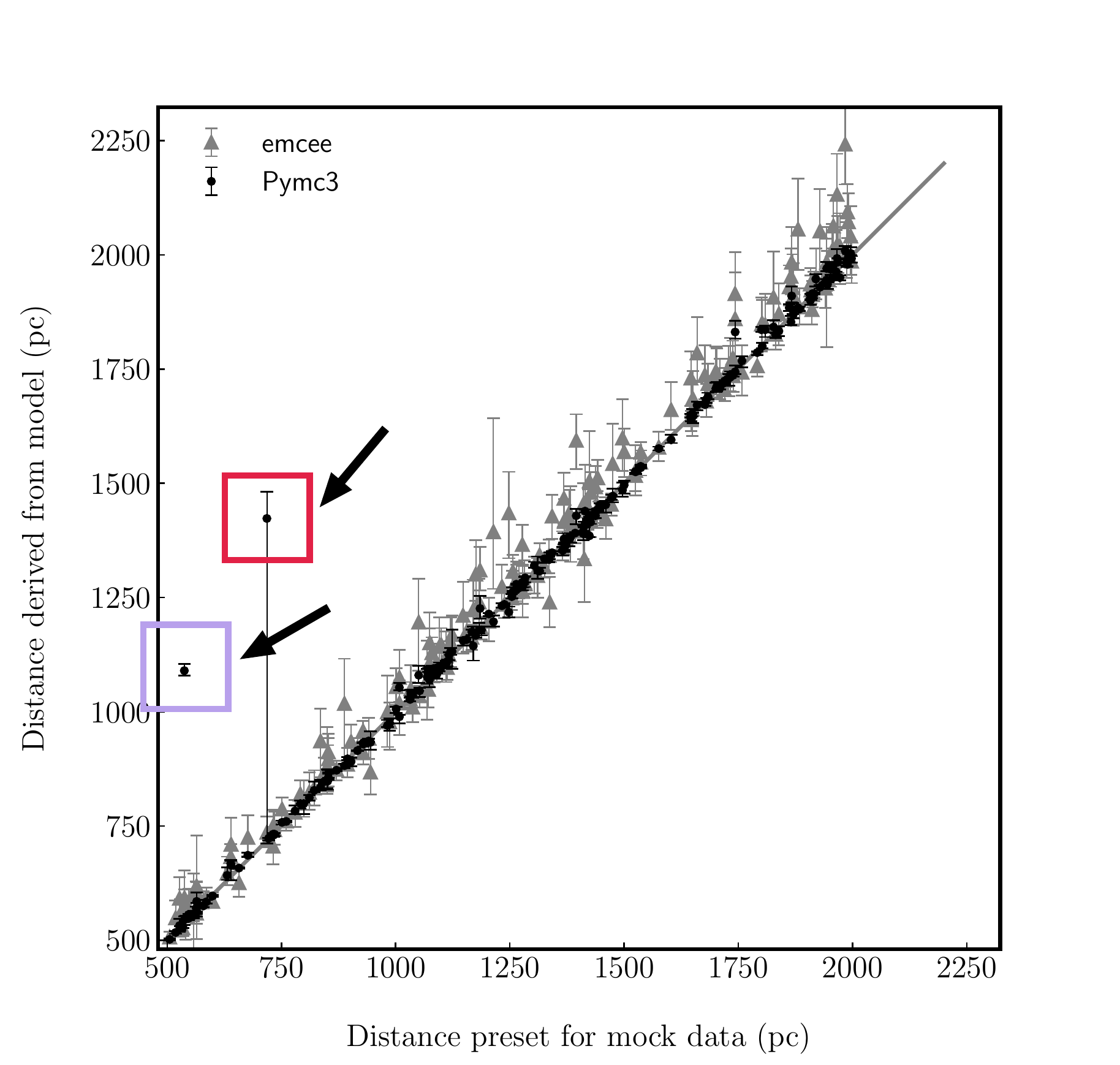}{0.48\textwidth}{(b)}
          }
\caption{Comparison of distance measurements with different sampling modules. 
(a) The emcee module vs. Pymc3 module in our samples.
(b) Distance derived from different modules vs. preset distance in mock data. 
\label{fig:figb4}}
\end{figure*}
\setcounter{figure}{3}

\section{star distances estimated from parallaxes} \label{app:appc}
In this work, we simply use MCMC sampling from the parallaxes of stars to estimate their distances. 
In comparison, we follow the method of \citet{2015PASP..127..994B}, \citet{2018A&A...616A...9L} to calculate median/uncertainty of the posterior based on the exponentially decreasing space density (EDSD) prior, the posterior of the UD prior, the method using naive inverse, and the transformation method \citep{1996MNRAS.281..211S}. 

The likelihood function of parallax ($\varpi$), distance ($r$), and uncertainty ($\sigma$) is 
\begin{equation}
P(\varpi \mid r, \sigma)=\frac{1}{\sqrt{2 \pi} \sigma} \exp \left(-\frac{1}{2 \sigma^2}\left(\varpi-\frac{1}{r}\right)^2\right)
\end{equation}

The transformation Methods are applied by modifying the parallax $\varpi$: \\
\begin{equation}
    \quad r^*=\frac{1}{\varpi^*}, \quad  
     \varpi^*=\beta \sigma \phi g_\phi  \\
\end{equation}

where
\begin{equation} 
    \phi=\frac{1}{0.8} \ln \left(1+e^{\frac{0.8 \varpi}{\sigma}}\right) \\
\end{equation}

and 
\begin{equation}
\left\{\begin{array}{l}
        g_\phi=1, \quad \varpi>0,  \\ 
        g_\phi=e^{-0.605 \frac{\varpi^2}{\sigma^2}}, \quad \varpi \leq 0, 
        \quad \beta=1.01
\end{array}\right.
\end{equation}

Therefore, the distance module $\mu^*$ can be defined as 
\begin{equation}
\mu^*=m-\hat{M}=-(5 \log (\hat{\varpi})+5) \quad \\
\end{equation}

Finally, the modified parallax $\varpi$ satisfy 
\begin{equation}
\hat{\varpi}=\beta \sigma\left(\frac{1}{e^\phi+e^{\frac{-5 \omega}{\sigma}}}+e^\phi\right)
\end{equation}

Because our star samples are in a smaller parallax uncertainties ($\leqslant$ 20\%), the discrepancies between different methods can be ignored. Our distances are just like the naive inverse from parallaxes, but the median value and uncertainty are reproduced when we calculate the distance. The correlation of median distances and errors between different methods are presented in Figure \ref{fig:figc1}.

\renewcommand\thefigure{\Alph{section}\arabic{figure}}
\setcounter{figure}{0}
\begin{figure*}

\gridline{\fig{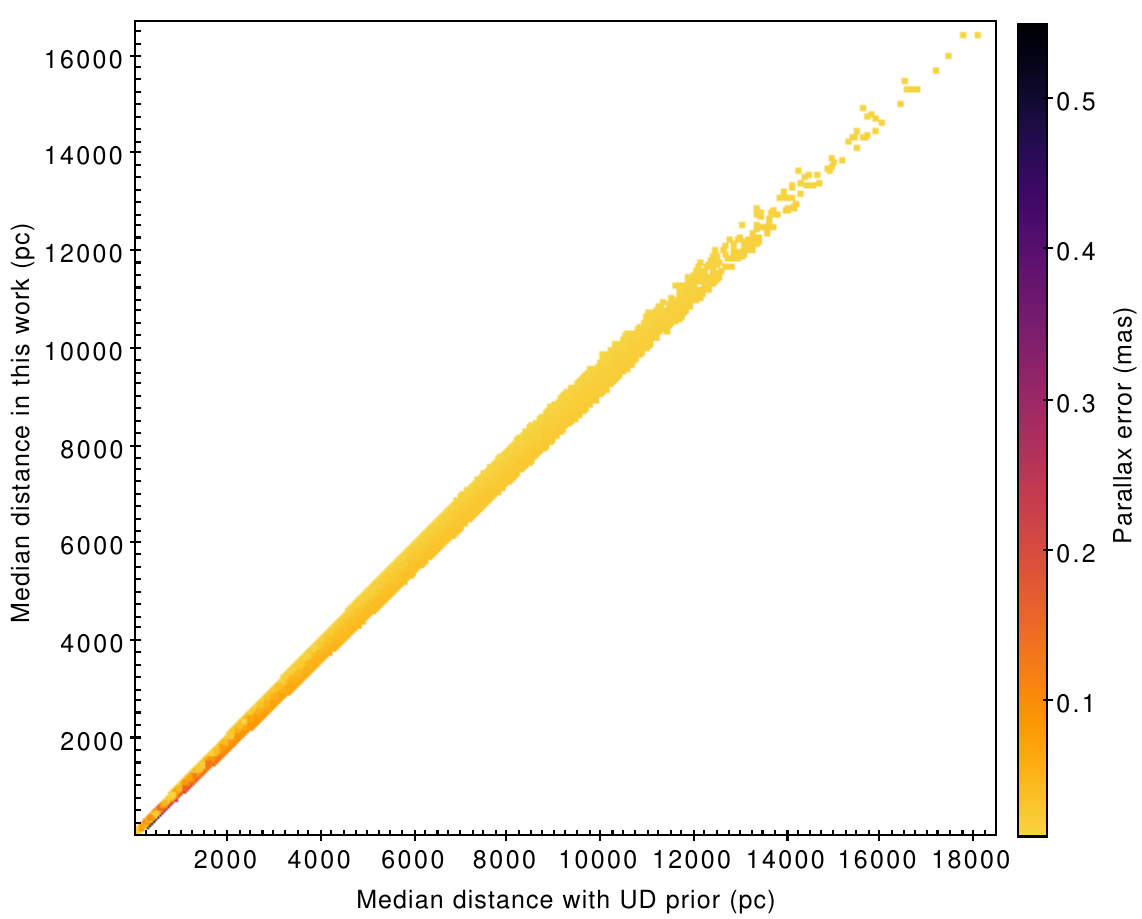}{0.48\textwidth}{(a)}
          \fig{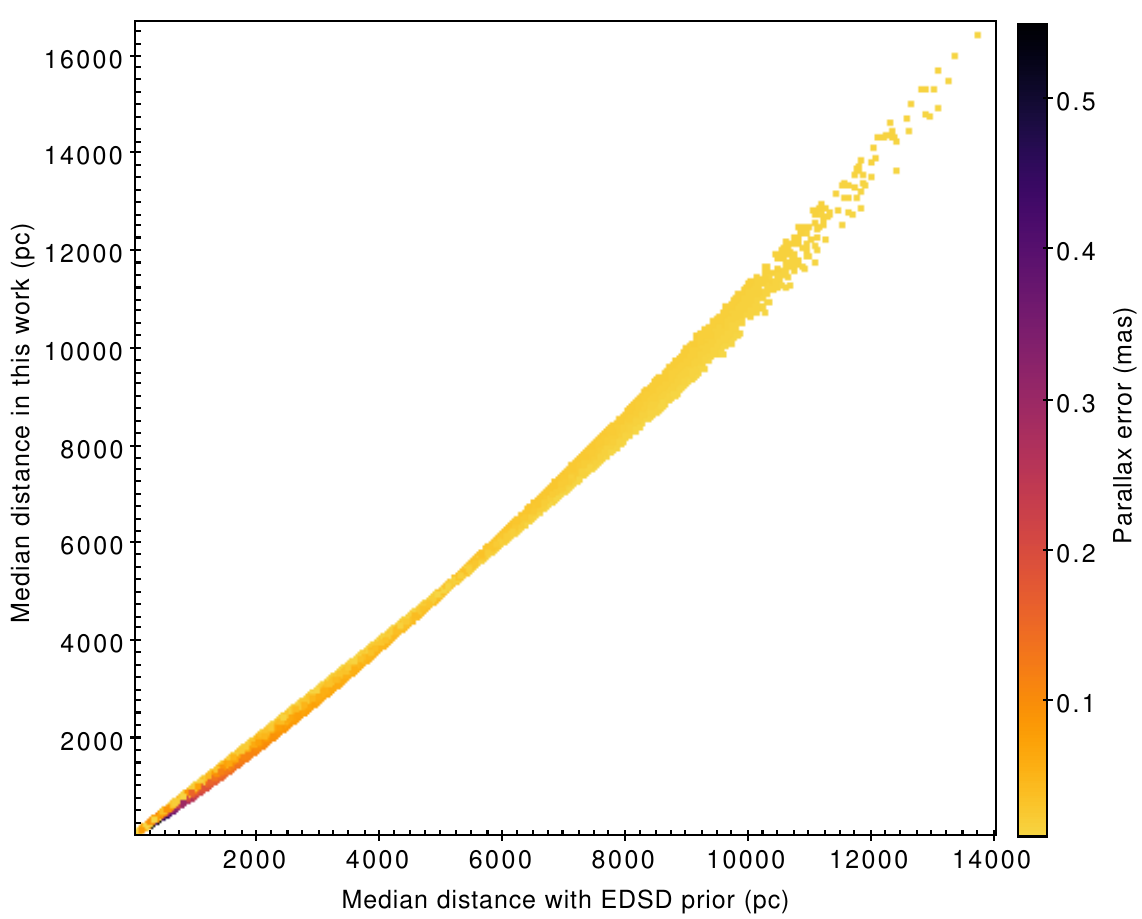}{0.48\textwidth}{(b)}
          }
\gridline{\fig{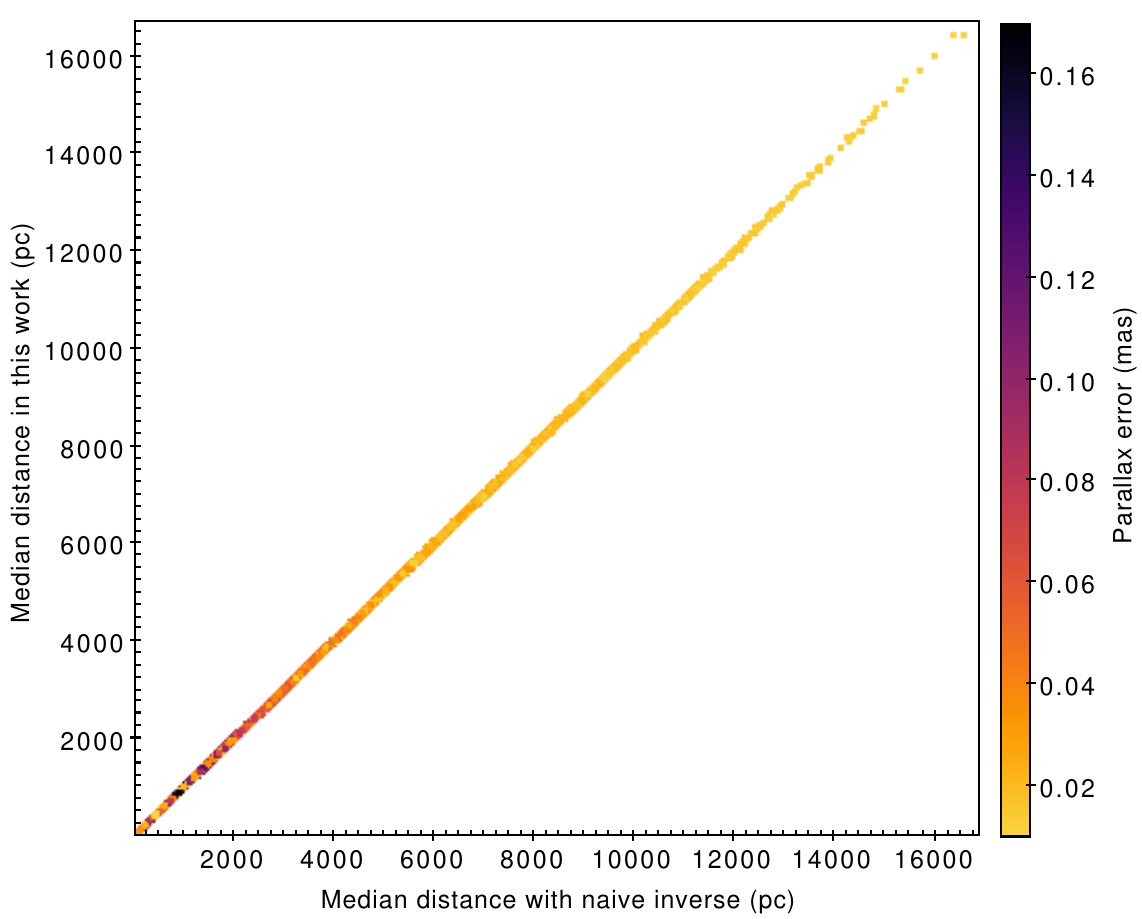}{0.48\textwidth}{(c)}
          \fig{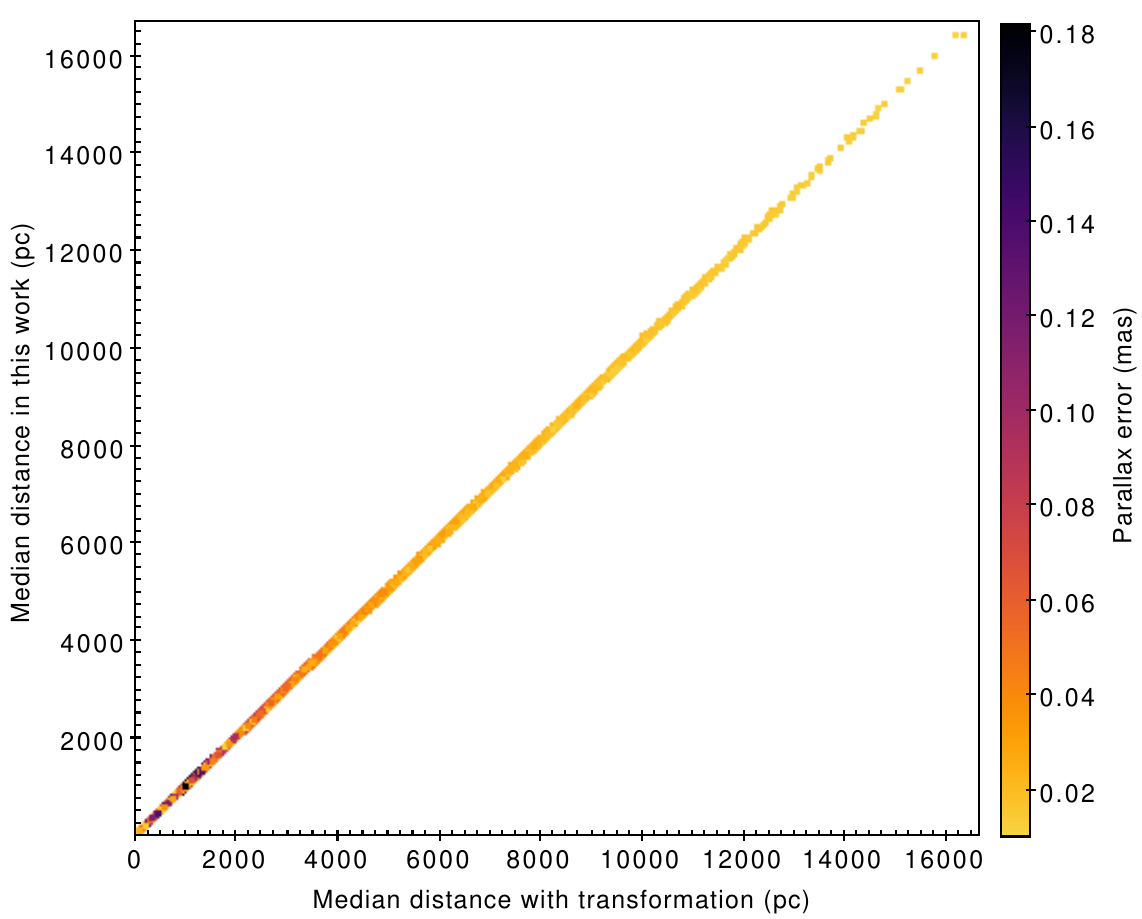}{0.48\textwidth}{(d)}
          }
\gridline{\fig{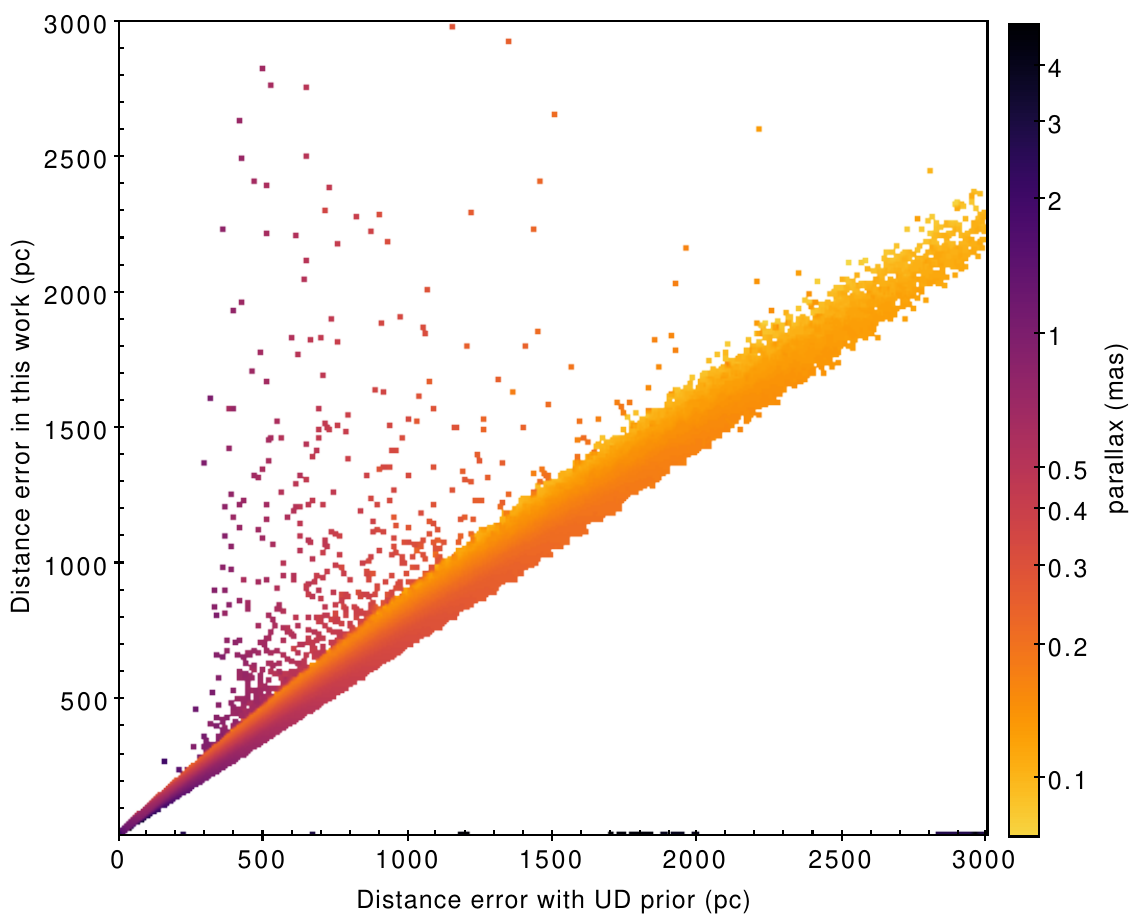}{0.48\textwidth}{(e)}
          \fig{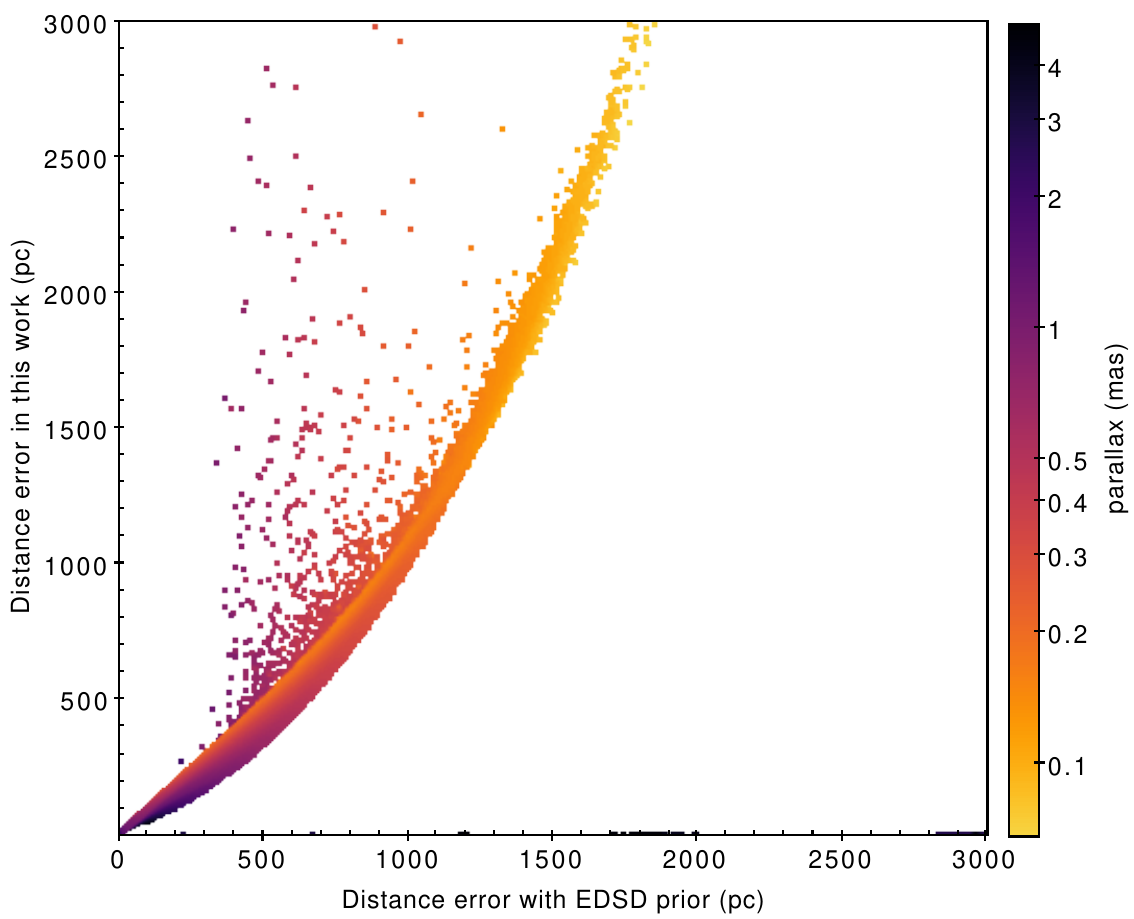}{0.48\textwidth}{(f)}
          }
\caption{The median and uncertainty of distances derived by different methods compared to this work.
\label{fig:figc1}}
\end{figure*}
\setcounter{figure}{0}

\section{Individual cases in complicated extinction environment} \label{app:appd}
In addition to the robust distance result confirmed by both Models A and B (class I), some cases failed to include enough field stars in Model A. For these clouds in the complicated extinction environment, we successfully measured distances of 22 special cases by Model B because of the prominent extinction on the main body of the clouds. On the other hand, referenced stars near the measured clouds (see orange region in Figure \ref{fig:fig4}b) may share the same but smaller background extinction along the LOS (see Figure \ref{fig:fig6}b). We think distance measurements for these samples alone by Model B are also reliable.

From Figure \ref{fig:figd1}, we can see these clouds match well with the identified gas layers in the text. %More clouds are confirmed to be in 1.3 kpc layer, and it suggests clouds in 1.3 kpc are distributed more extended, especially in Cygnus South. 
More clouds with large-scale extensions are confirmed to be in the 1.3 kpc layer. In particular, toward the Cygnus X South region, these clouds have more of a contribution on masses, which is consistent with the 1.35 kpc overdensity in extinction of the Cygnus X South region \citep{2022A&A...658A.166D}.

\renewcommand\thefigure{\Alph{section}\arabic{figure}}
\setcounter{figure}{0}
\begin{figure}[htbp]
%\plotone{c311.pdf}
\centering 
\includegraphics[width=18cm,angle=0]{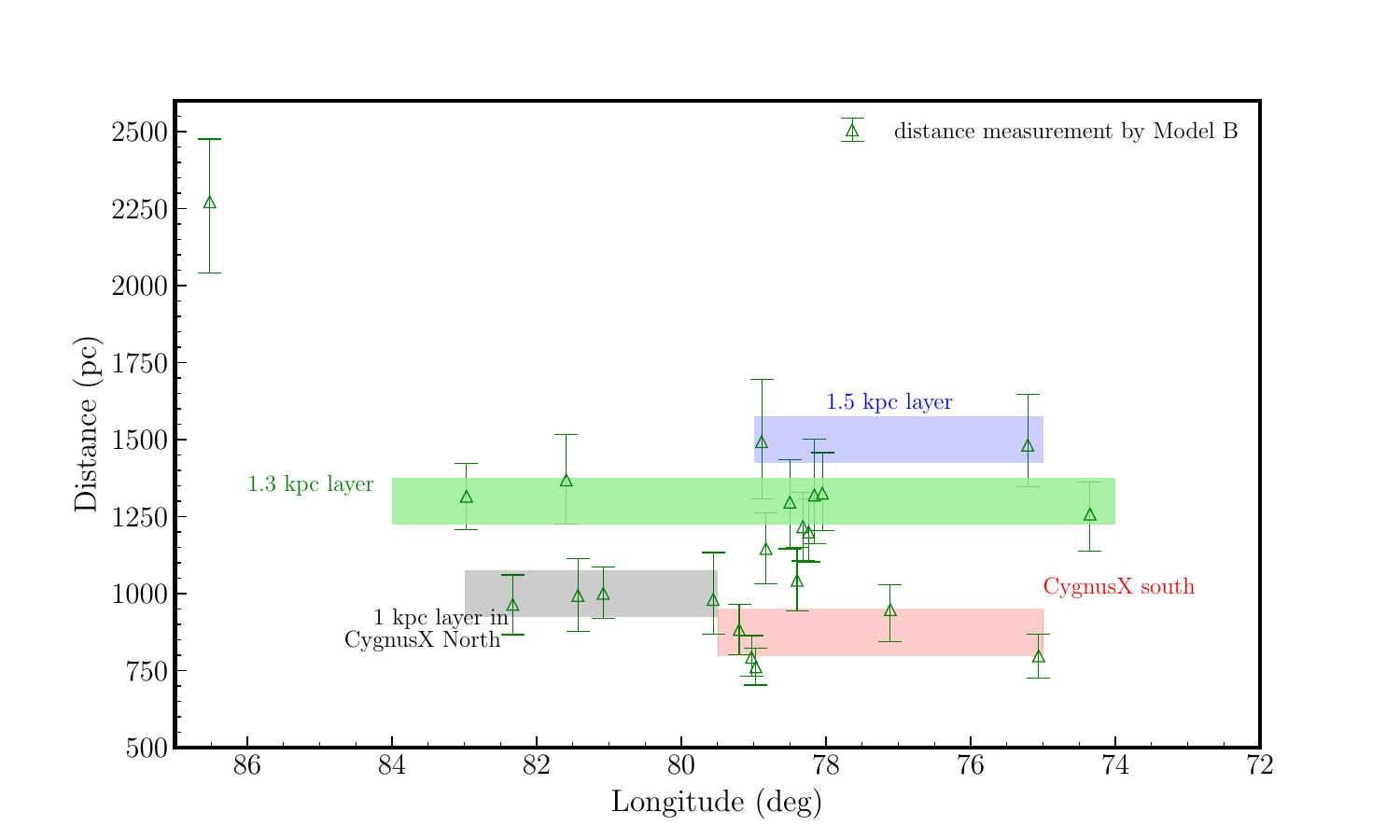}
\caption{Distances measured by Model B for those 22 clouds in complicated extinction environments. The colorful bands denote gas layers we identified previously. \label{fig:figd1}}
\end{figure}
\setcounter{figure}{0}

\section{Retrieve part of flux loss during clustering}
\label{app:appe}

When we make a reconstruction using fully fledged clusters from ACORNS, a very small portion of the flux missed in the process of clustering. See panels (a) and (b) in Figure \ref{fig:fig2}; the flux loss indicates some Gaussian components are dropped after clustering. We found the dropped components are mainly from three parts. The first is due to large amplitude error from Gaussian fitting. These components are excluded from catalog before the first loop in clustering. We do not retrieve those components, as it has little influence to the main structure of cloud. The second part is from the components isolated in spatial Euclidean distance or velocity. They could be false identifications by Gaussian fitting, while some are probably from line wings with smaller amplitudes. We drop them because they fail to meet our criteria for fledged clusters. Finally, some extra components are difficult for clustering because of the variance of centroid velocity and line width. We find these unassigned components are originated from Gaussian fitting and clustering processes.
%keep them away from the hierarchy. 
For some pixels, the overdecomposition in GaussPy+ might lead to failure of clustering with a uniform criteria. 
%in areas with multiple components, when the line profile appears to be complex (with many spikes), it's hard to judge whether it is multiple components or self absorption. 
%And to achieve the smallest Chi-square value of residuals of fitting, automatic fitting tools tend to  though spatial coherence is applied. 

%based on the premise that we accept the present situation of our data and Gaussian decomposition result, since we don't know the reason of variations to main structure near them, we don't do extra reduction to data cube and GaussPy+. 
%We don't do extra reduction to raw data cube and GaussPy+ results. 
We try to retrieve those unassigned components by adding another loop in the clustering process. Fully fledged clusters are already formed during the first three loops in the clustering process, so further development of the hierarchy is avoided. Next, we relax the criteria of velocity and line width to a large extent before the last loop, and remove the extra conditions when linking clusters. By using the same strategies of finding the most similar cluster in velocity and line width statistically, those unassigned components are merged into the nearest level of the hierarchy. To avoid developing the hierarchy and merging existing clusters together, we skip the ``resolving ambiguities'' method of ACORNS in the final loop we added, but restore the components with the smallest variation in equal weight of velocity and line width. Finally, an additional 1.5\% flux is retrieved in our reconstruction map.

\section{YSO candidates toward the Cygnus Region}
\label{app:appf}
Our YSO candidates are selected based on infrared data collected from three surveys: the Spitzer Cygnus-X Legacy Survey \citep[CXLS;][] {2009AAS...21335601H}, the GLIMPSE360 program \citep{2008sptz.prop60020W}, and the Wide-field Infrared Survey Explorer \citep[WISE;][]{2010AJ....140.1868W} survey. We summarise the selection criteria here briefly for each data set, and the analysis of these YSOs will be presented in detail in a future paper (X.-L. Wang, et al. in preparation).

For the CXLS dataset, we select YSO candidates following the prescription in \citet{2009ApJS..184...18G}, but update the criteria with the new extinction law from \citet{2019ApJ...877..116W}. The new criteria are \\
(1) $[I2-I3]>0.7$ and $[I1-I2]>0.7$; \\
(2) $[I2-I4]-\sigma_{[I2-I4]}>0.5$, $[I1-I3]-\sigma_{[I1-I3]}>0.35$, $[I1-I2]-\sigma_{[I1-I2]}>0.15$, and $[I1-I3]-\sigma_{[I1-I3]}\leq18\times([I2-I4]-\sigma_{[I2-I4]}-0.5)+0.5$; \\
(3) $[K-I1]_{0}-\sigma_{[K-I1]}>0$, $[I1-I2]_{0}-\sigma_{[I1-I21]}>0.101$, and $[K-I1]_{0}-\sigma_{[K-I1]}>-2.85714\times([I1-I2]_{0}-\sigma_{[I1-I2]}-0.101)+0.5$; \\
(4) $[I3-M1]>2.5$ for sources without $I4$ measurements;  \\
(5) $[I2-M1]>2.5$ for sources with no $I3$ and $I4$ measurements.

With these criteria, we selected 24,757 as YSO candidates.

For data from the GLIMPSE360 program, we select YSO candidates following \citet{2020AJ....160...68W}; again, the extinction law is updated with the \citet{2019ApJ...877..116W} law. The new criteria are \\
(1) $0.9375\times([J-H]-0.6+\sigma_{[J-H]})+1.0+\sigma_{[H-I2]}<[H-I2]$ and $[J-H]>0$;\\ 
(2) $0.9811\times([H-K]+\sigma_{[H-K]})+0.4+\sigma_{[K-I2]}<[K-I2]$, $[H-K]>0$, and $[K-I2]>0.2+\sigma_{K-I2}$; \\
(3) $[I1-I2]_{0}-\sigma_{[I1-I2]}>0$, $[K-I1]_{0}-\sigma_{[K-I1]}>0.2\times[I1-I2]_{0}+0.3$, and $[K-I1]_{0}-\sigma_{[K-I1]}>-([I1-I2]_{0}-\sigma_{[I1-I2]})+0.8$.

With the above criteria, we select 6168 candidates.

For the WISE data, only the first two channels are used in the selection procedure. Since the first two WISE channels have a similar central wavelength as the first two IRAC bands, we utilize similar criteria as for the GLIMPSE360 data, but replace $I1$ with $W1$ and $I2$ with $W2$. The selection criteria are \\
(1) $0.9375\times([J-H]-0.6+\sigma_{[J-H]})+1.0+\sigma_{[H-W2]}<[H-W2]$ and $[J-H]>0$; \\
(2) $0.9811\times([H-K]+\sigma_{[H-K]})+0.4+\sigma_{[K-W2]}<[K-W2]$, $[H-K]>0$, and $[K-W2]>0.2+\sigma_{[K-W2]}$; \\
(3) $[W1-W2]_{0}-\sigma_{[W1-W2]}>0$, $[K-W1]_{0}-\sigma_{[K-W1]}>0.2\times[W1-W2]_{0}+0.3$, and $[K-W1]_{0}-\sigma_{[K-W1]}>-([W1-W2]_{0}-\sigma_{[W1-W2]})+0.8$.

With the above criteria, we select 25,853 candidates.

These three lists of YSO candidates are combined, and the duplicates are removed, resulting in a total of 50,101 YSO candidates. In the current work, we are comparing the distribution of YSOs with $\rm ^{13}CO$ traced molecular gas (as displayed in Figure \ref{fig:figf1} and details in discussion). Therefore, we focus on the subsample (22,158 candidates) that have Gaia parallaxes. Following \citet{2020ApJ...899..128K}, possible giant star contaminants are removed, leaving 19,220 candidates in our FOV. Finally, 11,244 candidates are found to be within the interval [0.5, 3] kpc.

\renewcommand\thefigure{\Alph{section}\arabic{figure}}
\setcounter{figure}{0}
\begin{figure}[htbp]
%\plotone{c311.pdf}
\centering 
\includegraphics[width=18cm,angle=0]{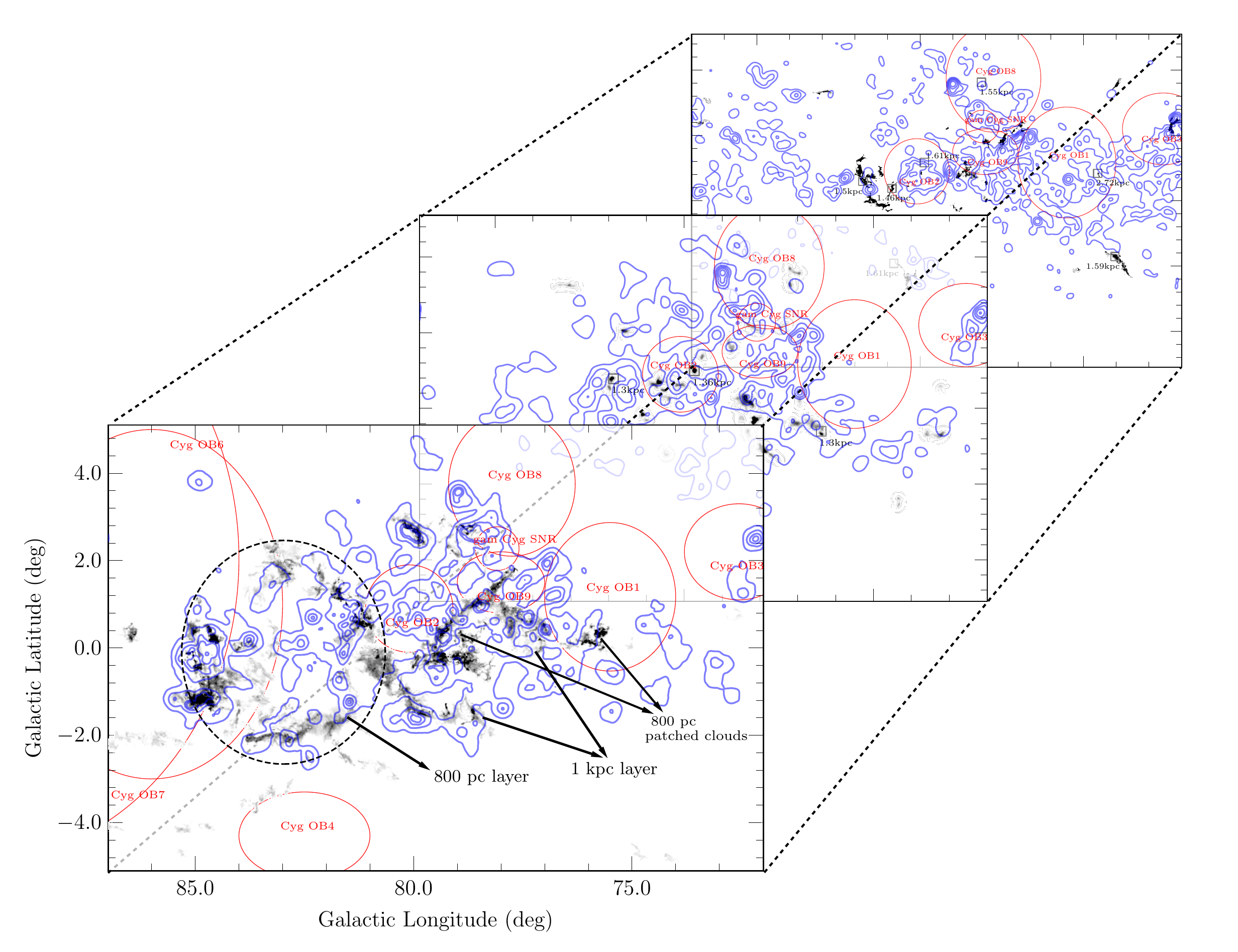}
\caption{The multiple layers of gas structure with YSO candidates toward the Cygnus region. Similar to Figure \ref{fig:fig17}, gray-scale maps are reconstructed by identified MC structures. The dashed ellipse shows a large-scale molecular loop with diameter of $\sim$56 pc at a distance of 800 pc. Red ellipses show OB associations and SNR, while boxes show the masers in Table \ref{tab:tab2}. The overdensity of YSOs' candidates is shown as blue contours in each interval. The levels in different intervals are presented as follows: 3$\sigma$ to 99.5\% of the maximum value with an increased step of 25\% in a logarithm scale for [500, 1000] pc. 3$\sigma$ to 99.5\% of the maximum value with an increased step of 20\% in a logarithm scale for [1000, 1400] pc. 10$\sigma$ to 99.5\% of the maximum value with an increased step of 20\% in a logarithm scale for $\gtrsim$1400 pc. $\sigma$ ($\sim$15 $\rm deg^{-1}$) is the noise level of equivalent density map for each interval. \label{fig:figf1}}
\end{figure}
\setcounter{figure}{0}

\bibliography{reference2}{}
\bibliographystyle{aasjournal}



\end{document}